\documentclass[useAMS,usenatbib]{mn2e}
\usepackage[pdftex]{graphicx}
\usepackage{txfonts}

\usepackage{epstopdf}
\usepackage{url}
\usepackage{subfig}
\usepackage{longtable,lscape}

\usepackage[amssymb,mediumspace,thickqspace]{SIunits}
\addunit{\mol}{mol}
\addunit{\erg}{erg}
\addunit{\cm}{cm}
\addunit{\gramm}{g}

\def\arcsec{\hbox{$^{\prime\prime}$}}

\newcommand{\um}{\,\micro\metre}
\newcommand{\umm}{\,\micro\metre\,}

\newcommand{\Rx}{R^\mathrm{R}_\mathrm{X}}

\newcommand{\nh}{N_\mathrm{H}}

\newcommand{\fnn}{F^\mathrm{nuc}(12\,\mu\mathrm{m})}
\newcommand{\fspft}{F^\mathrm{Spi}(15\,\mu\mathrm{m})}
\newcommand{\lnn}{L^\mathrm{nuc}(12\,\mu\mathrm{m})}
\newcommand{\lmt}{L^\mathrm{tot}(5.8\,\mu\mathrm{m})}
\newcommand{\lsf}{L^\mathrm{Spi}(15\,\mu\mathrm{m})}
\newcommand{\lnt}{L^\mathrm{tot}(12\,\mu\mathrm{m})}

\newcommand{\fqn}{F^\mathrm{nuc}(18\,\mu\mathrm{m})}
\newcommand{\lqn}{L^\mathrm{nuc}(18\,\mu\mathrm{m})}

\newcommand{\lx}{L_\mathrm{X}}
\newcommand{\lxi}{L^\mathrm{int}(\textrm{2-10\,keV})}
\newcommand{\fxi}{F^\mathrm{int}(\textrm{2-10\,keV})}
\newcommand{\lxo}{L^\mathrm{obs}(\textrm{2-10\,keV})}
\newcommand{\fxo}{F^\mathrm{obs}(\textrm{2-10\,keV})}
\newcommand{\lxh}{L^\mathrm{obs}(\textrm{14-195\,keV})}
\newcommand{\lxhi}{L^\mathrm{obs}(\textrm{17-60\,keV})}
\newcommand{\fxh}{F^\mathrm{obs}(\textrm{14-195\,keV})}

\newcommand{\lfive}{L(\textrm{4.9\,GHz})}
\newcommand{\lone}{L(\textrm{1.4\,GHz})}
\newcommand{\lpeight}{L(\textrm{0.8\,GHz})}

\newcommand{\ratmx}{R^\mathrm{M}_\mathrm{X}}

\newcommand{\aratmx}{\langle R^\mathrm{M}_\mathrm{X}\rangle}

\newcommand{\sigmx}{\sigma^\mathrm{M}_\mathrm{X}}
\newcommand{\sigint}{\sigma_\textrm{int}}

\newcommand{\dks}{D_\mathrm{KS}}
\newcommand{\pks}{p_\mathrm{KS}}

\newcommand{\rhoS}{\rho_\mathrm{S}}
\newcommand{\rhop}{\rho_\mathrm{p}}
\newcommand{\pS}{p_\mathrm{S}}

\newcommand{\spitzerr}{{\it Spitzer}\ }                 % w/ space behind
\newcommand{\spitzer}{{\it Spitzer}}                    % w/o space behind

\newcommand{\akarii}{{\it AKARI}\ }
\newcommand{\akari}{{\it AKARI}}
\newcommand{\irass}{{\it IRAS}\ }
\newcommand{\iras}{{\it IRAS}}
\newcommand{\isoo}{{\it ISO}\ }

\newcommand{\xmm}{\textit{XMM-Newton}}

\newcommand{\asca}{\textit{ASCA}}
\newcommand{\chandra}{\textit{Chandra}}
\newcommand{\suzaku}{\textit{Suzaku}}
\newcommand{\beppo}{\textit{BeppoSAX}}
\newcommand{\integral}{\textit{INTEGRAL}}
\newcommand{\xmmm}{\textit{XMM-Newton}\ }

\newcommand{\ascaa}{\textit{ASCA}\ }
\newcommand{\chandraa}{\textit{Chandra}\ }
\newcommand{\suzakuu}{\textit{Suzaku}\ }
\newcommand{\beppoo}{\textit{BeppoSAX}\ }
\newcommand{\integrall}{\textit{INTEGRAL}\ }
\newcommand{\swiftt}{\textit{Swift}\ }
\newcommand{\swift}{\textit{Swift}}
\newcommand{\wise}{\textit{WISE}}
\newcommand{\wisee}{\textit{WISE}\ }
\newcommand{\nustar}{\textit{NuSTAR}}
\newcommand{\nustarr}{\textit{NuSTAR}\ }

\newcommand{\nev}{[Ne\,V]\ }
\newcommand{\oiv}{[O\,IV]\ }
\newcommand{\oiii}{[O\,III]\ }

 \title[The mid-infrared--X-ray correlation]{The subarcsecond mid-infrared view of local active galactic nuclei:\\ II. The mid-infrared--X-ray correlation 
}

   \author[D. Asmus et al.]{D.~Asmus,$^{1,2,3}$\thanks{E-mail: dasmus@eso.org}
          P.~Gandhi,$^{4,5,6}$ 
          S.~F.~H\"onig,$^{4,7,8}$       
          A.~Smette$^1$
          and
          W.~J.~Duschl$^{3,9}$\\
             $^1$European  Southern Observatory, Casilla 19001, Santiago 19, Chile
             \\          
         $^2$Max-Planck-Institut f\"ur Radioastronomie, 
           Auf dem H\"ugel 69, 53121 Bonn, Germany              
         \\
         $^3$Institut f\"ur Theoretische Physik und Astrophysik,
             Christian- Albrechts-Universit\"at zu Kiel, Leibnizstr. 15, 24098 Kiel, Germany
         \\
         $^4$ School of Physics \& Astronomy, University of Southampton, Hampshire SO17 1BJ, Southampton, United Kingdom         
         \\
             $^5$Department of Physics, Durham University, South Road, Durham, DH1 3LE, United Kingdom 
         \\
             $^6$Institute of Space and Astronautical Science (ISAS), Japan, Aerospace Exploration Agency,  3-1-1 Yoshinodai, chuo-ku,\\ Sagamihara, Kanagawa 252-5210, Japan
         \\
             $^7$Dark Cosmology Center, Niels Bohr Institute, University of Copenhagen, Juliane Maries Vej 30, 2100 Copenhagen, Denmark
         \\
             $^8$UCSB Department of Physics, Broida Hall 93106-9530, Santa Barbara, CA, USA
         \\
             $^9$Steward Observatory, The University of Arizona, 933 N. Cherry Ave, Tucson, AZ 85721, USA
              }

   \date{Received July 31, 2015; accepted August 20, 2015}

\pagerange{\pageref{firstpage}--\pageref{lastpage}} \pubyear{2015}
\begin{document}

\label{firstpage}

\maketitle
 
\begin{abstract}
We present an updated mid-infrared (MIR) versus X-ray correlation for the local active galactic nuclei (AGN) population based on the high angular resolution 12 and 18$\um$ continuum fluxes from the AGN subarcsecond mid-infrared atlas and 2-10\,keV and 14-195\,keV data collected from the literature. 
We isolate a sample of 152 objects with reliable AGN nature and multi-epoch X-ray data and minimal MIR contribution from star formation.
Although the sample is not homogeneous or complete, we show that our results are unlikely to be affected by significant biases.
The MIR--X-ray correlation is nearly linear and within a factor of two independent of the AGN type and the wavebands used. 
The observed scatter is $<0.4$\,dex. 
A possible flattening of the correlation slope at the highest  luminosities probed ($\sim 10^{45}\,$erg/s) towards low MIR luminosities for a given X-ray luminosity is indicated but not significant. 
Unobscured objects have, on average, an MIR--X-ray ratio that is only $\le0.15\,$dex higher than that of obscured objects.
Objects with intermediate X-ray column densities ($22 < \log \nh < 23$) actually show the highest MIR--X-ray ratio on average.
Radio-loud objects show a higher mean MIR--X-ray ratio at low luminosities, while the ratio is lower than average at high luminosities. 
This may be explained by synchrotron emission from the jet contributing to the MIR at low-luminosities and additional X-ray emission at high luminosities.
True Seyfert~2 candidates and double AGN do not show any deviation from the general behaviour.
Finally, we show that the MIR--X-ray correlation can be used to verify the AGN nature of uncertain objects.
Specifically, we give equations that allow to determine the intrinsic 2-10\,keV luminosities and column densities for objects with complex X-ray properties to within 0.34\,dex.
These techniques are applied to the uncertain objects of the remaining AGN MIR atlas, demonstrating the usefulness of the MIR--X-ray correlation as an empirical tool.

\end{abstract}


%
%________________________________________________________________

\section{Introduction}
Supermassive black holes in the centres of galaxy grow through various phases of accretion during the cosmic evolution. 
During these phases large amounts of radiation are emitted by the infalling material leading to dramatic brightening of the galaxy cores, which are then called active galactic nuclei (AGN).
Observing this emission allows us to study the structure of the objects up to the highest redshifts known for galaxies. 
The primary emission from the hot gas accretion disk peaks in the ultraviolet (UV) and then is partly reprocessed and reemitted at other wavelengths.
In particular, it is commonly assumed that at least in radio-quiet AGN part of the UV is reprocessed in a hot corona above the accretion disk, where the X-ray emission is produced through Compton up-scattering  of the UV photons \citep{haardt_x-ray_1993}.
Part of the UV emission from the accretion disk is absorbed by the dust in the material at larger distances (beyond the sublimation radius). 
This dust reemits the absorbed emission thermally in the infrared. 
Usually the mid-infrared (MIR) spectra of AGN are dominated by a warm component approximately consistent with a black body of a temperature $\sim300\,$K (e.g., \citealt{edelson_spectral_1986}). 
In total, these processes explain the observed spectral energy distributions (SEDs) of AGN, 
which often peak in the X-ray and MIR regimes in the $\nu F_\nu$ units (e.g., \citealt{prieto_spectral_2010}).

Furthermore, the common origin of the MIR and X-ray emission from reprocessed UV emission, lets us expect a correlation between the luminosities at both wavelengths.
The exact relation will depend heavily on the structure of obscuring dust.
Usually a thick disk or torus-like geometry is assumed for the latter.
Such structures introduce an orientation dependence for the obscuration, and we thus expect the MIR emission to be at least mildly anisotropic \citep{pier_infrared_1992}.
Specifically unobscured, type~I, objects should have higher MIR luminosities compared to obscured, type~II, objects at the same intrinsic power.  
At the same time, the X-ray emission is likely also anisotropic, depending on the corona geometry (e.g., \citealt{liu_are_2014}).
Furthermore, the amount of obscuration or opening angle of the torus possibly decreases with increasing accretion disk luminosity because the dust is sublimated, or pushed to larger distances by the radiation pressure. 
These effects would lead to a decrease of the ratio of the MIR to bolometric luminosity with increasing bolometric luminosity.
Finally, in radio-loud objects, a strong jet is present in addition to the other components.
The synchrotron emission of this jet might contribute or even dominate both the X-ray emission and the mid-infrared  for low accretion rates (e.g., \citealt{falcke_scheme_2004}; \citealt{yuan_radio-x-ray_2005};\citealt{perlman_mid-infrared_2007}).
Therefore, radio-loud objects should  behave differently in the MIR--X-ray plane than the radio-quiet objects.

For the reasons mentioned above, measuring the relation between MIR and X-ray emission can constrain the proposed scenarios and reveal information about the structure of the AGN.  
Apart from the AGN physics, this relation is also relevant for studying the connection between the AGN contributions to the infrared and X-ray cosmological background (e.g., \citealt{gandhi_x-ray_2003}).

Indeed, a correlation between the MIR and X-ray emission was already found observationally by \cite{elvis_seyfert_1978} and \cite{glass_mid-infrared_1982} based on ground-based MIR bolometer observations of small sample of local AGN and the first generation of X-ray telescopes.
\cite{krabbe_n-band_2001} then remeasured the correlation using one of the first ground-based MIR imagers, MANIAC, mounted on the ESO MPI 2.2\,m to obtain nuclear $N$-band ($\sim10\um$) photometry of a sample of eight nearby AGN.
The X-ray fluxes were in the 2-10\,keV energy band and absorption corrected, from \asca, \beppo, and \textit{Ginga}. 
\citeauthor{krabbe_n-band_2001} attempted to correct for extinction in the MIR, which is possibly significant in many of the Compton-thick (CT) objects. 
The MIR--X-ray correlation was then investigated with a much larger sample by \cite{lutz_relation_2004} and \cite{ramos_almeida_mid-infrared_2007} using low angular resolution MIR data ($>1$\,arcsec) from \isoo and 2-10\,keV measurements also from \xmmm and \chandra.
The former work decomposed the \isoo spectra to isolate the AGN continuum emission at $6\um$, still a large scatter affected their correlation results as in the latter work.
In parallel, a number of other works made use of the now available MIR instruments on eight-meter class telescopes like VISIR and T-ReCS to study the MIR--X-ray correlation at subarcsecond resolution at $\sim12\um$ \citep{horst_small_2006,horst_mid_2008, gandhi_resolving_2009,levenson_isotropic_2009, honig_dusty_2010-1}.
Isolating the AGN emission to a much better degree, these works found, in general, a very small scatter in the correlation ($\sim 0.3\,$dex).
In particular, \cite{gandhi_resolving_2009} demonstrated that even the CT obscured AGN follow the same correlation as the unobscured AGN without large offsets or scatter.  
This made the MIR--X-ray correlation the tightest among the other multiple wavelength correlations found for AGN and especially intriguing because of its applicability to all different AGN types.
Using \spitzer, the correlation was at the same time extended towards higher luminosities \citep{fiore_chasing_2009,lanzuisi_revealing_2009} and shown to be valid even for radio-loud AGN \citep{hardcastle_active_2009}.
Further high angular resolution imaging allowed then to extend the correlation into the low-luminosity regime \citep{asmus_mid-infrared_2011,mason_nuclear_2012}, where some indication was found that radio-loud low-luminosity objects show on average higher MIR--X-ray emission ratios.
Finally, with the advent of the X-ray telescopes \swiftt and \integral, sensitive to very hard X-rays, the correlation could be shown to extend also the 14-195\,keV X-ray regime \citep{mullaney_defining_2011,matsuta_infrared_2012,ichikawa_mid-_2012,sazonov_contribution_2012}.
These works mainly used low angular resolution MIR data from \iras, \spitzer, \akari, and \wisee  and could demonstrate that the correlation also extends towards longer wavelengths in the MIR. 
\cite{mullaney_defining_2011} and \cite{sazonov_contribution_2012} applied spectral decomposition to isolate the AGN continuum at 12 and $15\um$ respectively. 
They found rather shallow slopes for the MIR--X-ray correlation in particular at odds with \cite{fiore_chasing_2009} and \cite{lanzuisi_revealing_2009}.
In general, the findings of all these works remained inconclusive or contradicting with respect to the predictions of the models described above, owing to either small sample sizes, selection effects or the lack of sufficient angular resolution.
Therefore, the correlation needs to be investigated at high angular resolution for a large sample covering the whole luminosity range of AGN.

The first paper of this series \citep{asmus_subarcsecond_2014} presented high angular resolution MIR photometry for a large number of local AGN obtained with ground-based instruments on eight-meter class telescopes. 
Here, we use this large well suited data set to accurately redetermine the MIR--X-ray correlation and address the still open issues with a detailed analysis.

\section{Data Acquisition \& Sample selection}\label{sec:sam}
We start with the total sample from the AGN MIR atlas of 253 objects \citep{asmus_subarcsecond_2014}.
Note that sample is a combination of a hard X-ray (14-195\,keV) selected and a quasi MIR selected sample as explained in \cite{asmus_subarcsecond_2014}. 
Thus, this combined sample does not suffer from any obvious (single) bias against highly obscured sources.
The fraction of Compton-thick obscured sources is 18 per cent (confirmed plus candidates).
We adopt optical classifications, distances (using the same cosmology) and the MIR continuum fluxes at 12 and $18\,\mu$m, $\fnn$ and $\fqn$ from \cite{asmus_subarcsecond_2014}.
As done in that work we also divide the various AGN types into the following rough optical groups to facilitate discussion:
\begin{description}
 \item -- \textbf{type~I}: Seyferts with broad unpolarised emission lines (includes Sy\,1, Sy\,1.2, Sy\,1.5, Sy\,1/1.5 and Sy\,1.5/L);
 \item -- \textbf{NLS1}: (narrow-line Sy\,1);
 \item -- \textbf{type~Ii}: intermediate type Seyferts (includes Sy\,1.8, Sy\,1.9 and Sy\,1.5/2);
 \item -- \textbf{type~II}: Seyferts without broad emission lines (in unpolarised light; includes Sy\,1.8/2, Sy\,1.9/2 and Sy\,2:); 
 \item -- \textbf{LINERs}: low-ionization nuclear emission line regions (includes L, L:);
 \item -- \textbf{AGN/SB comp}: AGN/starburst composites (includes Cp, Cp:, and L/H; see \citealt{yuan_role_2010}).
\end{description}
Here, for example Sy\,1.8/2 means that the object has both Sy\,1.8 and Sy\,2 classifications in the literature, i.e. the existence of broad emission line components is controversial.
Note that different to \cite{asmus_subarcsecond_2014}, we keep the NLS1 as an individual class.
There are four clear NLS1 in the atlas sample (I\,Zw\,1, IRAS\,13349+2438,  Mrk\,1239, and NGC\,4051).

The MIR continuum fluxes are the unresolved nuclear fluxes extracted from ground-based multi-filter photometry obtained with the instruments VISIR \citep{lagage_successful_2004}, T-ReCS \citep{telesco_gatircam:_1998}, Michelle \citep{glasse_michelle_1997}, and COMICS; \citep{kataza_comics:_2000}. 
The angular resolution of these data is of the order of $0.35\arcsec$ or 120\,pc for the median sample distance of 72\,Mpc at 12$\,\um$.

We collect observed 14-195\,keV fluxes, $\fxh$, measured with \swift/BAT by combining the data of the 54 and 70\,month source catalogues with preference for the latter \citep{cusumano_palermo_2010,baumgartner_70_2013}. 
Of the 253 AGN, 120 are in  the 70\,month catalogue, comprising about $\sim 20$\,per cent of all known AGN detected by \swift/BAT after 70\,months.
Six additional sources missing in the 70\,month are in the 54\,month catalogue (3C\,98, ESO 500-34, NGC\,4579, PG\,0844+349 and UGC\,12348). 
For all 127 sources without any detection in the BAT catalogues, we adopt an upper limit on $\fxh$ equal to the nominal sensitivity limit of the 70\,month catalogue ($\log \fxh = -10.87$; \citealt{baumgartner_70_2013}). 

In addition, we search the literature for the 2-10\,keV properties of the individual objects using the most recent X-ray observations with satellites like \chandra, \xmm, \suzaku, and \nustar.
In particular, we compile observed and intrinsic 2-10\,keV fluxes, $\fxo$ and $\fxi$, and hydrogen column densities, $\nh$ based upon X-ray spectral modelling (compiled in Table~\ref{tab:sam}). 
In case of multiple measurements, the average of all detections is used, while for multiple upper limits, the lowest upper limit is taken.
For 19 objects we analysed in addition archival \swift/XRT data \citep{burrows_swift_2005} as further described in Appendix~\ref{sec:xrt}.
In order to avoid non-AGN contamination, we exclude all objects optically classified as uncertain (38 objects) or AGN/starburst composites (18 objects; see \citealt{asmus_subarcsecond_2014}) and discuss these separately in the Appendices~\ref{sec:can} and \ref{sec:CP}.
Of the remaining 197 object from the atlas, no 2-10\,keV properties could be collected or computed for three objects (NGC\,3166, PKS\,1932-46, and UGC\,12348), mainly because of the lack of suitable observations.
For further 42 objects, the X-ray data turned out to be problematic because of too few counts for reliable analysis, contradicting results ($>1$\,dex difference in $\lxi$), or Compton-thick  obscured objects, for which no recent pointed hard X-ray observations with satellites like \xmm, \suzakuu and \nustarr are available. 
These objects are referred to as "X-ray unreliable".  
All these unobserved or problematic sources are excluded from the general analysis and are further discussed as a group in Appendix~\ref{sec:CT} and individually in Appendix~\ref{app:obj}.
This leaves 152 reliable AGN for the general analysis, called the "\textbf{reliable}" sample. 
It contains 55 type~I, 4 NLS1, 14 type~Ii, 58 type~II, 19 LINERs, and 2 unclassified (NGC\,4992 and NGC\,6251; see \citealt{asmus_subarcsecond_2014}).
The individual objects and the relevant properties mentioned above are listed in Table~\ref{tab:sam}.

%Furthermore, we define those 99 objects with at least three independent epochs of 2-10\,keV measurements as the "best" sample.
For objects with at least three independent epochs of X-ray data (100 objects), we use the standard deviation (STDDEV) between the individual measurements as uncertainty for $\fxo$ and $\fxi$ in logarithmic space (similarly for the luminosities).
For the other objects, we use the median + STDDEV of all 100 uncertainties above as flux uncertainty (0.3\,dex) unless they are CT obscured, in which case we use 0.6\,dex as uncertainty for $\fxi$ (5 objects).
Note that 0.3\,dex also corresponds to the typical long-term variability found in individual AGN (e.g., \citealt{mchardy_combined_2004}).
For the CT objects, we choose 0.6\,dex because {\it a)} individual $\fxi$ estimates easily differ by an order of magnitude in this regime depending on the model used (e.g., \citealt{gandhi_compton-thick_2015}), and {\it b)} by this means, the corresponding objects will statistically be weighted only half during the fitting (where included). 
The values for $\nh$ and their uncertainties are computed exactly in the same way as the fluxes, and the corresponding median + STDDEV uncertainty is 0.45\,dex.

Finally, for all MIR and X-ray bands, we compute the corresponding luminosities under the assumption of isotropic emission.

\section{Methods}\label{sec:meth}
We measure correlation strengths by using the Spearman rank correlation coefficient, $\rhoS$, and the corresponding null-hypothesis probability $\log \pS$, applied to the detections of a sample. 
The correlations are tested in both luminosity and flux space to show how much correlation is artificially introduced by the distance.
Furthermore, we compute the partial correlation rank in luminosity space, $\rhop$, which measures the correlation of the residuals of the correlations of the luminosities with the object distance (\texttt{p\_correlate} in IDL).
To measure the functional description of the correlations, we perform linear regression in logarithmic space with the Bayesian based \texttt{linmix\_err} algorithm and a normal prior distribution \citep{kelly_aspects_2007}.
We prefer \texttt{linmix\_err} for this work over the canonical \texttt{fitexy} algorithm \citep{press_numerical_1992} because the former can handle upper limits and also takes into account the intrinsic scatter, $\sigint$, in the sample (see \citealt{park_recalibration_2012} for a thorough comparison of fitting algorithms).
In all correlations, we choose the X-ray component as independent variable, which is physically motivated by the fact that the X-ray emission is a more direct proxy of the primary accretion than the MIR emission. 
Owing to the nature of the sample which originates from a combined indirect MIR and X-ray selection, there is no preference for the choice of independent variable from the statistical point of view.
In Sect.~\ref{sec:fit}, we further discuss how our results are affected by possible biases and the choice of methods.
Furthermore, we use the logarithmic ratio of the MIR over X-ray luminosity, $\ratmx = \log L_\mathrm{MIR} - \log L_\mathrm{X}$ and its standard deviation, $\sigmx$, to describe the various relations ($L_\mathrm{MIR}$ and $L_\mathrm{X}$ are to be substituted with the corresponding subbands used in that section).
Finally, we use the two-sided one-dimensional Kolmogorov-Smirnov (KS) test to quantify the statistical difference in the MIR--X-ray ratio between various subsamples.
The test provides a maximum sample difference, $\dks$ and a null-hypothesis probability, $\pks$.
To better address the robustness of the KS test results, we  perform Monte-Carlo resampling with $10^5$ random draws. 
Here, we vary the individual MIR--X-ray ratios by a normal distribution with the STDDEV of the individual uncertainties and then perform simple bootstrapping with drawing and replacement \citep{press_numerical_1992}.
The combination of both allows us to take into account the effects of both measurement uncertainties and sample incompleteness.
%The resulting distributions of  $\dks$ and $\pks$ are roughly normal distributions in our cases so that they are well characterised by stating their median and standard deviations.
Apart from the linear regression analysis, non-detections are always excluded from the quantitative analysis.
However, the consistency of the corresponding upper limits with the obtained results is discussed where suitable.

\section{Results}\label{sec:res}
In the following, the relations between the MIR and X-ray emission of the reliable sample are investigated in the four different wavelength ranges, $12\um, 18\um$, 2-10\,keV and 14-195\,keV.

% \subsection{12$\,\mu$m--14-195\,keV correlation}\label{sec:12:14}
% 
% \subsection{18$\,\mu$m--14-195\,keV correlation}\label{sec:18:14}
% 
% - in flux space
% 
% \subsection{12$\,\mu$m--2-10\,keV correlation}\label{sec:12:2}
% 
% \subsection{18$\,\mu$m--2-10\,keV correlation}\label{sec:18:2}

\subsection{12$\,\mu$m--2-10\,keV correlation}\label{sec:cor}
\begin{figure}
%    \centering
%    \sidecaption
   \includegraphics[angle=0,width=8.5cm]{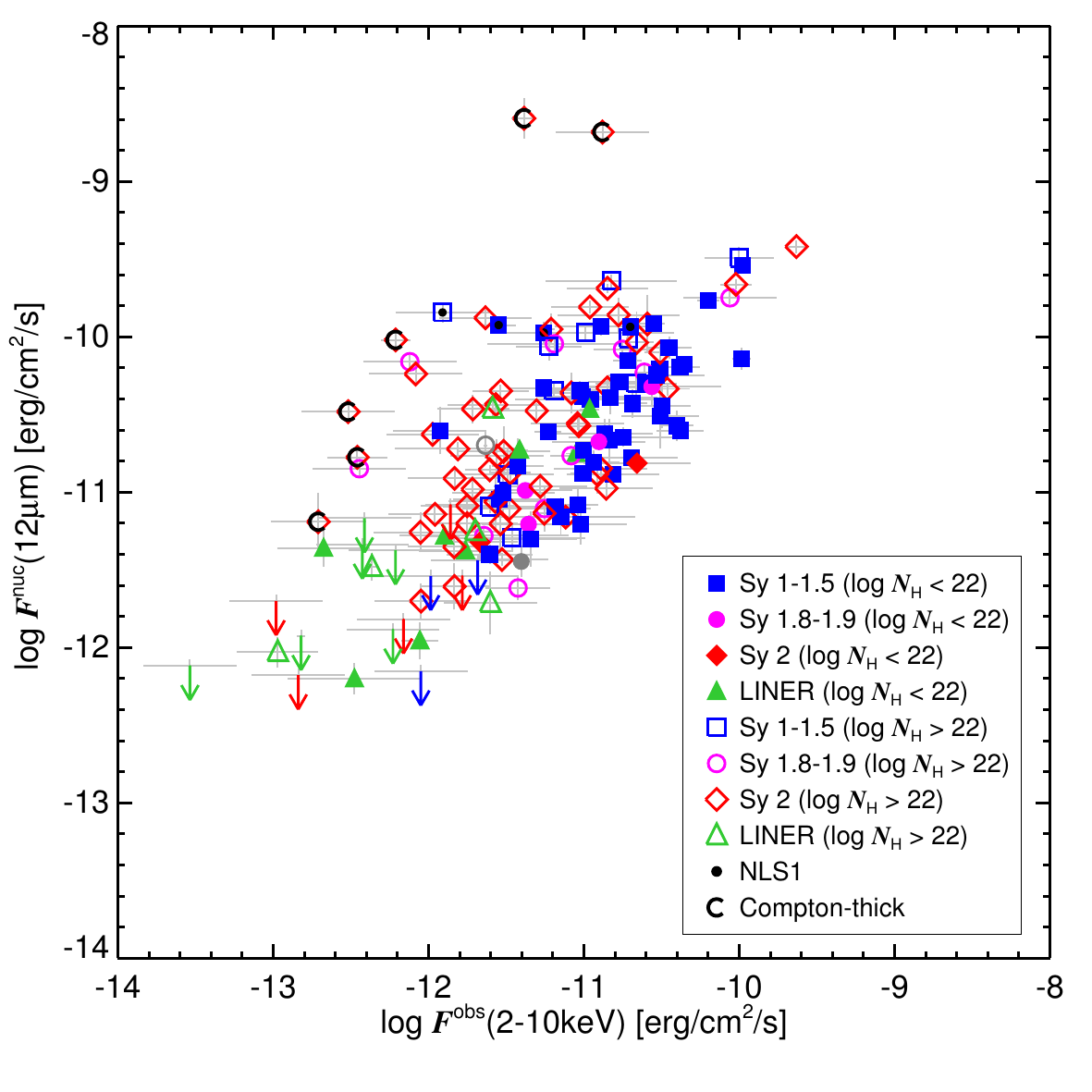}
    \caption{
             Relation of the $12\um$ and observed 2-10\,keV fluxes for the reliable sample.
             Blue squares are type~I AGN, magenta circles type~Ii, red diamonds type~II, green triangles LINERs, and grey filled circles objects without a clear optical type (NGC\,4992 and NGC\,6251). 
             Filled symbols mark X-ray unabsorbed objects ($\log \nh \le 22$) while empty symbols are the obscured ones..
             The arrows mark upper limits of MIR non-detections.
             Objects marked with a small black filled circle are NLS1 and objects marked with a C are Compton-thick obscured. 
            }
   \label{fig:f12_f2obs}
\end{figure}

\begin{figure}
%    \centering
%    \sidecaption
   \includegraphics[angle=0,width=8.5cm]{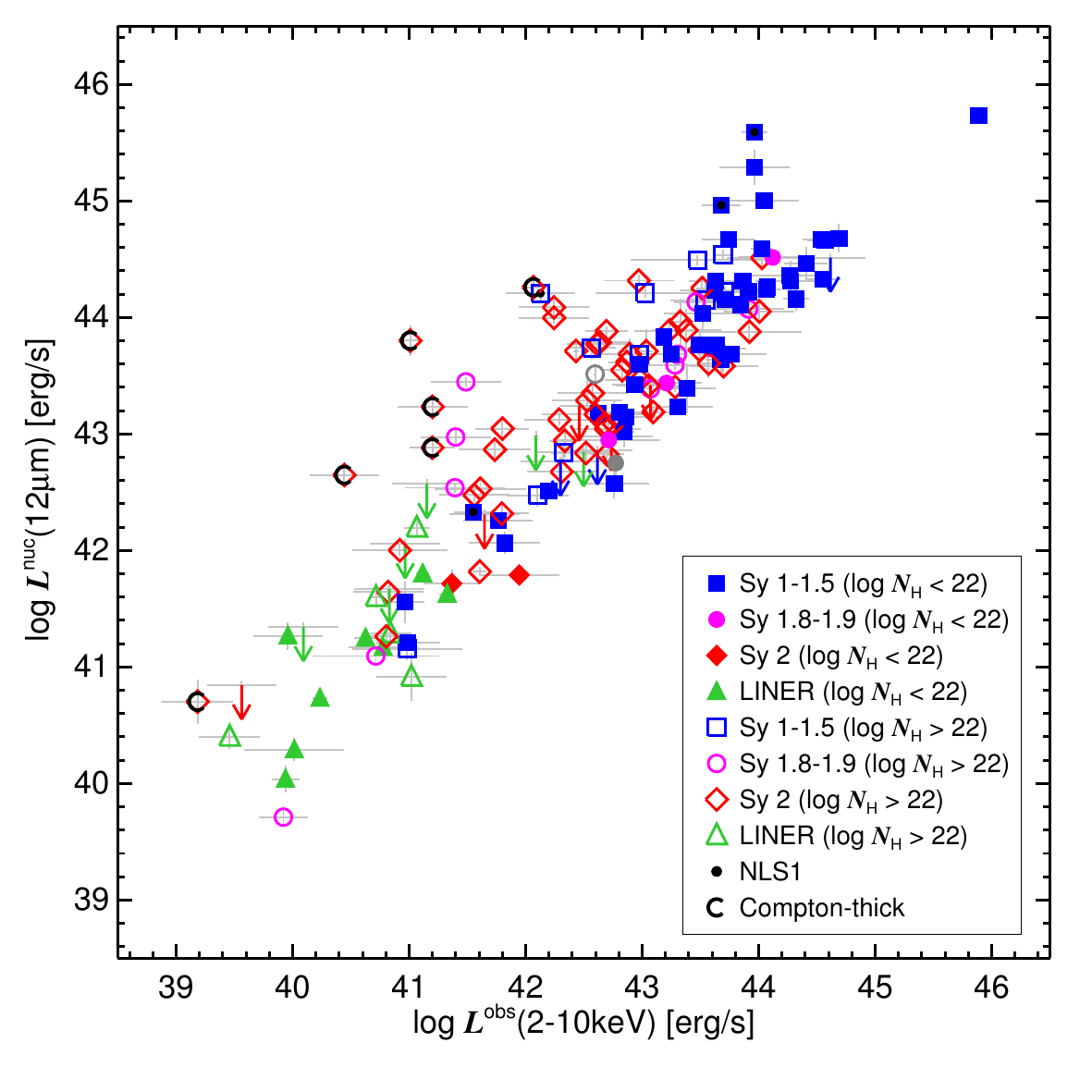}
    \caption{
             Relation of the nuclear $12\um$ and observed 2-10\,keV luminosities for the reliable sample.
             Symbols and colours are as in Fig.~\ref{fig:f12_f2obs}.
            }
   \label{fig:l12_l2obs}
\end{figure}

Figs.~\ref{fig:f12_f2obs} and \ref{fig:l12_l2obs} show the relation of the observed $12\um$ and 2-10\,keV bands in flux and luminosity space respectively.  
As expected from previous works, a significant correlation is present between $\fnn$ and $\fxo$.
The Spearman rank correlation coefficient for all detected sources is $\rhoS=0.59$ and the null-hypothesis probability $\log \pS= -13.7$ with highly obscured objects introducing large scatter towards high MIR--X-ray ratios (standard deviation $\sigmx = 0.57$).
In particular,  most of the CT objects exhibit large offsets owing to their strong suppression in X-rays. 
Interestingly, three of the four NLS1 also exhibit high MIR--X-ray ratios despite low obscuration.
We come back to this observation in Section~\ref{sec:typ}.
If one only regards the 68 X-ray unabsorbed AGN of the reliable sample ($\log \nh \le 22$), the correlation strength increases to $\rhoS = 0.77$ and $\log \pS = -13.5$. 
A corresponding linear regression of the unabsorbed sources yields $\log \fnn \propto (0.79\pm0.08) \log \fxo$ with \texttt{linmix\_err}.
The detailed fitting results are given in Table~\ref{tab:cor}.
In luminosity space (Fig.~\ref{fig:l12_l2obs}), the sample is spread over seven orders of magnitude leading to a formally stronger correlation ($\rhoS = 0.94$ and $\log \pS = -27.8$ for the unabsorbed AGN).
The partial correlation rank is $\rhop = 0.73$ for the unabsorbed AGN, showing that the correlation strength is not dominated by the effect of the distance.

\begin{figure}
%    \centering
%    \sidecaption
   \includegraphics[angle=0,width=8.5cm]{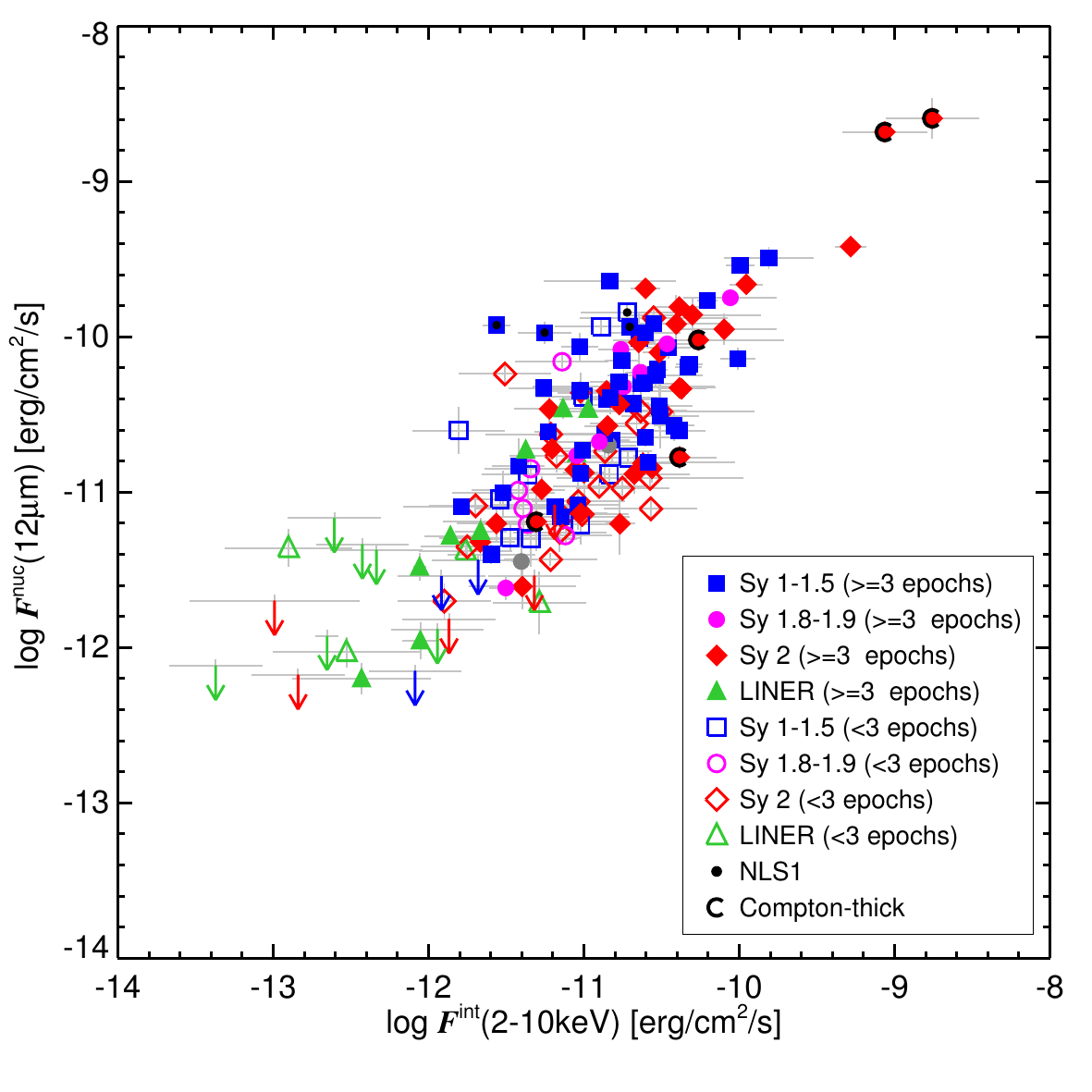}
    \caption{
             Relation of the nuclear $12\um$ and intrinsic 2-10\,keV fluxes for the reliable sample.
             Symbols and colours are as in Fig.~\ref{fig:f12_f2obs} apart from the filled symbols which here mark objects with at least three epochs of 2-10\,keV data.
            }
   \label{fig:f12_f2int}
\end{figure}

\begin{figure}
%    \centering
%    \sidecaption
   \includegraphics[angle=0,width=8.5cm]{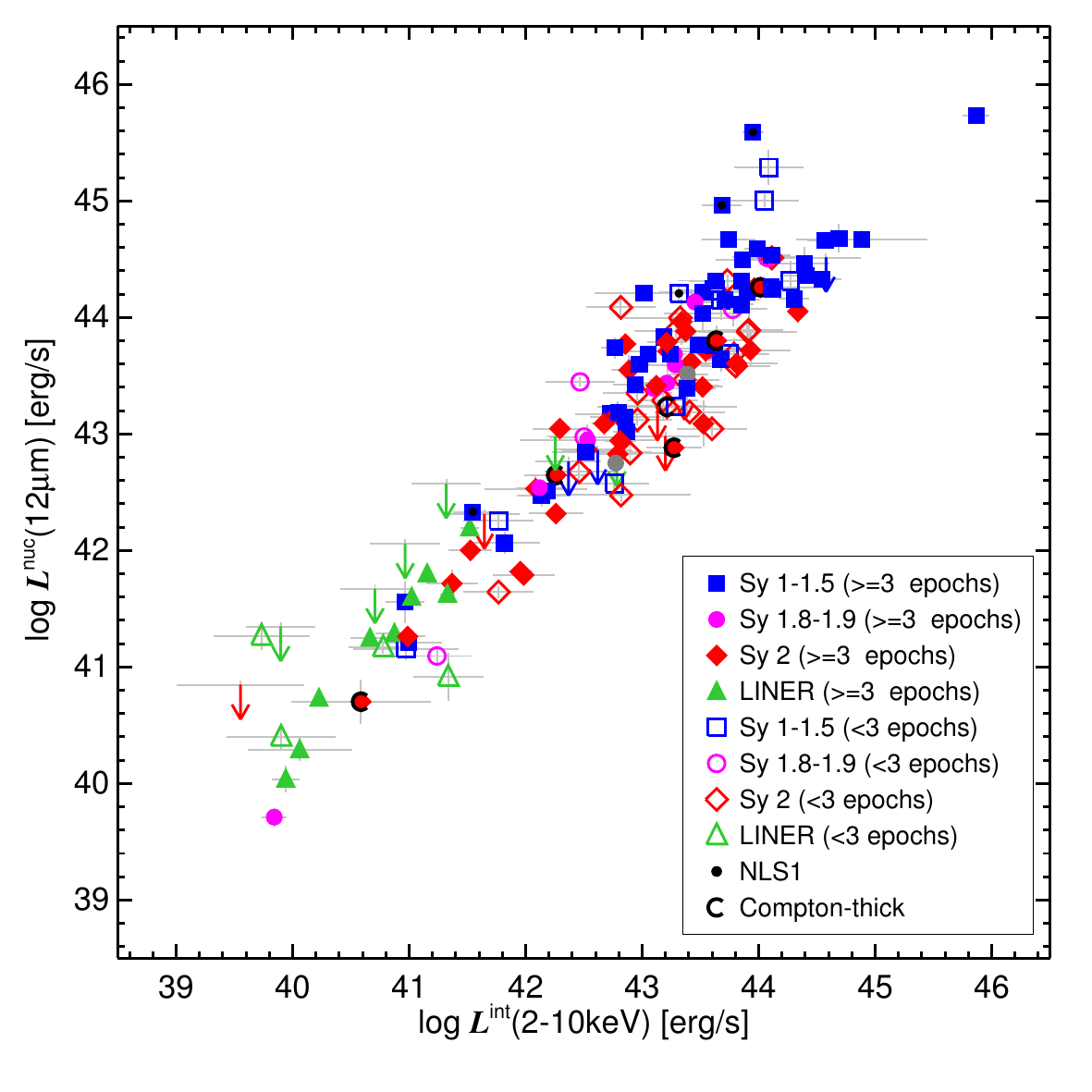}
    \caption{
             Relation of the nuclear $12\um$ and intrinsic 2-10\,keV luminosities for the reliable sample.
             Symbols and colours are as in Fig.~\ref{fig:f12_f2int}.
            }
   \label{fig:l12_l2int}
\end{figure}

Owing to the absorption corrections available for the 2-10\,keV emission of the reliable objects, we can also investigate the correlation of $\fnn$ and $\lnn$ with the intrinsic X-ray fluxes and luminosities, $\fxi$ and $\lxi$, shown in Figs.~\ref{fig:f12_f2int} and \ref{fig:l12_l2int}.
As expected the correlation strength increases in flux space when using an absorption correction for the X-rays ($\rhoS = 0.73$ and $\log \pS = -23.2$ for all 152 reliable AGN), and the slope of the linear fit is $1.02\pm0.06$.
Note that two of the NLS1 are still outliers (I\,Zw\,1 and IRAS\,13349+2438).
In luminosity space, the partial correlation rank increases to 0.79 and the slope remains consistent at $0.98\pm0.03$ (Table~\ref{tab:cor}).
The observed scatter is $\sigmx=0.39$\,dex, while the intrinsic scatter  is smaller, $\sigint = 0.33$\,dex.

Many AGN show significant variability in the 2-10\,keV band, while any MIR variability appears to be much weaker and only occurs on much longer time scales \citep{neugebauer_variability_1999,asmus_subarcsecond_2014}.
Therefore, one would expect a further decrease in the scatter of the correlation when regarding only those 100 AGN for which our $\lxi$ value is based on at least three epochs of X-ray observations in the last decade. 
For this sub-sample, the partial correlation rank indeed increases ($\rhop = 0.83$) and the observed scatter in the $\lnn/\lxi$ ratio decreases ($\sigmx = 0.35$\,dex).
However, the intrinsic scatter estimate stays the same ($\sigint = 0.32$\,dex).
The similarity of observed and intrinsic scatter indicate that the former is dominated by the latter. 
The corresponding fit represents the best estimate of the "true" $12\um$--2-10\,keV correlation so far:
\begin{eqnarray}\label{eq:best}
 \log \left( \frac{\lnn}{10^{43}\textrm{erg}\,\mathrm{s}^{-1}}\right)   &=& ( 0.33 \pm 0.04) \cr &+& ( 0.97 \pm 0.03 )  \log \left( \frac{\lxi}{10^{43}\textrm{erg}\,\mathrm{s}^{-1}}\right).
\end{eqnarray}

\begin{figure}
%    \centering
%    \sidecaption
   \includegraphics[angle=0,width=8.5cm]{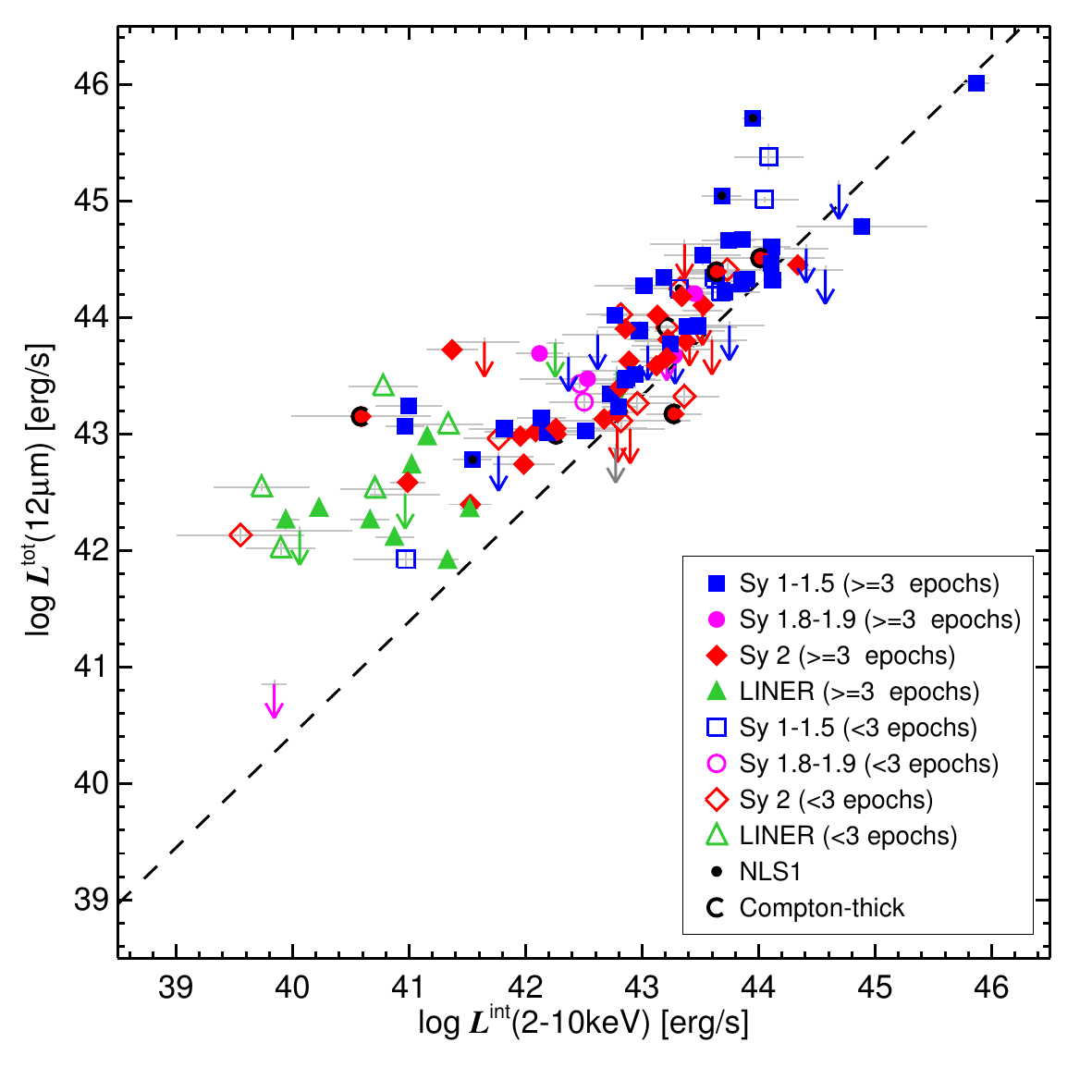}
    \caption{
             Relation of the total $12\um$ (\textit{IRAS}) and intrinsic 2-10\,keV luminosities for the reliable sample.
             Symbols and colours are as in Fig.~\ref{fig:f12_f2int}.
             In addition, the dashed line shows the correlation for all reliable objects with multi-epoch X-ray data (Eq.~\ref{eq:best}).
            }
   \label{fig:l12tot_l2int}
\end{figure}

For comparison purposes, in Fig.~\ref{fig:l12tot_l2int}, we also show the $12\um$--2-10\,keV relation when using low angular resolution MIR data, here 12\umm luminosities, $\lnt$, obtained with the \irass satellite \citep{neugebauer_infrared_1984}, which correspond rather to the total emission of the galaxies  than the nuclear emission.
The corresponding data were collected in \cite{asmus_subarcsecond_2014} from the literature, mainly from NED (\url{http://ned.ipac.caltech.edu}), \cite{sanders_iras_2003}, \cite{golombek_iras_1988}, \cite{rush_extended_1993}, \cite{rice_catalog_1988} and \cite{sanders_luminous_1996}.  
The resulting luminosity correlation has a significantly higher observed scatter ($\sigmx = 0.65$) and the fitted slope is also much flatter ($b = 0.63$).
This is obviously caused by host emission dominating the total MIR emission in many objects up to luminosities of $\approx 10^{44}$\,erg/s \citep{asmus_subarcsecond_2014}.
But even if one fits only the higher luminosity part of the population ($> 10^{43}$\,erg/s), the slope is still flatter compared to the high angular resolution data ($b = 0.82$).
Therefore, $\lxi$ estimates based on MIR luminosities that include the entire galaxy are very unreliable.

\begin{table*}
\caption{Correlation properties.} % title of Table
\label{tab:cor} % is used to refer this table in the text
\centering % used for centering table
\begin{tabular}{	c		c		l		c		c		c				c				c		c	c				}
\hline\hline																										
$X$	&	$Y$	&	Sample	&	$N$	&	$\rho$	&	$a$			&	$b$			&	$\overline{\ratmx}$	&	$\sigmx$	&	$\sigint$			\\
(1)	 &	(2)	&	(3)	&	(4)	&	(5)	&	 (6) & (7)	&			(8) &				(9)	&	(10)				\\
																									
%	&		&	sample ID	&	# objects	&	Rank	&	a	$\pm$	error	&	b	$\pm$	error	&	ratio	&	stddev	&	intrinsic scatter	$\pm$	its stddev	\\

\hline																									
$\fxo$	&	$\fnn$	&	$\log \nh \le 22$	&	68	&	0.77	&	0.35	$\pm$	0.05	&	0.79	$\pm$	0.08	&	0.40	&	0.42	&	0.34	$\pm$	0.16	\\
$\lxo$	&	$\lnn$	&	$\log \nh \le 22$	&	68	&	0.73	&	0.34	$\pm$	0.05	&	0.96	$\pm$	0.04	&	0.36	&	0.39	&	0.35	$\pm$	0.18	\\
$\fxi$	&	$\fnn$	&	All (reliable)	&	152	&	0.73	&	0.29	$\pm$	0.03	&	1.02	$\pm$	0.06	&	0.32	&	0.39	&	0.33	$\pm$	0.14	\\
$\lxi$	&	$\lnn$	&	All (reliable)	&	152	&	0.79	&	0.30	$\pm$	0.03	&	0.98	$\pm$	0.03	&	0.32	&	0.39	&	0.33	$\pm$	0.14	\\
$\fxi$	&	$\fnn$	&	$N$(X-ray epochs)$\ge 3$	&	100	&	0.73	&	0.34	$\pm$	0.04	&	0.94	$\pm$	0.07	&	0.33	&	0.35	&	0.32	$\pm$	0.14	\\
$\lxi$	&	$\lnn$	&	$N$(X-ray epochs)$\ge 3$	&	100	&	0.83	&	0.33	$\pm$	0.04	&	0.97	$\pm$	0.03	&	0.33	&	0.35	&	0.32	$\pm$	0.14	\\
$\lxi$	&	$\lnt$	&	\textit{IRAS}	&	116	&	0.57	&	0.72	$\pm$	0.05	&	0.63	$\pm$	0.04	&	0.97	&	0.65	&	0.45	$\pm$	0.19	\\
$\lxi$	&	$\lnt$	&	\textit{IRAS} \& $\log \lxi > 43$	&	53	&	0.48	&	0.64	$\pm$	0.12	&	0.82	$\pm$	0.14	&	0.64	&	0.39	&	0.43	$\pm$	0.24	\\
$\fxi$	&	$\fnn$	&	$18\um$	&	38	&	0.83	&	0.75	$\pm$	0.05	&	0.88	$\pm$	0.07	&	0.69	&	0.30	&	0.23	$\pm$	0.14	\\
$\lxi$	&	$\lqn$	&	$18\um$	&	38	&	0.90	&	0.53	$\pm$	0.06	&	1.00	$\pm$	0.05	&	0.51	&	0.30	&	0.25	$\pm$	0.16	\\
$\lxi$	&	$\lnn$	&	$18\um$	&	38	&	0.90	&	0.42	$\pm$	0.06	&	0.99	$\pm$	0.05	&	0.38	&	0.31	&	0.23	$\pm$	0.15	\\
$\fxh$	&	$\fnn$	&	14-195\,keV \& $\log \nh < 23.7$	&	101	&	0.62	&	-0.10	$\pm$	0.08	&	0.82	$\pm$	0.10	&	-0.19	&	0.41	&	0.40	$\pm$	0.16	\\
$\lxh$	&	$\lnn$	&	14-195\,keV \& $\log \nh < 23.7$	&	101	&	0.62	&	-0.15	$\pm$	0.05	&	0.91	$\pm$	0.04	&	-0.19	&	0.41	&	0.39	$\pm$	0.16	\\
$\lxi$	&	$\lxh$	&	14-195\,keV \& $\log \nh < 23.7$	&	101	&	0.85	&	0.47	$\pm$	0.03	&	1.01	$\pm$	0.02	&	0.47	&	0.23	&	0.17	$\pm$	0.09	\\
$\lxi$	&	$\lnn$	&	Type~I	&	54	&	0.70	&	0.35	$\pm$	0.05	&	0.96	$\pm$	0.05	&	0.35	&	0.34	&	0.25	$\pm$	0.15	\\
$\lxi$	&	$\lnn$	&	Type~II	&	58	&	0.75	&	0.18	$\pm$	0.06	&	0.97	$\pm$	0.07	&	0.19	&	0.37	&	0.31	$\pm$	0.18	\\
$\lxi$	&	$\lnn$	&	Type~I+II	&	112	&	0.74	&	0.26	$\pm$	0.04	&	0.99	$\pm$	0.04	&	0.27	&	0.36	&	0.29	$\pm$	0.14	\\
$\lxi$	&	$\lnn$	&	LINER	&	19	&	0.71	&	0.22	$\pm$	0.44	&	0.92	$\pm$	0.19	&	0.46	&	0.43	&	0.30	$\pm$	0.31	\\
$\lxi$	&	$\lnn$	&	$\log \nh < 22$ 	&	63	&	0.79	&	0.30	$\pm$	0.04	&	0.95	$\pm$	0.03	&	0.32	&	0.35	&	0.22	$\pm$	0.14	\\
$\lxi$	&	$\lnn$	&	$22  \le \log \nh < 23$ 	&	32	&	0.88	&	0.43	$\pm$	0.07	&	1.09	$\pm$	0.06	&	0.36	&	0.31	&	0.23	$\pm$	0.17	\\
$\lxi$	&	$\lnn$	&	$\log \nh \ge 23$ 	&	52	&	0.77	&	0.20	$\pm$	0.07	&	1.00	$\pm$	0.09	&	0.23	&	0.41	&	0.37	$\pm$	0.22	\\
$\lxi$	&	$\lnn$	&	Radio-quiet	&	41	&	0.88	&	0.27	$\pm$	0.05	&	1.01	$\pm$	0.05	&	0.26	&	0.28	&	0.21	$\pm$	0.14	\\
$\lxi$	&	$\lnn$	&	Radio-loud	&	48	&	0.83	&	0.20	$\pm$	0.06	&	0.90	$\pm$	0.04	&	0.28	&	0.41	&	0.28	$\pm$	0.17	\\
$\lxi$	&	$\lnn$	&	BAT9	&	80	&	0.74	&	0.23	$\pm$	0.04	&	1.02	$\pm$	0.05	&	0.21	&	0.32	&	0.24	$\pm$	0.13	\\

\hline						                              
\end{tabular}
\begin{minipage}{1.0\textwidth}
 
{\it -- Notes:} 
(1) and (2) the quantities, of which the correlation is measured;
(3) sample used for the \texttt{linmix\_err} fitting; 
(4) $N$: number of objects used for the analysis; 
(5) $\rho$: linear correlation coefficient (Spearman rank $\rhoS$ for fluxes and partial correlation rank $\rhop$ for luminosities); 
(6) and (7) $a,b$: fitting parameters of $\log(Y)-c = a + b (\log(X)-c)$ with c being -11 for fluxes and 43 for luminosities; 
(8) $\langle R^Y_X\rangle$: average of the ratio $R^Y_X$ with (9) $\sigma^Y_X$ its standard deviation;
(10) $\sigma_\mathrm{int}$: intrinsic scatter (from \texttt{linmix\_err}) .
        
\end{minipage}

\end{table*}

\subsection{MIR--X-ray correlation using 18\,$\mu$m or 14-195\,keV}\label{sec:altcor}
\begin{figure}
%    \centering
%    \sidecaption
   \includegraphics[angle=0,width=8.5cm]{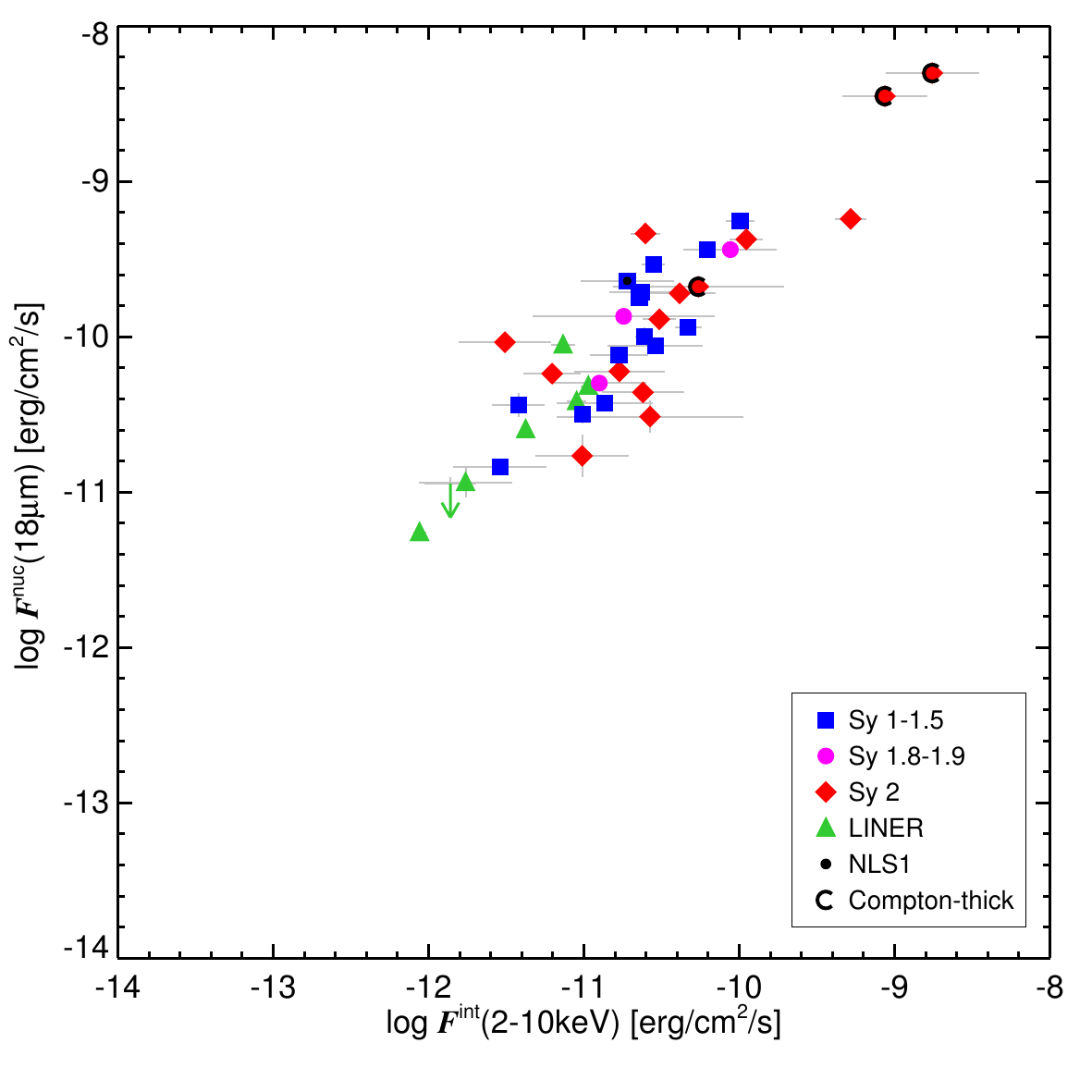}
    \caption{
             Relation of the nuclear $18\um$ and intrinsic 2-10\,keV fluxes for the reliable sample.
             Symbols and colours are as in Fig.~\ref{fig:f12_f2obs} apart from all objects having filled symbols.
            }
   \label{fig:f18_f2int}
\end{figure}

\begin{figure}
%    \centering
%    \sidecaption
   \includegraphics[angle=0,width=8.5cm]{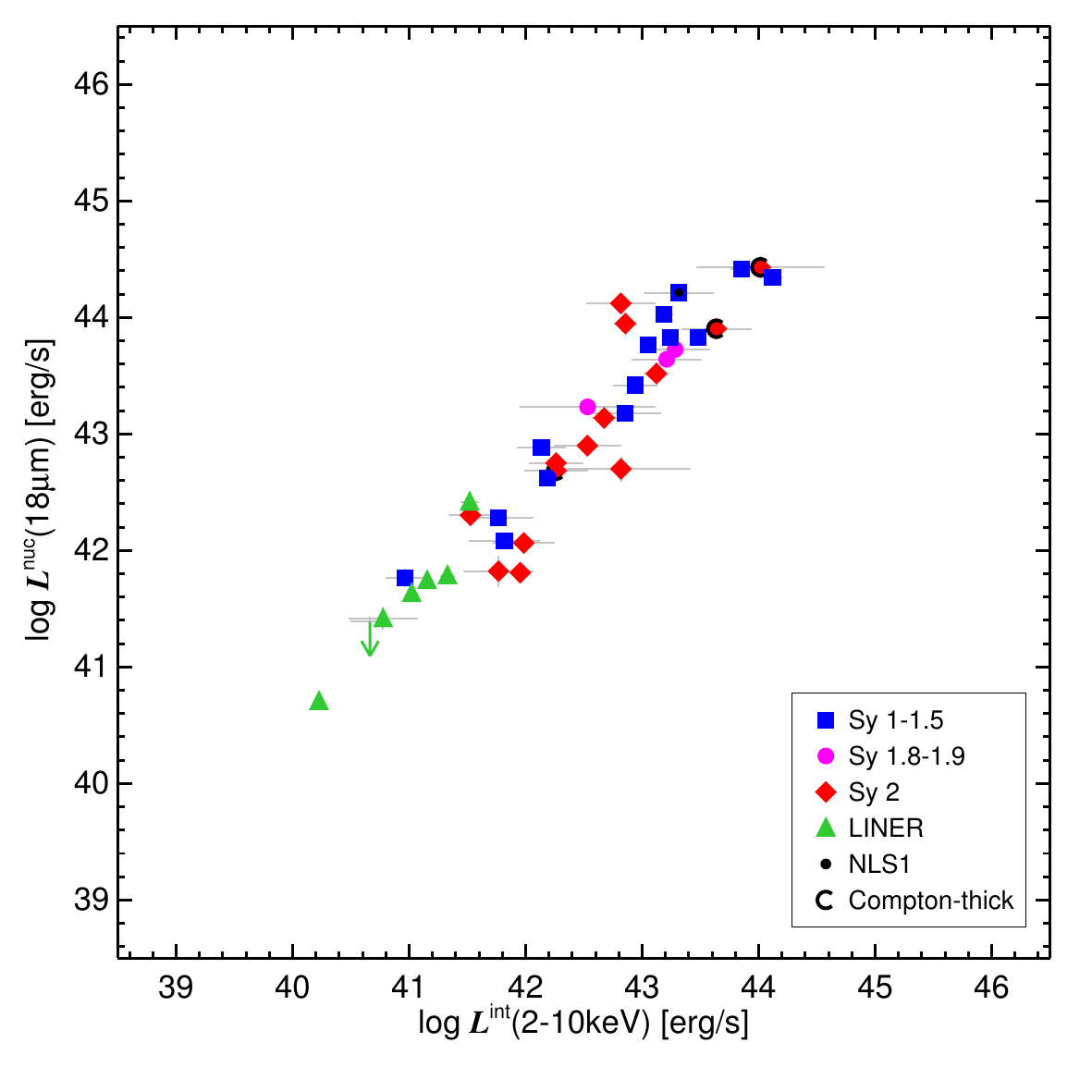}
    \caption{
             Relation of the nuclear $18\um$ and intrinsic 2-10\,keV luminosities for the reliable sample.
             Symbols and colours are as in Fig.~\ref{fig:f18_f2int}.
            }
   \label{fig:l18_l2int}
\end{figure}

The AGN MIR atlas also contains nuclear $18\um$ flux measurements for 38 of the reliable AGN.
This wavelength region has two advantages compared to the $12\um$ region, it is even less affected by obscuration and it contains the MIR emission peak of most AGN in $\nu F\nu$, in particular type~1 sources (e.g., \citealt{asmus_subarcsecond_2014}).
Note however that at this wavelength, the subarcsecond resolution is even more important to isolate the AGN from the host emission which is stronger at $18\,\um$ than at $12\,\um$.
We plot the corresponding $18\um$--2-10\,keV relations in flux and luminosity space in Figs.~\ref{fig:f18_f2int} and \ref{fig:l18_l2int}.
Despite the lower object number and coverage in orders of magnitude, the correlation is surprisingly strong already in flux space  ($\rhoS = 0.83$ and $\log \pS = -9.8$).
The slope of the correlation is flatter compared to that with the $12\um$ band, $\log \fqn \propto (0.88\pm0.07) \log \fxi$.
In luminosity space, the partial correlation rank is higher than for   using $12\um$ ($\rhop = 0.90$) while the observed and intrinsic scatter measures are smaller ($\sigmx = 0.3$ and $\sigint = 0.25$).
Thus, the correlation appears to be significantly tighter for $18\um$ compared to $12\um$.
However, this is just a selection effect because the scatter becomes as small for $12\um$ if exactly the same objects are used as for the $18\um$ analysis.
Note that the slope of the linear fit is higher than in flux space  and consistent with the $12\um$-based fitted slopes, $\log \lqn \propto (1.00\pm0.05) \log \lxi$.

\begin{figure}
%    \centering
%    \sidecaption
   \includegraphics[angle=0,width=8.5cm]{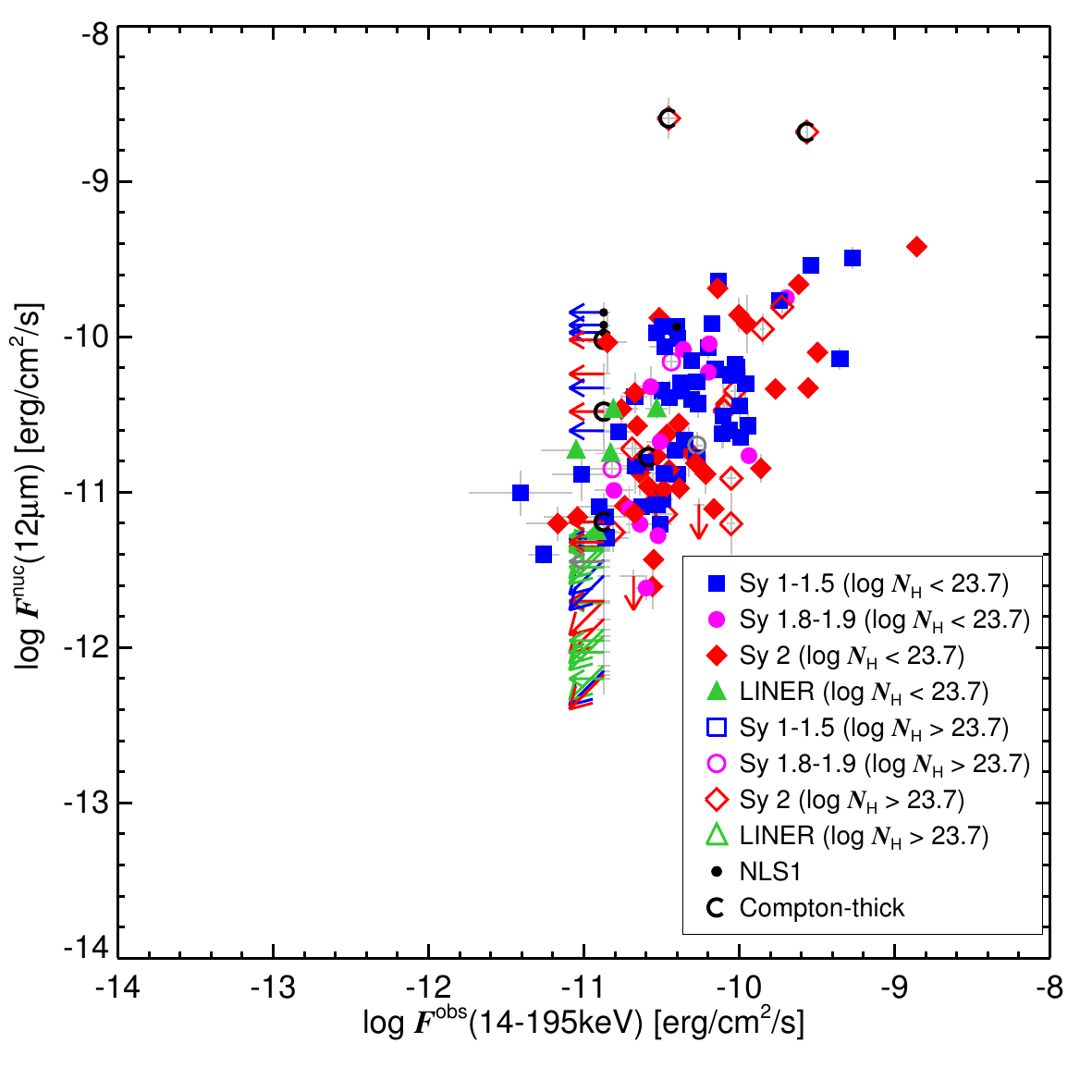}
    \caption{
             Relation of the nuclear $12\um$ and observed 14-195\,keV fluxes for the reliable sample.
             Symbols and colours are as in Fig.~\ref{fig:f12_f2obs} apart from filled symbols here marking objects with  $\log \nh \le 23.7$. 
            }
   \label{fig:f12_f14}
\end{figure}

\begin{figure}
%    \centering
%    \sidecaption
   \includegraphics[angle=0,width=8.5cm]{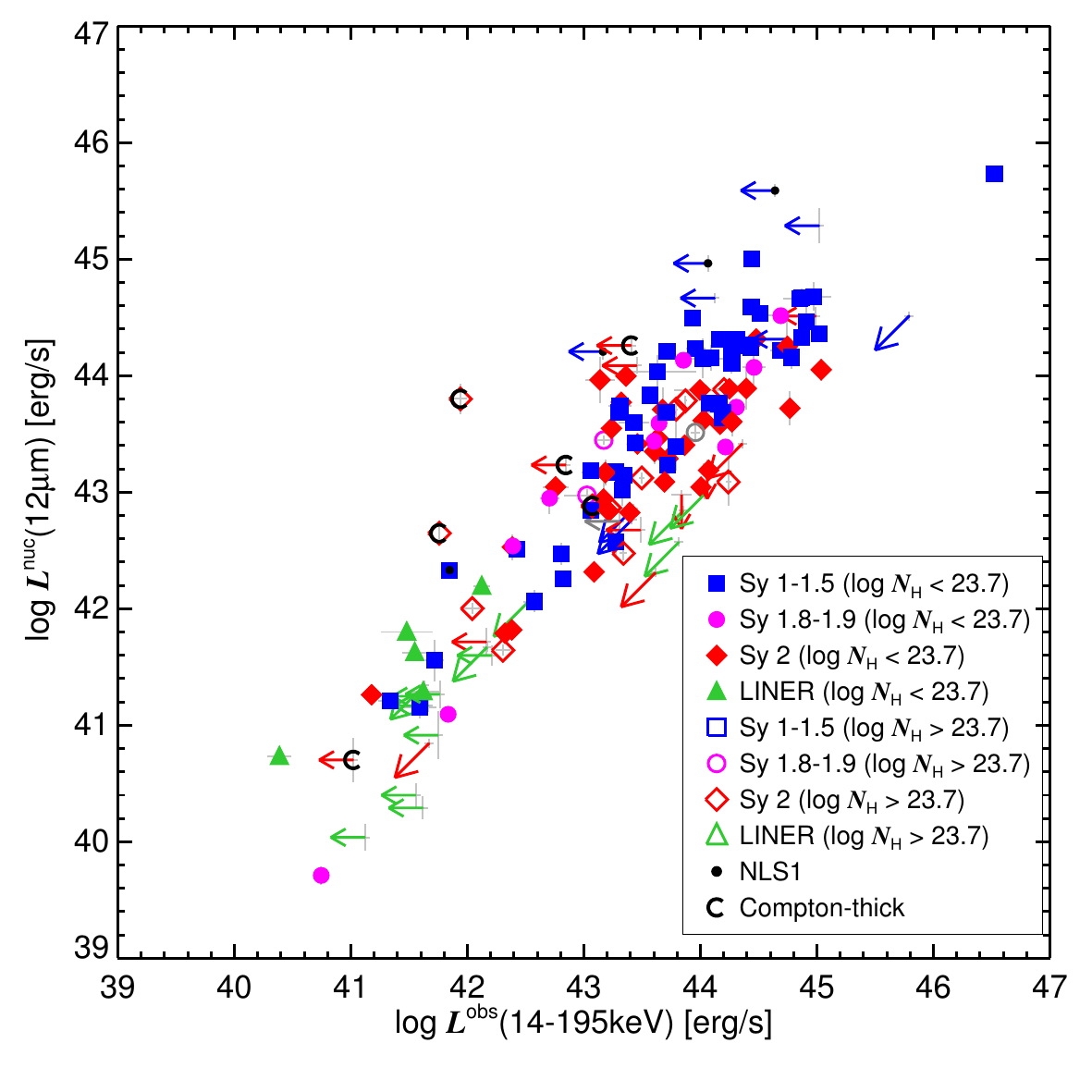}
    \caption{
             Relation of the nuclear $12\um$ and observed 14-195\,keV luminosities for the reliable sample.
             Symbols and colours are as in Fig.~\ref{fig:f12_f14}.
            }
   \label{fig:l12_l14}
\end{figure}

Instead of using the substantially absorption-affected 2-10\,keV energy range, one can now use the 14-195\,keV long-term data obtained with \swift/BAT over the last years, which is available for 118 of the 152 reliable AGN.
This energy range is less affected by absorption, and the data are also less prone to variability because they are long-term averages. 
The relation of the observed 14-195\,keV to $12\um$ fluxes and luminosities is shown in Figs.~\ref{fig:f12_f14} and \ref{fig:l12_l14} respectively.
Similar to the 2-10\,keV band, a correlation is present ($\rhoS = 0.60$ and $\log \pS = -11.9$ in flux space and $\rhop = 0.81$ in luminosity space).
The correlation strength further increases in luminosity space if we regard only those 101 objects that are unaffected by absorption in the 14-195\,keV band, i.e. $\log \nh < 23.7$ ($\rhop = 0.62$).
The correlation strength still remains lower than that of the $12\um$--2-10\,keV correlation however. 
In addition, the observed and intrinsic scatter are higher as well ($\sigmx = 0.41$ and $\sigint = 0.39$) while the fitted slope is slightly lower: $\log \lnn \propto (0.91\pm0.04) \log \lxh$.
As a consistency check, we also compute the 14-195\,keV--2-10\,keV correlation, which yields $\log \lxh \propto (1.01\pm0.02) \log \lxi$, also listed in Table~\ref{tab:cor}. 
Thus, taking this slight non-linearity into account, the different luminosity correlations are consistent to each other within $1\sigma$.

In the following, we concentrate on the $12\um$ versus intrinsic 2-10\,keV luminosity correlation because the corresponding data are available for all objects of the reliable sample.

\subsection{Dependence on luminosity}\label{sec:ldep}
\begin{figure}
%    \centering
%    \sidecaption
   \includegraphics[angle=0,width=8.5cm]{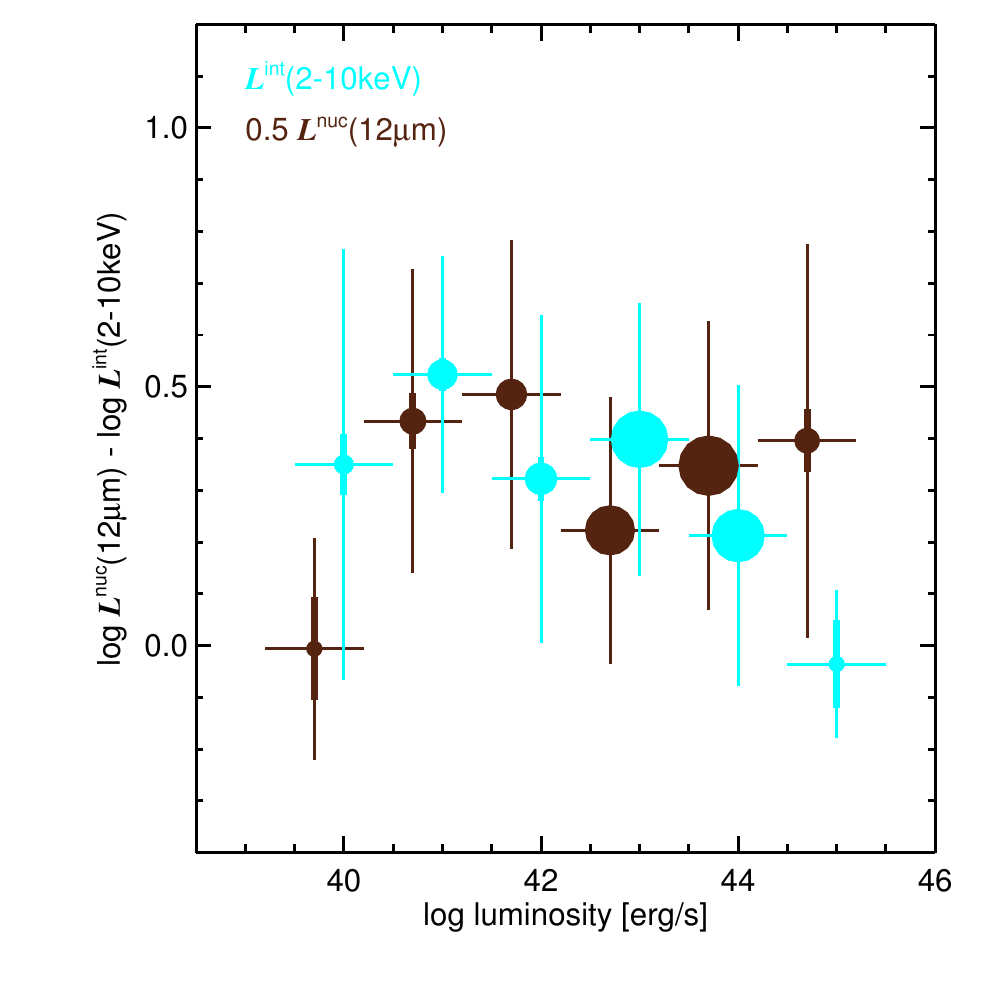}
    \caption{
             MIR--X-ray ratio over luminosity for the reliable sample.
             The data are binned with 1\,dex bin width.
             Cyan symbols mark the error-weighted mean for a binning with $\lxi$ as $x$-axis while brown symbols mark a binning along $\lnn$. 
             The latter is shifted by $-0.3$\,dex to compensate for the constant $a$ in the correlation. 
             The area of the filled circles is proportional to the number of objects in that bin with the smallest circle corresponding to six and the largest to 61 objects.            
             The error bars mark the bin width and error-weighted standard deviation of the MIR--X-ray ratio in that bin.
             In addition, the uncertainty of the weighted means are marked with thick vertical error bars which in most cases are smaller than the filled circles.
            }
   \label{fig:rat_lum}
\end{figure}

The increased number of objects and improved accuracy of the MIR and X-ray data motivates the search for fine structure in the MIR--X-ray correlation based on other AGN properties.
First, we investigate whether there is a luminosity dependence in the logarithmic MIR--X-ray ratio, $\ratmx$, and whether the slope of the correlation changes with luminosity. 
For this purpose, we compute the error-weighted mean MIR--X-ray ratio of the detected objects from the reliable sample for different binnings in $\lxi$ and $\lnn$, shown in Fig.~\ref{fig:rat_lum}.
We exclude the NLS1 sources from this analysis for the reasons given in Sect.~\ref{sec:typ}, as well as 3C\,273.
The latter object is presumably beamed (e.g., \citealt{soldi_multiwavelength_2008}) and would have an disproportionally large effect on the fitting because of its isolated position at very high luminosities ($\approx 10^{46}$\,erg/s), separated from the main population by more than an order of magnitude.
Note that 3C\,273 however follows the general behaviour found below.
There is a weak global trend of decreasing $\ratmx$ with increasing luminosity for binning along the $\lxi$ axis. 
%and along the $\lnn$ axis although with large scatter.
For the X-ray binning, the ratio changes from approximately 0.4\,dex to 0\,dex over approximately five orders of magnitude in luminosity, which is consistent with the fitted slopes $< 1$ that we found in Sect.~\ref{sec:cor}.
Out of the 14 upper limits, only LEDA\,013946 is not consistent with the general behaviour by more than 1$\sigma$ ($\log \lxi = 43.2$; $\ratmx \le 0.08$).
%If the binning is performed along the $\lxi$ axis, then a decreasing trend of $\ratmx$ with increasing $\lxi$ becomes obvious. 
In particular, at the highest probed luminosities, $\log \lxi \gtrsim 44$, a significant decrease seems to be present.
The two-sided KS test provides maximum difference for separating the population at a threshold of $\sim 44.1$ ($\dks=0.66$; $\log \pks = -4.2$).
We perform also the Monte-Carlo sampling with the above found threshold as described in Sect.~\ref{sec:meth} and get a 68 per cent confidence interval for $\pks$ of  0.004 to 3 per cent.
\begin{figure}
%    \centering
%    \sidecaption
   \includegraphics[angle=0,width=8.5cm]{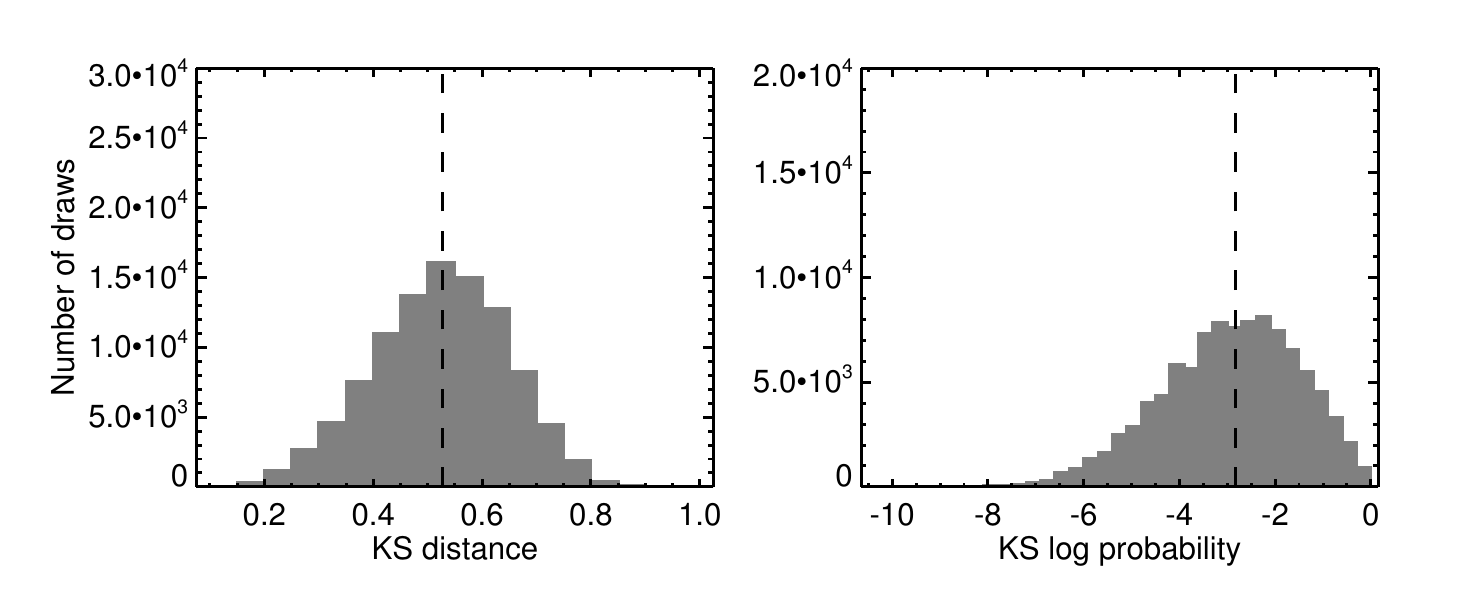}
    \caption{
             Results of the KS test for an MIR--X-ray ratio change above the luminosity $\log \lxi = 44.1$.
             The dashed vertical lines mark the median. 
            }
   \label{fig:KS_lum}
\end{figure}
However, such a decrease is not visible when the binning is done along the $\lnn$ axis (brown data points in Fig.~\ref{fig:rat_lum}) and thus the corresponding KS test shows no significant difference. 
%If a real trend would be present it would have to show up consistently for both binnings. 

\begin{figure}
%    \centering
%    \sidecaption
   \includegraphics[angle=0,width=8.5cm]{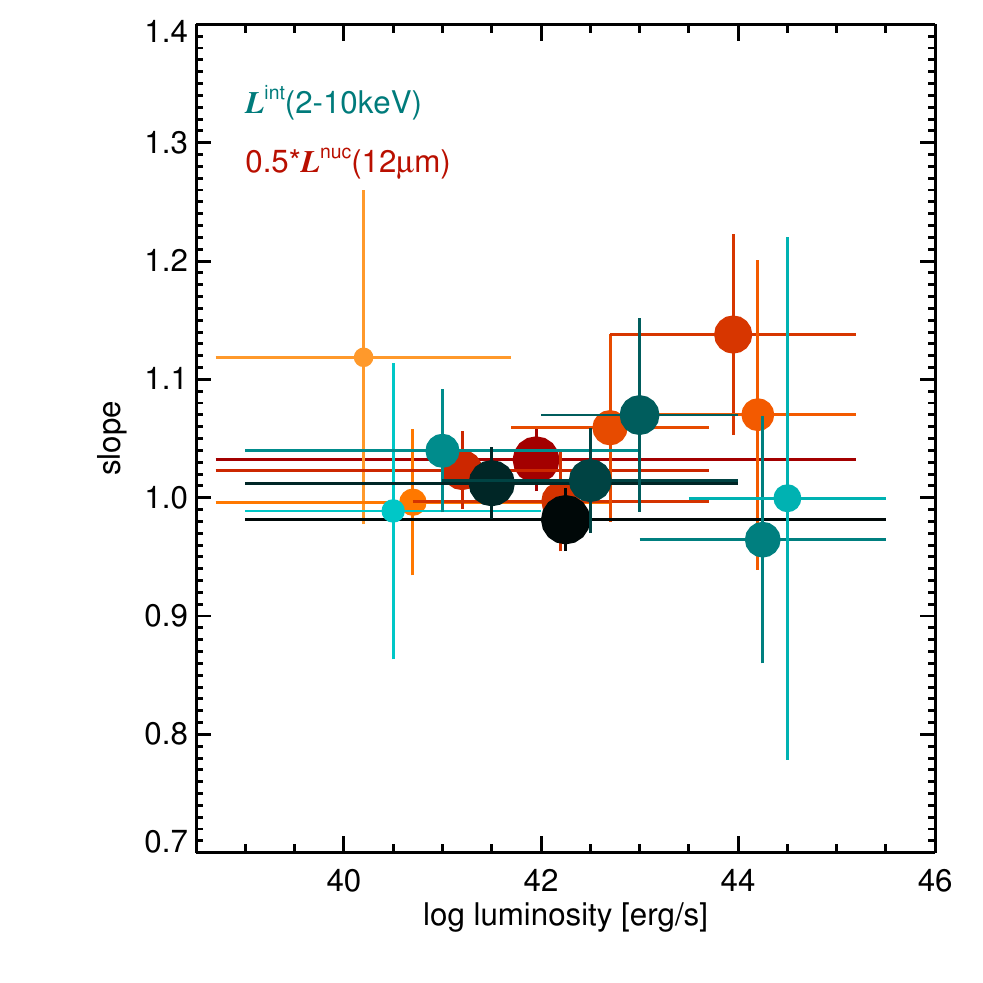}
    \caption{
             Fitted linear slopes, $b$ of $\log(\lnn)-43 = a + b (\log(\lxi)-43)$ for the reliable sample.
             Each data point represents the slope of the \texttt{linmix\_err} fit for a sub-sample with luminosities in between the thresholds given by the width of the horizontal error bars.
             The vertical error bar gives the 1-$\sigma$ uncertainty of the fitted slope.
             The area of the filled circles and darkness in color are proportional to the number of objects in that sub-sample with the smallest circle corresponding to 23 and the largest to 147 objects.            
             Cyan to black coloured symbols mark sub-samples with thresholds in $\lxi$ while orange to red symbols mark sub-samples with thresholds in $\lnn$.
             For the latter, $\log(\lxi)-43 = a' + b' (\log(\lnn)-43)$ was fitted and $b'$ then inverted for direct comparison.  
             In addition, the $\lnn$-based data points are shifted by $-0.3$\,dex to compensate for the constant $a$ in the correlation.             
            }
   \label{fig:slope}
\end{figure}

%A similar effect is seen for the slope of the linear fit for different luminosity cuts in $12\um$ and 2-10\,keV space (Fig.~\ref{fig:slope}).
Therefore, we employ another diagnostic, namely to compare the slopes of linear fits for different luminosity cuts in $12\um$ and 2-10\,keV space in Fig.~\ref{fig:slope}.
The individual fitted slopes, $b$, for using thresholds in either $\lxi$ or $\lnn$ space are between 0.95 and 1.15 and are in general consistent with the unity. 
Only the highest luminosity bin for $\lnn$ has a steep slope of $\sim1.14$ that differs from unity by more than one $\sigma$. 
On the other hand, the corresponding $\lxi$ bin has the lowest slope of $\sim 0.96$. 
Again, this is caused by the insufficient object numbers above $10^{44}$\,erg/s.
However, the two highest bins of both threshold directions are consistent with the slopes at lower luminosities and do not indicate any trend. 
%Interestingly, even the slopes fitted to the whole sample are affected by this inconsistency at the high luminosity end.
%It  leads to a flatter slope when $\lxi$ is used as independent variable (0.99; (black circle in Fig.~\ref{fig:slope}) and a steeper slope when $\lnn$ is treated as independent variable (1.05; red circle) compared to those fits that exclude the high-luminosity end.
%This indicates that the "true" global slope is between 0.99 and 1.05, which is however at odds with the result from the luminosity ratio investigation above.
Therefore, we conclude that there is no conclusive evidence for any local changes in the MIR--X-ray ratio or correlation slope with luminosity (see Sect.~\ref{sec:lit} for further discussion and also \citealt{mateos_revisiting_2015}; but see \citealt{stern_x-ray_2015}).

\subsection{Dependence on optical type}\label{sec:typ}
\begin{figure}
%    \centering
%    \sidecaption
   \includegraphics[angle=0,width=8.5cm]{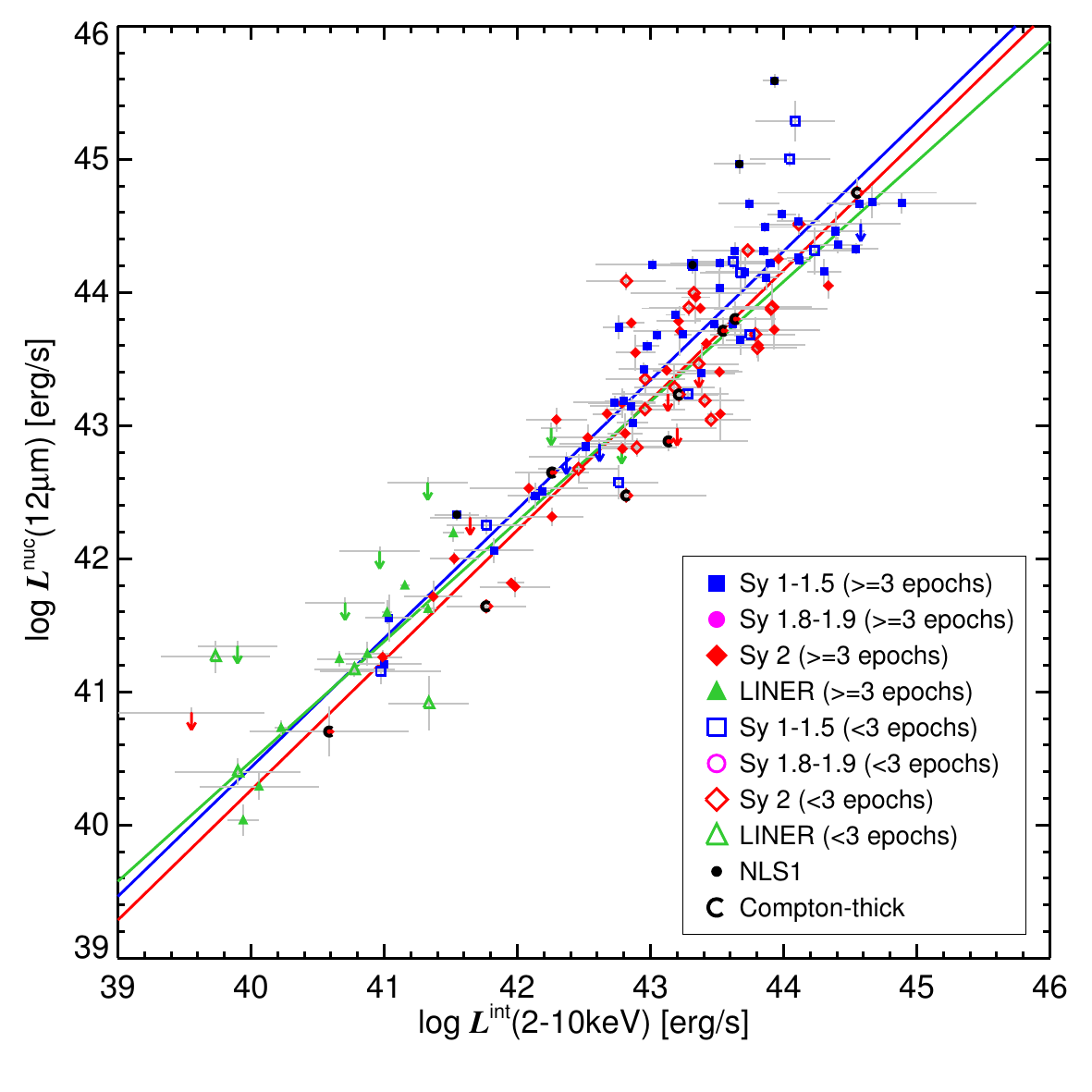}
    \caption{
             Relation of the nuclear $12\um$ and intrinsic 2-10\,keV luminosities for the different optical AGN types.
             Symbols and colours are as in Fig.~\ref{fig:f12_f2int}.
             In addition, the blue, red and green lines mark the \texttt{linmix\_err} fits to the type~1, type~2 and LINER objects in the reliable sample.
            }
   \label{fig:l12_l2int_type}
\end{figure}

Next, we investigate differences with the optical type by fitting the $12\um$--2-10\,keV luminosity distributions separately for type~I, type~II and LINER AGN (Fig.~\ref{fig:l12_l2int_type}).
Note that we treat NLS1 not as type~I AGN and that we also exclude 3C\,273 from this analysis.
This leaves 54 type~I AGN of the reliable sample, while there are 58 type~II and 19 LINER objects.
The corresponding linear fits are listed in Table~\ref{tab:cor} and shown in Fig.~\ref{fig:l12_l2int_type} as well.
The slopes of the type~I and II correlations are similar to each other and the total population fit. 
However, type~I objects display a systematically higher constant $a$ and MIR--X-ray ratio than type~II objects ($0.15$\,dex difference at 2$\sigma$ significance).
Note that this offset is consistent to the finding for more powerful AGN \citep{honig_quantifying_2011}. 
In addition, the observed and intrinsic scatter is lower in the type~I population ($\sigint = 0.25$) compared not only to the type~II ($\sigint = 0.31$) but also to the whole reliable sample ($\sigint = 0.32$). 
Although the uncertainties of these values are larger than the differences, it indicates that the scatter in the total population is dominated by the type~II objects.
LINERs exhibit the highest average MIR--X-ray ratio (0.46) and a flatter but consistent fitted slope compared to the Seyferts.
We note that all 14 upper limits except again LEDA\,013946 are consistent with the obtained fits within the 1$\sigma$ X-ray uncertainty.
Finally, there are too few NLS1 for a reasonable fit but their average MIR--X-ray ratio is very high compared to the other AGN classes (1.16).

\begin{figure}
%    \centering
%    \sidecaption
   \includegraphics[angle=0,width=7.5cm]{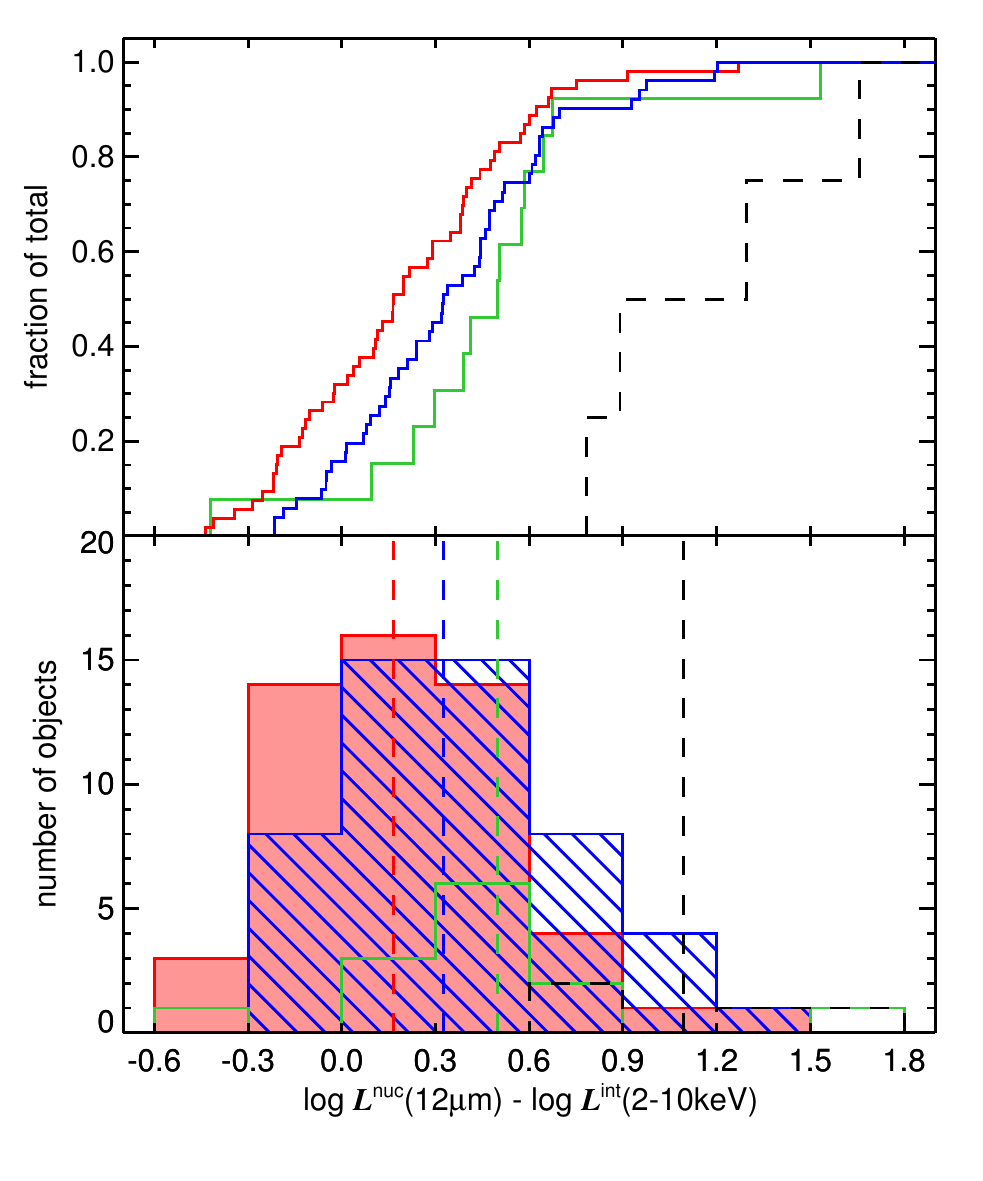}
    \caption{
    Distribution of the 12$\um$--2-10\,keV ratio for the different optical AGN types.
    \textit{Top:} Normalised cumulative distribution with the blue solid line being type~1, the red type~2, the green LINER, and the black dashed line NLS1.
    \textit{Bottom:} Histogram distribution with blue hatched being type~1, red solid type~2, green empty LINER, and black empty dashed NLS1. 
    The dashed vertical lines indicate the corresponding median  12$\um$--2-10\,keV ratios.
            }
   \label{fig:rat_hist}
\end{figure}

\begin{figure}
%    \centering
%    \sidecaption
   \includegraphics[angle=0,width=8.5cm]{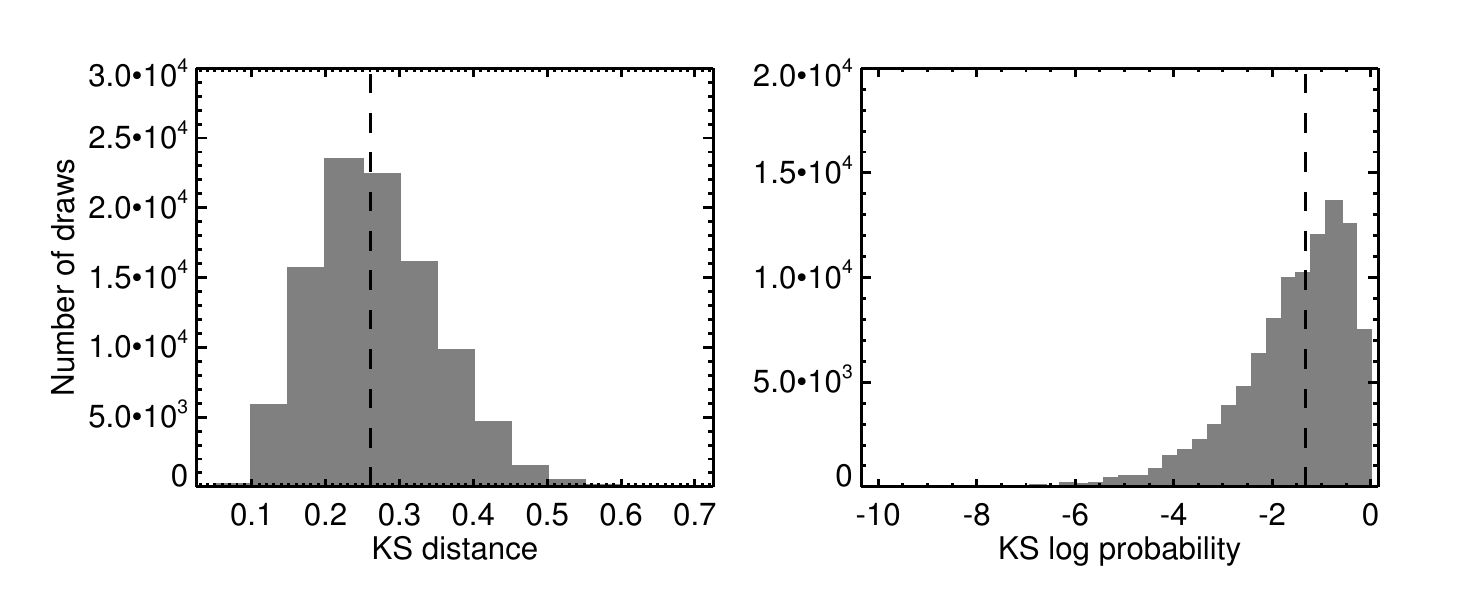}
    \caption{
             Results of the KS test for an MIR--X-ray ratio difference between type~I and II AGN.
             The dashed vertical lines mark the median. 
            }
   \label{fig:KS_type}
\end{figure}

To further investigate if the differences between the different optical classes are statistically significant we employ the two-sided KS test on the distribution of MIR--X-ray ratios of the MIR-detected reliable AGN shown in Fig.~\ref{fig:rat_hist}.
Note that here the upper limits of all MIR non-detected AGN are within the central 68 per cent interval of the $\ratmx$ distributions for their corresponding optical classes.  
The difference between type~I and type~II AGN is not  significant at 3-$\sigma$ level according to this test, the resulting distribution shown in Fig.~\ref{fig:KS_type}.
The 68 per cent confidence interval for the null-hypothesis probability is from 0.25 to 30 per cent.

Similarly, the difference between LINERs to the combined type~I+II sample remains insignificant at 3-$\sigma$ level, the 68 per cent interval for $\pks$ is 0.5 to 36 per cent.
Still, we state a correlation fit excluding LINERs (type~I+II) in Table~\ref{tab:cor} for completeness.

The largest statistically significant difference is found for the NLS1 versus type~I according to the KS-test with a 68 per cent interval for $\pks$ of 0.06 to 2 per cent and a median $\dks = 0.86$.
A possible explanation for their high MIR--X-ray ratios is that owing to their steep X-ray spectral energy distributions, the 2-10\,keV emission is not representative of the bolometric luminosity in NLS1. 
On the other hand, NLS1 generally have complex X-ray spectra which make intrinsic luminosity estimates difficult. 
Thus, a larger sample of NLS1 is required to verify our finding.

\subsection{Dependence on X-ray column density}\label{sec:n_h}
\begin{figure}
%    \centering
%    \sidecaption
   \includegraphics[angle=0,width=8cm]{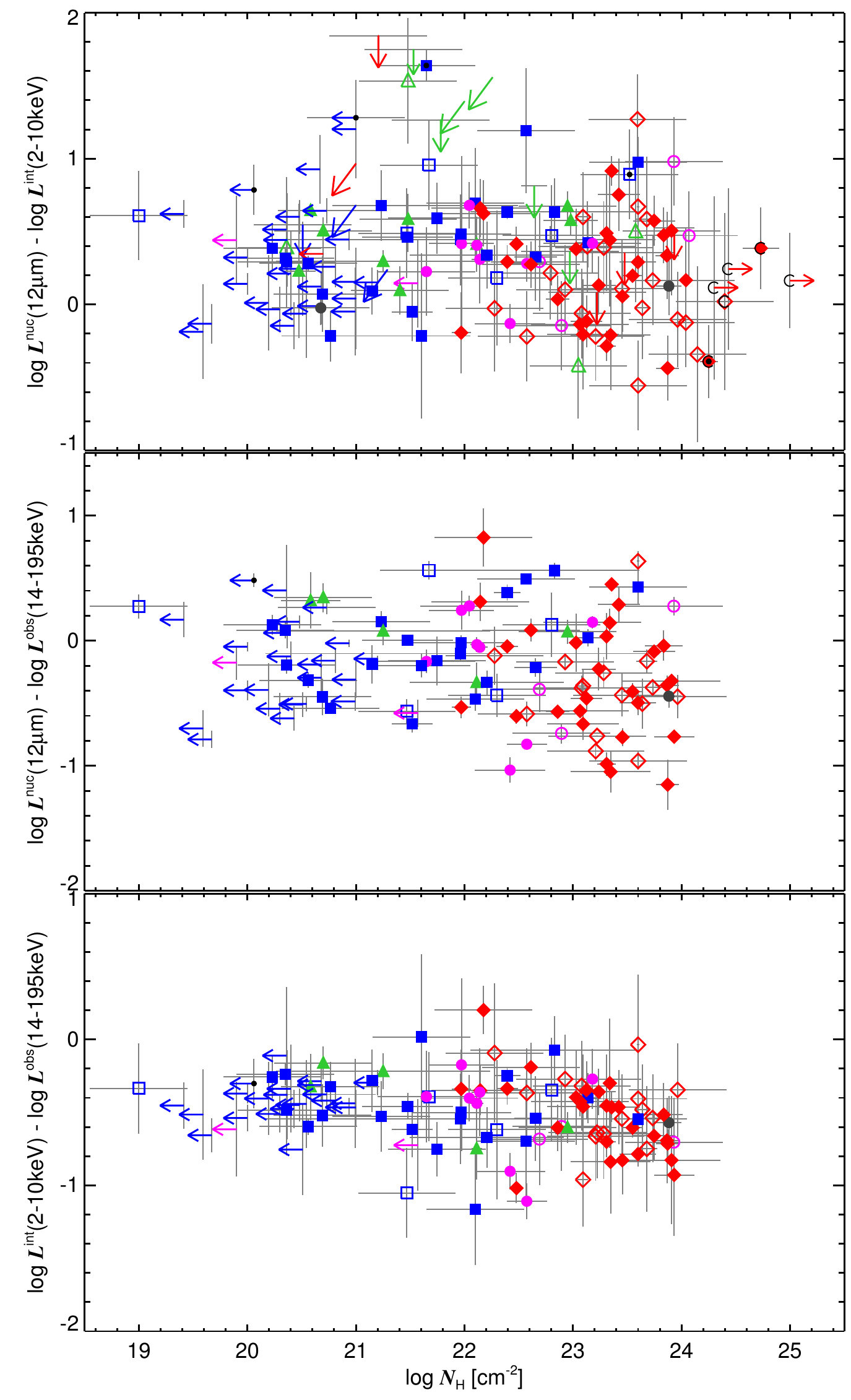}
    \caption{MIR--X-ray ratio versus the X-ray column density for the reliable sample. 
    \textit{Top:} 12$\um$--2-10\,keV ratio.
    Symbols and descriptions are as in Fig.~\ref{fig:f12_f2int}.
    \textit{Middle:} 12$\um$--14-195keV ratio for all 14-195\,keV detected non-CT objects in the reliable sample. 
    \textit{Bottom:} 2-10--14-195 ratio for all objects as in the middle plot. 
            }
   \label{fig:ratio_nh}
\end{figure}

To investigate possible differences between obscured and unobscured AGN, it is presumably superior to use the X-ray column density, $\nh$, rather than the optical type because $\nh$ probes the nuclear obscuration more directly and in a quantitative way and neutralises possible selection effects against obscured AGN. 
According to results of the previous sections, we exclude the NLS1 sources here.
Fig.~\ref{fig:ratio_nh} displays the relation of the MIR--X-ray ratio with the column density.
As expected, type~I AGN have in general $\log \nh \lesssim 22$ and type~II AGN $\log \nh \gtrsim 22$. 
However, there are a few exceptions, namely two type~I AGN with $\log \nh > 23$ (3C\,445 and ESO\,323-77).
In both cases, the absorption is highly variable and/or partly caused by broad-line region clouds \citep{reeves_chandra_2010,miniutti_properties_2014}.
In addition, there are three type~II AGN with $\log \nh < 22$, namely the true Sy\,2 candidate NGC\,3147 \citep{pappa_x-ray_2001} and the borderline Sy\,2/LINER objects 3C\,317 and NGC\,4374. 
LINERs, intermediate type AGN (and NLS1) are distributed over most of the whole $\nh$ range.
No global trend of the MIR--X-ray ratio with $\nh$ is obvious in Fig.~\ref{fig:ratio_nh} when using either the intrinsic 2-10\,keV or the observed 14-195\,keV band,
 and none of the objects with limits on either $\ratmx$ or $\nh$ sticks out. 
On the other hand, there seems to be a decline of the MIR--X-ray ratio for column densities $\gtrsim 10^{23}$\,cm$^{-2}$. 
For example, the median $\nh$ is $>10^{23}$\,cm$^{-2}$ for all 32 objects with a $12\um$--2-10\,keV ratio $<0$.
In particular, all four objects with $12\um$--2-10\,keV ratio $<-0.3$ have $\nh >10^{23}$\,cm$^{-2}$ (NGC\,1144, NGC\,3169, ESO\,297-18, and NGC\,5728).
However, the result of the KS test splitting the whole sample at $10^{23}$\,cm$^{-2}$ does not indicate significant differences (68 per cent confidence interval on $\pks$ is from 0.2 to 27 per cent). 

\begin{figure}
%    \centering
%    \sidecaption
   \includegraphics[angle=0,width=8.5cm]{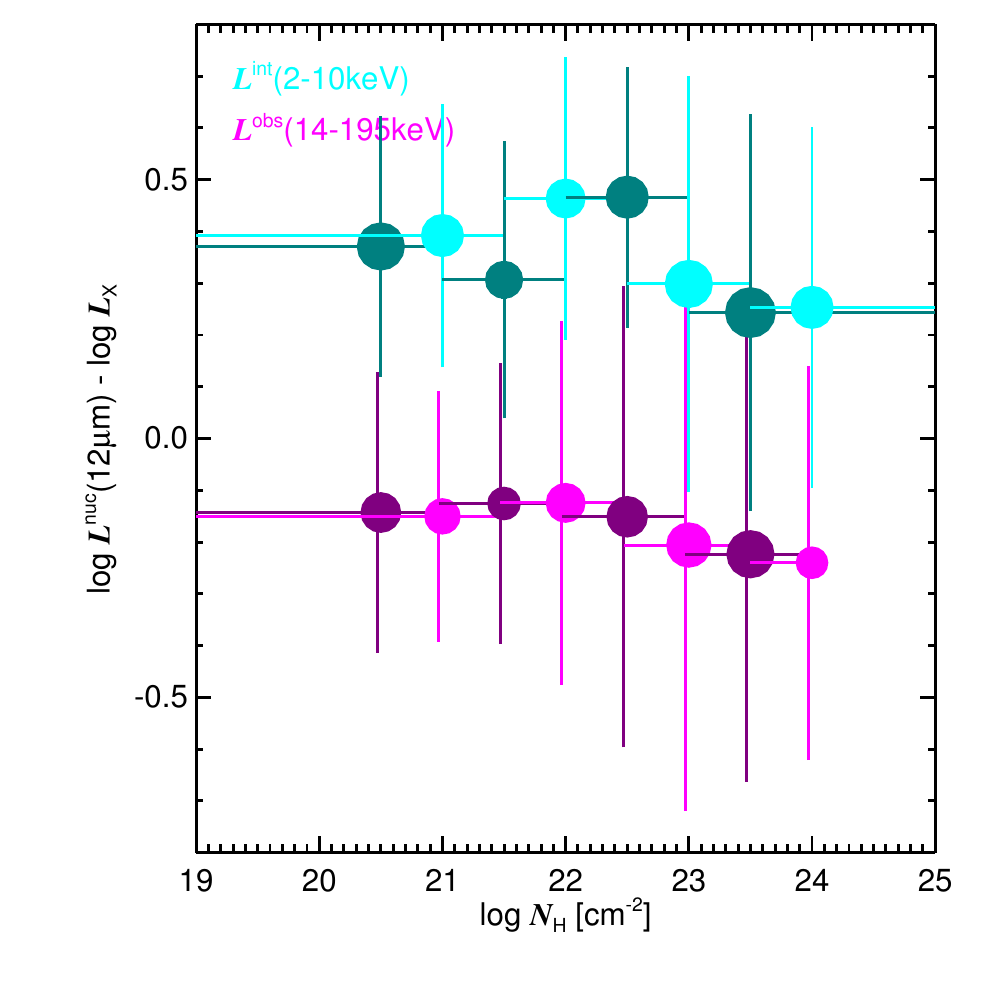}
    \caption{
             MIR--X-ray ratio over X-ray column density for the reliable sample.
             The data are binned with 1\,dex bin width and two different choices of bin centres, shifted by 0.5\,dex.
             Cyan symbols mark the error-weighted mean for a binning  using  $\lxi$ whereas lighter and darker color distinguish the two binnings.
             Magenta symbols mark a binning using $\lxh$ for the MIR--X-ray ratio again with lighter and darker color distinguishing the two binnings. 
             The area of the filled circles is proportional to the number of objects in that bin with the smallest circle corresponding to 18 and the largest to 40 objects.
             The error bars mark the bin width and weighted standard deviation of the MIR--X-ray ratio in that bin.
             In addition, the uncertainty of the weighted means are smaller than the filled circles in all cases.
            }
   \label{fig:rat_bin_nh}
\end{figure}

In order to quantify a decrease of $\ratmx$  at high $\nh$ better, we first look at the weighted mean MIR--X-ray ratio for both 2-10\,keV and 14-195\,keV using different binnings of the column density in Fig.~\ref{fig:rat_bin_nh}.
Here, objects with only limits on $\ratmx$ are excluded, which includes 14-195\,keV upper limits for the binning in $\lxh$.  Limits in $\nh$ are treated as belonging to the bin of their limit value for the sake of simplicity. 
This does not affect the results obtained below.  
The binnings using both $\lx$ and $\lxh$ for the MIR--X-ray ratio show qualitatively a similar behaviour, namely a $\sim 0.1$\,dex increase of $\ratmx$ for $22 \lesssim \log \nh \lesssim 23$, followed by a decrease of $\sim 0.15$\,dex for $\log \nh \gtrsim 23$.
Note that this trends occurs independently of the exact bin locations chosen as demonstrated in Fig.~\ref{fig:rat_bin_nh} and also when removing the outliers with the most extreme MIR--X-ray ratios at intermediate column densities visible in Fig.~\ref{fig:ratio_nh}.  
Thus, there is indeed a weak global decreasing trend present with, however, a maximum MIR--X-ray ratio at intermediate column densities, instead of at  the lowest column densities as naively expected. 
The KS test, yields a median $\pks = 6$ per cent (68 per cent confidence interval: 0.2 to 30 per cent) between the intermediate and highly obscured objects using the thresholds above.
Furthermore, the ten objects with highest obscuration ($\log \nh > 24$), exhibit the lowest average MIR--X-ray ratio of $0.07$ ($\sigmx = 0.27$) compared to $0.36$ ($\sigmx= 0.41$) for objects with $\log \nh < 24$.
The median $\pks$ is  6 per cent (68 per cent confidence interval: 0.3 to 35 per cent) from a corresponding KS test.
Therefore, the observed trend might be real since consistent for both $\lx$ and $\lxh$ but is not significant at a 3-$\sigma$ level. 
The upper limits are consistent with the result except for two of the 19 objects with non-detections in $\lxh$ but detections in $\lnn$, PG\,2130+099 and MCG-2-8-39, which exhibit exceptionally high lower limits on the logarithmic $12\um$ to 14-195\,keV ratios of $\sim 0.5$.

\begin{figure}
%    \centering
%    \sidecaption
   \includegraphics[angle=0,width=8.5cm]{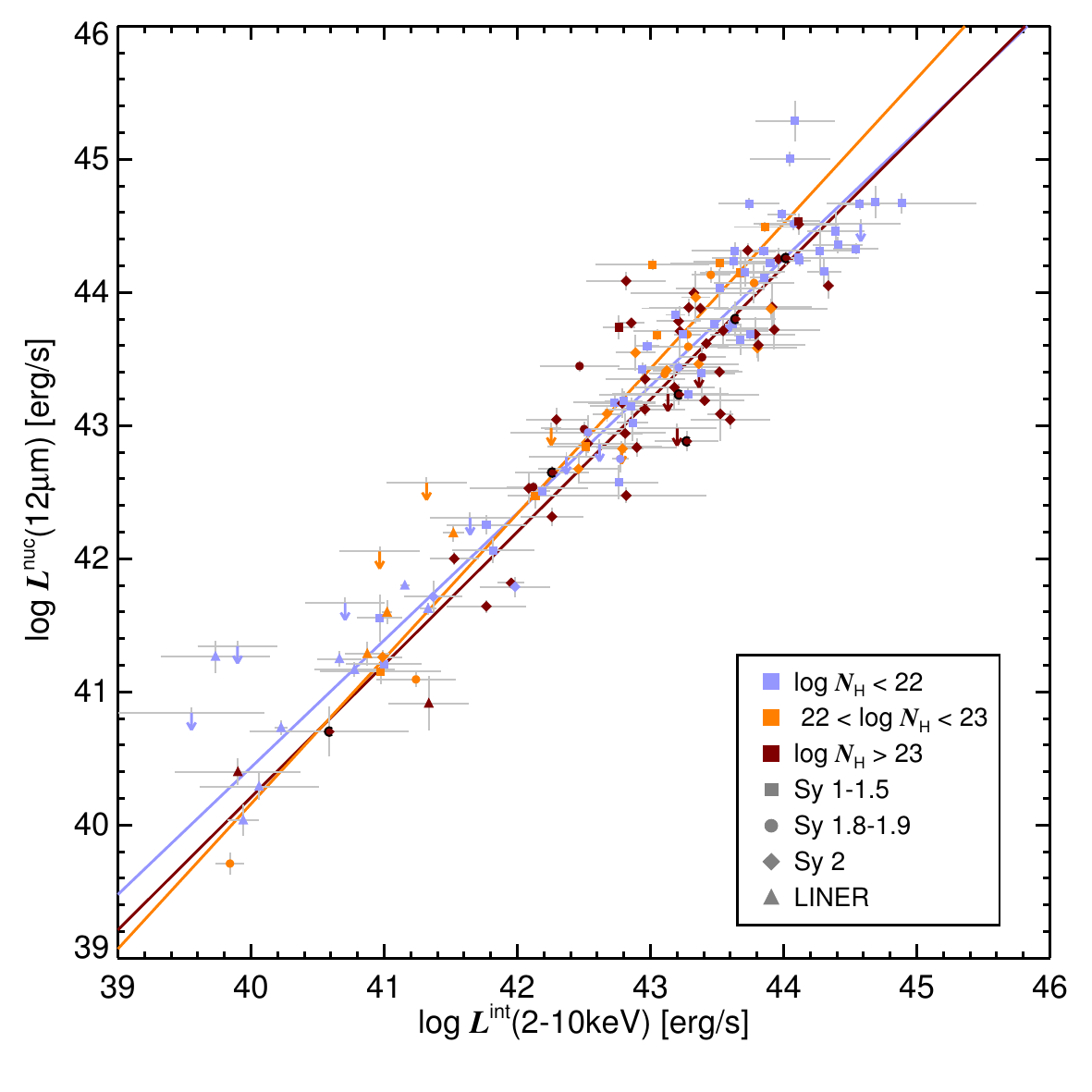}
    \caption{
             Relation of the nuclear $12\um$ and intrinsic 2-10\,keV luminosities for the different X-ray obscuration levels.
             Symbols are as in Fig.~\ref{fig:f12_f2int} while light blue marks objects with  $\log \nh < 22$, orange for $22 \lesssim \log \nh \lesssim 23$ and dark red for $\log \nh \gtrsim 23$.
             The corresponding \texttt{linmix\_err} fits are marked by solid lines in the same colors.
            }
   \label{fig:cor_nh}
\end{figure}

Second, we perform linear regression to the obscured and unobscured subsamples as displayed in Fig.~\ref{fig:cor_nh}.
Here, we exclude  3C\,273 from the analysis again. 
The correlation and fitting parameters listed in Table~\ref{tab:cor} show that the unobscured subsample shows a much stronger correlation and lower intrinsic scatter than the obscured  subsample ($\log \nh \gtrsim 23$), similar to the optical type division.
The linear fits of the unobscured and obscured AGN are consistent within the 1-$\sigma$ uncertainties with a minimal offset of the unobscured to a higher MIR--X-ray ratio by $\sim1$\,dex.
The intermediately obscured subsample ($22 \lesssim \log \nh \lesssim 23$) has a steeper slope ($1.09 \pm 0.06$) and higher intercept ($0.43 \pm 0.07$).
This result is insensitive against the exact selection of the $\nh$ threshold values, and removing the lowest luminosity objects would make the deviation from the fits of the other subsamples even larger.
Moreover, the partial correlation coefficient and scatter are tightest for the intermediately obscured subsample, making it stand out the most.
This is similar to the result of the previous test. 
We discuss the implications of these results in Sec.~\ref{sec:tor}.

\subsection{Dependence on radio-loudness}\label{sec:rad}
This section investigates whether there is any connection between MIR--X-ray correlation and the radio power of the AGN. 
This is interesting because non-thermal emission processes from a jet, if present, might significantly contribute or even dominate the MIR and/or the X-ray emission of the AGN.
For this purpose, we collect radio fluxes for all reliable objects from NED, of which 137 have detections reported in at least one frequency band. 
There is no single frequency band with observations of all targets. 
The most common is $\sim1.4\,$GHz with 121 objects (mainly from \citealt{white_new_1992, condon_nrao_1998, condon_radio_2002}), followed by $\sim4.9$\,GHz with 91 objects (mainly from \citealt{edelson_broad-band_1987,gregory_87gb_1991, gregory_parkes-mit-nrao_1994,barvainis_radio_1996,wright_parkes-mit-nrao_1996,nagar_evidence_2001,tingay_atca_2003,gallimore_parsec-scale_2004,nagar_radio_2005}).
Moreover, the data are highly heterogeneous with respect to the telescope used, angular-resolution, extraction apertures, and observing time. 
Therefore, rather than a direct comparison using the radio emission, we classify the objects into radio-quiet and radio-loud, following the definition by \cite{terashima_chandra_2003}, $\Rx = \log \lfive - \log \lxi$ and $\Rx > -4.5$ being radio-loud.
In order to classify objects with no available $\lfive$, we use the low-angular resolution measurements for  $\lone$ and $\lpeight$, the latter available for 33 objects \citep{mauch_sumss:_2003}.
These three bands cover all 137 objects with radio detections.
The average conversion factors from $\lfive$ to $\lone$ and $\lpeight$ are $\log \lone  = (0.35 \pm 0.85) + \log \lfive$ and $\log \lpeight = \log \lone (-0.14 \pm 0.12)$.
They are computed from the 85 and 19 objects with data in both bands respectively. 
Because the goal is to obtain robust radio-loudness determinations,
firstly, the standard deviations of the conversions are taken into account (e.g., only $\log \lone - \log \lxi > - 3.3$ is radio-loud). 
Secondly, high angular resolution $\lfive$ data are preferred, and $\Rx > -3$ as criterion is used when only low-angular resolution $\lfive$ is available.
Using these definitions, 48 objects can robustly be classified as  radio-loud, and 41 as radio-quiet. 
These classifications are listed in Table~\ref{tab:sam} for the individual objects.

\begin{figure}
%    \centering
%    \sidecaption
   \includegraphics[angle=0,width=8.5cm]{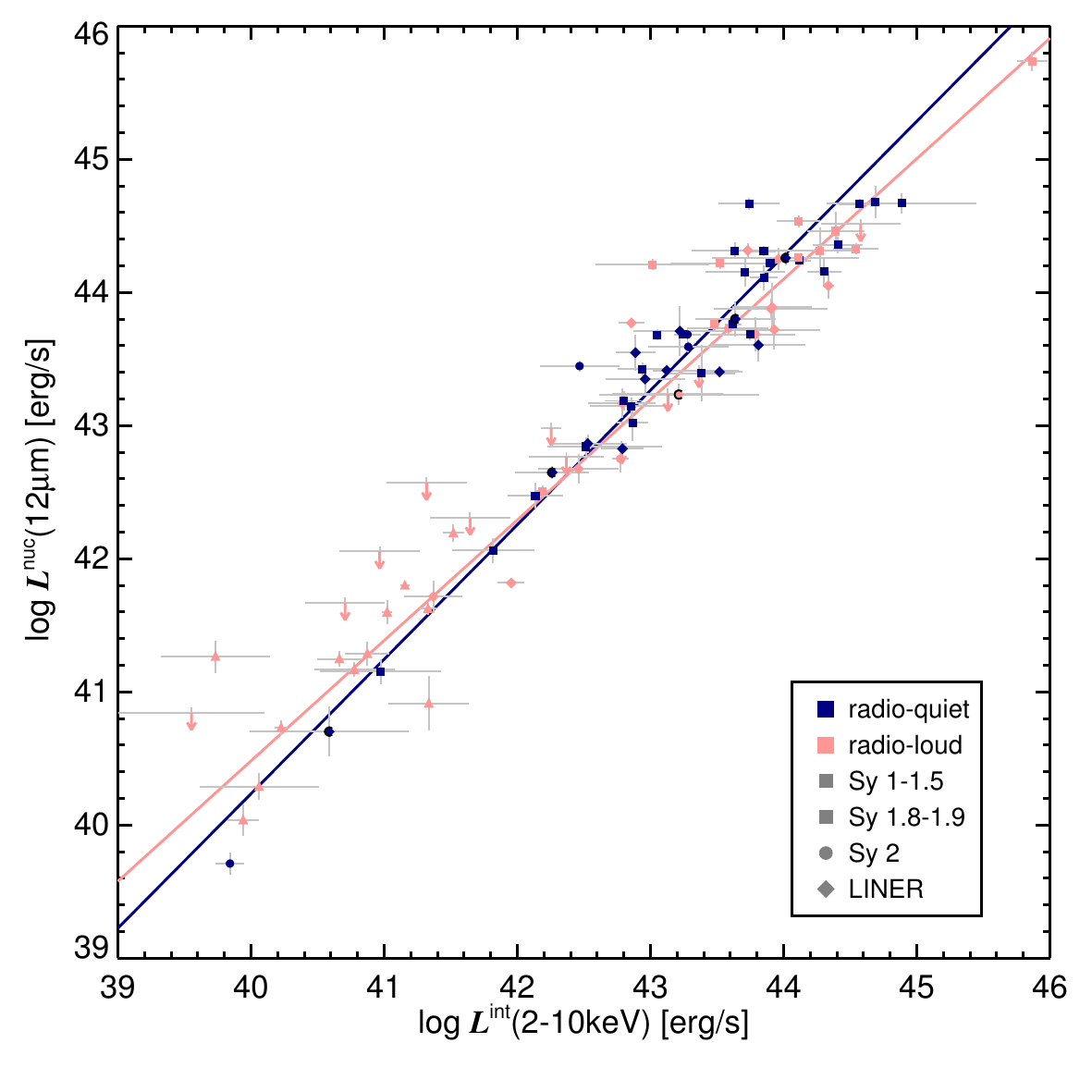}
    \caption{
             Relation of the nuclear $12\um$ and intrinsic 2-10\,keV luminosities for radio-quiet (dark-blue) and radio-loud (light-red) AGN.
             Symbols are as in Fig.~\ref{fig:f12_f2int}.
             The corresponding \texttt{linmix\_err} fits are marked by solid lines in the same colors.
            }
   \label{fig:cor_radio}
\end{figure}

Fig.~\ref{fig:cor_radio} shows the MIR--X-ray correlation for radio-loud and radio-quiet objects.
Radio-loud objects tend to exhibit on average higher MIR--X-ray ratios at low luminosities compared to the radio-quiet ones, while this trend inverts at high luminosities. 
The transition happens between $42 \lesssim \log \lxi \lesssim 43$.
Therefore, the corresponding radio-loud fit has a significantly flatter slope ($0.90 \pm 0.04$) than the radio-quiet fit ($1.01\pm0.05$; see Table~\ref{tab:cor}).
Note that this result does not depend on the low-luminosity objects ($ \log \lxi \le 42$). 
If these are removed from the fitting for the radio-loud subsample, the slope becomes even a little bit flatter ($0.88 \pm 0.09$). 
The radio-quiet sample shows a remarkable high partial correlation (0.88) and small scatter ($\sigmx = 0.28$; $\sigint = 0.21$).
Both are worse for the radio-loud subsample.

This result can be explained by the presence of a jet in the radio-loud objects which dominates the MIR emission at low-luminosities (e.g., \citealt{mason_nuclear_2012}).
Most of these objects are optically classified as LINERs, which showed a higher MIR--X-ray ratio already in Section~\ref{sec:typ}.
More puzzling is the situation at high luminosities where the jet apparently contributes much more to the X-rays (e.g., \citealt{hardcastle_active_2009}; see next section).

\section{Discussion}\label{sec:dis}
\subsection{Dependence of results on the fitting method}\label{sec:fit}
In order to investigate how the exact form of the correlation depends on the fitting method chosen, we concentrate on the $\lnn$--$\lxi$ correlation for the whole reliable sample.
First, we verify that the choice of input parameters for \texttt{linmix\_err} does not influence our results.
If instead of three Gaussians (the default) we choose a uniform prior distribution, the intercept, $a$, becomes  $\sim 2$ per cent smaller while the slope, $b$, increases by $\sim 0.1$ per cent. 
Both are much smaller than the uncertainties on $a$ and $b$.
Using two Gaussians leads to even smaller deviations.
The same applies to changing the algorithm to create the Markov Chains or the number of random draws as long as the latter is sufficiently large (here $10^4$). 

For the following comparisons, upper limits can not be taken into account.
Using only the 138 detected objects in the reliable sample, $a$ and $b$ become $0.33 \pm 0.03$ and $0.98 \pm 0.03$ respectively. 
The correlation parameters using \texttt{fitexy} instead are $a=0.37 \pm 0.01$ and $0.99 \pm 0.01$ and thus consistent to within 1-$\sigma$.
In other words, the difference in predicted $\lnn$ depending on the algorithm is smaller than $0.07$\,dex over the whole luminosity range ($39 \le \log \lxi \le 46$).
Therefore, we conclude that the choice of fitting algorithm or inclusion of non-detections does not have a significant effect on the results obtained here. 

However, we note that at least for using \texttt{linmix\_err}, the choice of independent variable makes a difference by more than 1 $\sigma$. 
Specifically, for using $\lnn$ as independent variable, $a$ and $b$ become $0.33 \pm 0.03$ and $1.06 \pm 0.03$ when inverting back to the $\log \lnn = a + b \log \lxi$ form. 
Therefore, for predicting $\lxi$ based on a $\lnn$ measurement, the following equation should be used:
\begin{eqnarray}\label{eq:inv}
 \log \left( \frac{\lxi}{10^{43}\textrm{erg}\,\mathrm{s}^{-1}}\right)   &=& ( -0.32 \pm 0.03) \cr &+& ( 0.95 \pm 0.03 )  \log \left( \frac{\lnn}{10^{43}\textrm{erg}\,\mathrm{s}^{-1}}\right).
\end{eqnarray}

\subsection{Sample biases}\label{sec:bias}
The sample of the 152 reliable AGN from the MIR atlas is neither uniform nor complete.
In order to test whether the MIR--X-ray correlations are representative for the local AGN population, we perform the following two tests.

\subsubsection{Uniform sample test}\label{sec:BAT}
\begin{figure}
%    \centering
%    \sidecaption
   \includegraphics[angle=0,width=8.5cm]{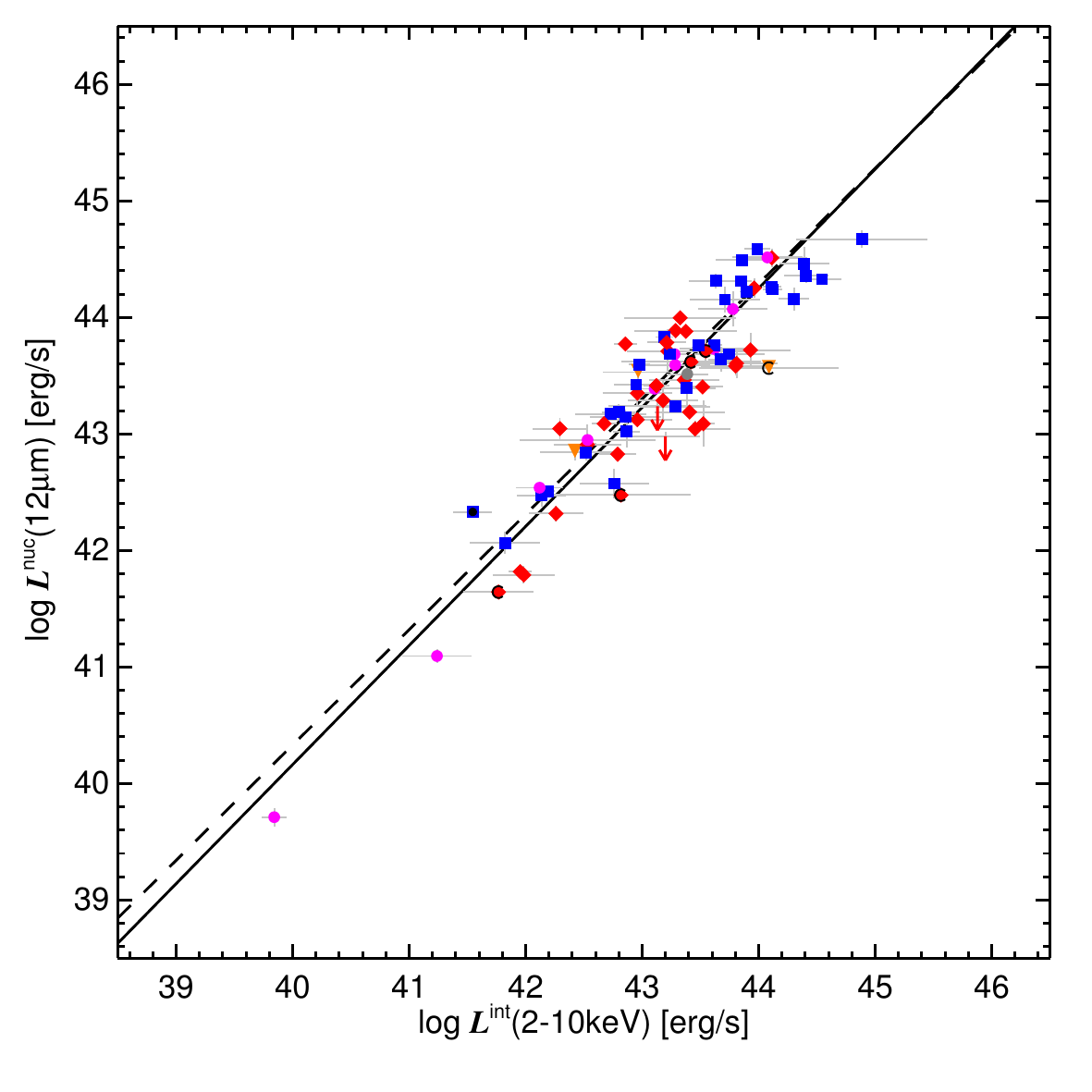}
    \caption{
             Relation of the nuclear $12\um$ and intrinsic 2-10\,keV luminosities for the nine-month BAT AGN sample.
             Symbols and colours are as in Fig.~\ref{fig:f12_f2int}.
             In addition, the solid black line marks the \texttt{linmix\_err} fit for the BAT sample while the dashed line marks the one for all reliable AGN. 
            }
   \label{fig:cor_BAT}
\end{figure}

First, the MIR--X-ray properties for the uniform flux-limited BAT 9\,month AGN sample as defined by \cite{winter_x-ray_2009} are measured.
The 14-195\,keV energy band is possibly best-suited for selecting AGN because emission in this band is dominated by AGN, and it is only weakly dependent on nuclear obscuration up to CT columns.
Out of the 102 objects in this sample, 80 have high angular resolution MIR photometry available. 
Note that four of these sources are not in the reliable AGN sample because they are AGN/starburst composites (Mrk\,520, NGC\,6240S and NGC\,7582) or are Compton-thick without reliable intrinsic 2-10\,keV  emission estimates (NGC\,3281).
In the $12\um$--2-10\,keV luminosity plane (Fig.~\ref{fig:cor_BAT}), the partial correlation rank for the 80 objects from the BAT 9\,month sample is 0.74 and the intrinsic scatter is 0.24, so slightly lower values than for the reliable sample.
The correlation according to \texttt{linmix\_err} is: 
\begin{eqnarray}
 \log \left( \frac{\lnn}{10^{43}\textrm{erg}\,\mathrm{s}^{-1}}\right)   &=& ( 0.23 \pm 0.04) \cr &+& ( 1.02 \pm 0.05 )  \log \left( \frac{\lxi}{10^{43}\textrm{erg}\,\mathrm{s}^{-1}}\right),
\end{eqnarray}
which is consistent with the fit for all reliable AGN. 
As visible in Fig.~\ref{fig:cor_BAT}, the rather high 14-195\,keV flux limit of the nine-month BAT AGN sample leads to a large under-density of low-luminosity AGN in that sample. 
Only the X-ray brightest low-luminosity objects are detected, which biases the correlation towards a steeper slope and lower intercept. 
For this reason, the correlation derived from reliable AGN sample is more representative of the whole local AGN population. 
Note that the more recent BAT AGN sample contains more low-luminosity objects but for these the high angular resolution MIR coverage becomes too small for investigating the MIR--X-ray correlation.
Finally, Compton-thick obscured objects are presumably under-represented in both the BAT and the reliable sample because of absorption in the 14-195\,keV band and the difficulty to estimate reliable intrinsic 2-10\,keV fluxes.

\subsubsection{Volume-limited sample test}\label{sec:vol}
\begin{figure}
%    \centering
%    \sidecaption
   \includegraphics[angle=0,width=8.5cm]{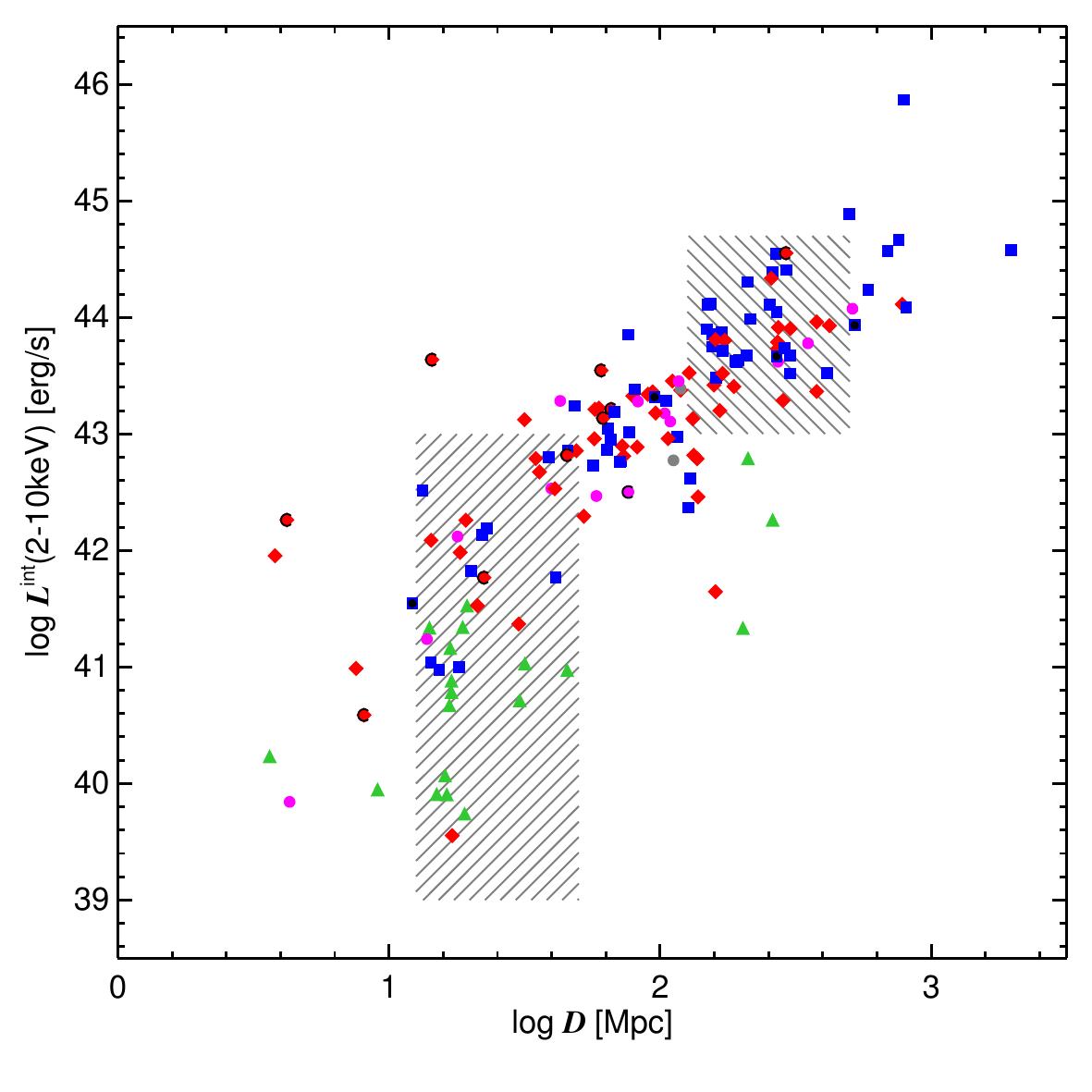}
    \caption{
             Intrinsic 2-10\,keV luminosity over distance for the reliable  sample.
             Symbols and colours are as in Fig.~\ref{fig:f12_f2int}.
             In addition, the hatched areas mark the two cubes used for the volume test in Sect.~\ref{sec:vol}. See text for explanation.
            }
   \label{fig:cuts}
\end{figure}

The second test is addressing biases caused by the incompleteness of the sample. 
Unfortunately, the fraction of local AGN with high angular resolution MIR photometry is too low to obtain significant coverage in representative volume-limited samples that contain low and high luminosity AGN. 
However, one can use distance and luminosity cuts of the reliable sample to form cubes in logarithmic distance--luminosity space with rather uniform sampling.
We select a cube with $1.1 < \log D < 1.7$, $39 < \log \lxi < 43$, representative of the low luminosity range (Fig.~\ref{fig:cuts}).
It contains 39 objects, and its corresponding $12\um$--2-10\,keV luminosity correlation has an intercept of $a = 0.38 \pm 0.12$ and a slope of $b = 1.00 \pm 0.07$, so similar to the values for the whole reliable sample. 
Another cube is ($2.1 < \log D < 2.7$, $43 < \lxi < 44.7$) with 45 objects and $a = 0.01 \pm 0.22$ and a slope of $b = 1.21 \pm 0.25$.
This fit is steeper than that for the reliable sample although consistent within the uncertainties. 
Although the chosen cuts are optimizing the sampling as well as possible, the resulting luminosity coverage of this cube is not large enough to provide valuable constraints.
This is indicated by the large uncertainties on $a$ and $b$.

In conclusion, the sample bias tests does not indicate that the results derived from the reliable sample of AGN in the MIR atlas are affected by significant biases.
However, we encourage further tests with larger volume-limited samples.

%In conclusion, 

\subsection{Comparison to the literature}\label{sec:lit}
\subsubsection{Previous studies based on high resolution MIR data -- Gandhi et al. (2009) and Levenson et al. (2009)} 
\begin{figure}
%    \centering
%    \sidecaption
   \includegraphics[angle=0,width=8.5cm]{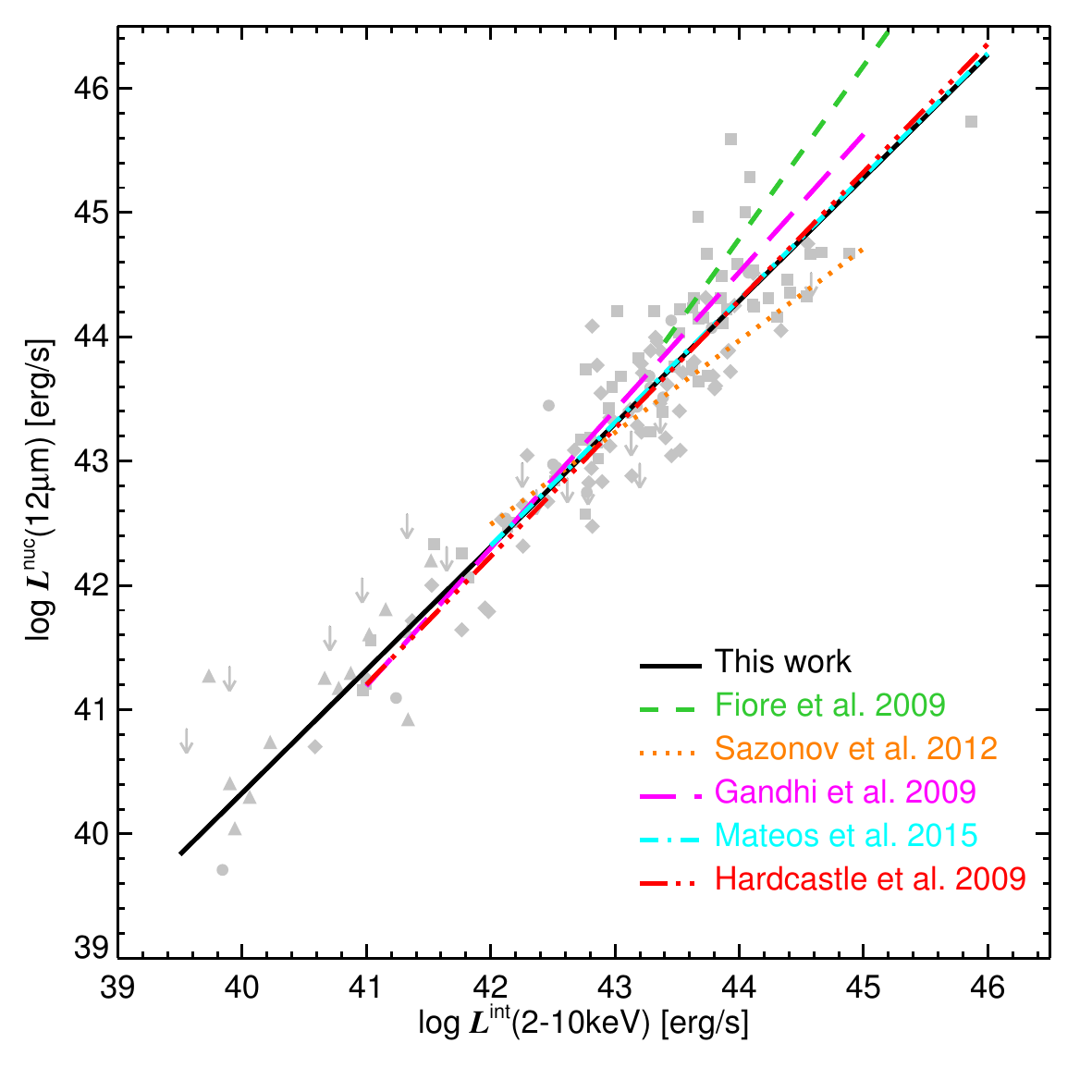}
    \caption{
             Comparison of different MIR--X-ray correlation fits in the literature. 
             Note that no correction is applied for the different subbands used. 
             The grey symbols mark the reliable AGN sample from this work.
            }
   \label{fig:fits}
\end{figure}

While a strong MIR--X-ray correlation was reported in \cite{elvis_seyfert_1978}, \cite{glass_mid-infrared_1982}, \cite{krabbe_n-band_2001} and \cite{lutz_relation_2004}, a fitting-based functional description of the correlation was only attempted in \cite{horst_small_2006,horst_mid_2008} and \cite{ramos_almeida_mid-infrared_2007}.
While the latter work used low angular resolution \isoo data, the former were based on high angular resolution ground-based MIR photometry.
\cite{horst_mid_2008} was then extended by \cite{gandhi_resolving_2009}, \cite{honig_dusty_2010-1} \cite{asmus_mid-infrared_2011} and finally this work.
Out of these studies, the correlation parameters from \cite{gandhi_resolving_2009} are the most widely used because they apply to all classes of AGN including CT objects. 
The parameters are based on the observed $12\um$  and intrinsic 2-10\,keV luminosities for a heterogeneous sample of 42 local AGN observed with VISIR.
For the fitting, the \texttt{fitexy} routine  was used \citep{press_numerical_1992} and provides as slope of $1.11 \pm 0.04$.
The corresponding correlation is shown in Fig.~\ref{fig:fits} in comparison to the results obtained here.
The slope for the Gandhi et al. sample depends on the choice of algorithm and becomes $1.00 \pm 0.08$ when using \texttt{linmix\_err} on the same data. 
The intercept is less sensitive to the fitting algorithm ($0.35\pm0.06$ for \texttt{linmix\_err}).
Therefore, our new results are fully consistent with \cite{gandhi_resolving_2009}.

A study similar to \cite{gandhi_resolving_2009} was presented in \cite{levenson_isotropic_2009} although based on a smaller sample of 17 objects spanning only two orders of magnitude in $\lxi$.
%Owing to these sample properties, the uncertainties in the obtained fits are quite large. 
They found a much steeper correlation $\log \lnn \propto (1.32 \pm 0.11) \log \lxi$ using bisector fitting. 
This fit is dominated by the type~II sources in their sample because the fit to the type~I objects only yields a slope of $1.0 \pm 0.2$ consistent with \cite{gandhi_resolving_2009} and this work. 
They attribute the difference to star formation contamination even on small scales to the objects in their sample.
In addition, half of their type~II sources are Compton-thick obscured and thus have uncertain $\lxi$ values. 
We attribute differences between our study and \cite{levenson_isotropic_2009} to this and the small sample size in the latter work, in particular because our sample incorporates all objects from \cite{levenson_isotropic_2009}.

\subsubsection{High luminosities -- Fiore et al. (2009), Lanzuisi et al. (2009), Mateos et al. (2015) and Stern (2015)}
Another often used version of the MIR--X-ray correlation is by \cite{fiore_chasing_2009} which focuses on the higher luminosity end.
It used the observed restframe $5.8\um$ and 2-10\,keV luminosities of $\sim80$ X-ray-selected type~I AGN in the COSMOS and CDF-S fields with \chandraa and \spitzerr observations.
Converted to our form of notation, their correlation is 
\begin{equation}
 \log \left( \frac{\lmt}{10^{43}\textrm{erg}\,\mathrm{s}^{-1}}\right)    = 0.40 + 1.39  \log \left( \frac{\lxo}{10^{43}\textrm{erg}\,\mathrm{s}^{-1}}\right),
\end{equation}
for $\log \lmt > 43.04$.
At the same time, \cite{lanzuisi_revealing_2009} performed a similar study but for luminous type~II AGN.
They obtain a slope of 1.24 in the same notation.
Thus, both slopes are much steeper than our findings as visible in Fig.~\ref{fig:fits}. 
Unfortunately, no uncertainties nor the algorithm used are stated in \cite{fiore_chasing_2009} and \cite{lanzuisi_revealing_2009}. 
However, the distribution of sources in the MIR--X-ray plane  (e.g., Fig.~5 in \citealt{fiore_chasing_2009}) shows that there is a clear trend of increasing MIR--X-ray ratio for high luminosities ($\log \lmt > 44$). 
Therefore, we can exclude error treatment and fitting technique as cause for the inconsistency. 
Furthermore, at these high luminosities, host related MIR contamination should not be an issue, in particular for the type~I AGN of \cite{fiore_chasing_2009} since these sources are presumably unobscured.
On the other hand, they excluded seven objects because of their outlying position in the MIR--X-ray plane, indicating heavy obscuration. 
Therefore, it is possible that also some of the objects showing less extreme MIR--X-ray ratios are affected by obscuration in X-rays.
Furthermore, the correlation using $5.8\um$ could be intrinsically different from the $12\um$ based correlation, in particular since the hot dust component dominates the MIR emission at $5.8\um$ in many type~I AGN rather than the warm component (dominating at $12\um$, if present; e.g., \citealt{edelson_spectral_1986, mor_hot_2012}).
Alternatively, it is possible that the MIR--X-ray correlation becomes steeper at high luminosities, although our sources overlapping with the \citeauthor{fiore_chasing_2009} fitting region are not consistent with such a steep slope (see also discussion about \citealt{sazonov_contribution_2012} below).

Very recently, \cite{mateos_revisiting_2015} and \cite{stern_x-ray_2015} re-examined the 6\,$\um$--2-10\,keV correlation.
The former uses a complete, flux limited sample of 232 AGN from the Bright Ultra hard XMM-Newton Survey.
The X-ray luminosities were corrected for absorption through spectral modelling while no Compton-thick sources were found in their sample.
The MIR luminosities were extracted from WISE through spectral decomposition of AGN and host contributions. 
Using also \texttt{linmix\_err} with the X-ray luminosity as independent variable, \citeauthor{mateos_revisiting_2015} find a correlation slope of $0.99\pm0.03$ applicable for the whole range up to luminosities of $10^{46}\,$erg/s in X-rays.
This agrees very well with our result (Fig.~\ref{fig:fits}) and contradicts as well the finding of \cite{fiore_chasing_2009}.

\cite{stern_x-ray_2015}, on the other hand, finds a steepening of the MIR--X-ray fit slope at luminosities above $\log \lxi \sim 44.5$ using a combination of several samples, with multiwavelength data obtained with various recent instruments.
Different, to all the other works, a second order polynomial is fitted to the data with the resulting fit bending at higher luminosities to higher MIR--X-ray ratios. 
This result agrees with our at lower luminosities ($\lesssim 10^{44}$\,erg/s) where our sampling is still high. 
At higher luminosities it agrees rather with \cite{fiore_chasing_2009} but not with \cite{mateos_revisiting_2015}. 
This disagreement is not discussed in \cite{stern_x-ray_2015}.
We do not attempt to solve this controversy at high luminosities here but we would like to point out that there is a mismatch between the two high luminosity samples used, "SDSS DR5" versus "high-luminosity" sample in \cite{stern_x-ray_2015}.
Namely, the median MIR luminosity of the former sample is 0.8\,dex lower than that of the latter sample in the overlapping X-ray luminosity range of both samples ($10^{45} \lesssim \lxi  \lesssim 10^{45.9}$\,erg/s). 
The fit is dominated by the "high-luminosity sample, owing to the much denser sampling in this region, and the difference between the bended fit of \cite{stern_x-ray_2015} and the straight fit of, e.g., \cite{mateos_revisiting_2015} is $\sim0.8$\,dex in MIR luminosity at $\lxi \sim 10^{45.5}\,$erg/s.
Clearly, a better sampling at these luminosities is required to resolve this mismatch.

\subsubsection{Radio-loud sources -- Hardcastle et al. (2009) and Mason et al. (2012)}
\cite{hardcastle_active_2009} investigate multiwavelength correlations for radio-loud AGN, namely the low-redshift 3CCR sources.
In this analysis, they use the  restframe $15\um$ global luminosity measured by \spitzer.
The absorption corrected restframe 2-10\,keV luminosities are split into an unabsorbed, jet-related and an absorbed, accretion-related component.
They find that the MIR luminosity is only weakly correlated with the jet-related X-ray luminosity, while the former correlates much stronger with the accretion-related component.
The correlation is fitted to 36 objects with an unspecified Bayesian code that takes into account errors and upper limits, similar to \texttt{linmix\_err}.
The resulting slope is $1.03 \pm 0.18$ and thus fully consistent with our results (Fig.~\ref{fig:fits}). 
Unfortunately, the data of \cite{hardcastle_active_2009} do not provide a good coverage at low luminosities and show a large scatter. 
Thus, it is not possible to use them to directly compare to our finding of radio-loud AGN differing from the correlation of radio-quiet sources.
However, the results of \cite{hardcastle_active_2009} are consistent with the jet increasingly contributing to the X-ray emission at high luminosities, which would explain the on average lower MIR--X-ray ratios of high luminosity radio-loud sources.

\cite{mason_nuclear_2012} used the correlation found from combining the samples of \cite{gandhi_resolving_2009} and \cite{levenson_isotropic_2009} to probe the position of the local low-luminosity AGN in the MIR--X-ray plane.
Different to \cite{asmus_mid-infrared_2011}, they find a significant MIR-excess for those sources, in particular the radio-loud ones.
Our work includes all the sources of these works and verifies this MIR-excess (Sect.~\ref{sec:rad}), although with a smaller strength. 
This difference is mainly caused by differences in the 2-10\,keV luminosities used, which can be very uncertain owing to the intrinsic faintness of low-luminosity AGN. 
For example, \cite{mason_nuclear_2012} uses $\log \lxi = 40.0$ for M\,87 (NGC\,4486) and a distance of 16\,Mpc based on the \cite{perlman_x-ray_2005} of \chandraa observations from 2000.  
However, \cite{donato_obscuration_2004} calculate  $\log \lxi = 40.6$ from the same observation for the same distance.
Furthermore, the \chandraa observation of 2002 analysed by \cite{gonzalez-martin_x-ray_2009} show an even higher value of $\log \lxi = 40.9$. 
Therefore, the long-term average X-ray luminosity is a factor of a few higher than the value used by \cite{mason_nuclear_2012}.
M\,87 is one of the objects with the strongest MIR-excess in that work.

\subsubsection{Correlations with hardest X-rays -- Mullaney et al. 2011, Matsuta et al. (2012), Ichikawa et al. (2012) and Sazonov et al. (2012)}
\cite{mullaney_defining_2011} was the the first to make use of the \swift/BAT data to probe the MIR--X-ray correlation in the hardest X-ray regime.
They cross-matched the nine-month BAT AGN catalogue with the \irass source catalogue and used those 44 objects with detections in all four \irass bands for fitting the correlation.
Before, they applied an SED fitting routine to isolate the AGN contribution in the large aperture \irass data.
They obtain a flatter but still marginally consistent correlation slope of $0.74 \pm 0.13$ and a significantly higher intercept of $0.37 \pm 0.08$ compared to our result. 
Unfortunately, no detailed information is given about the fitting algorithm or sources used for the fitting. 
Therefore, it is difficult to address the cause for the different correlation parameters.
However, as \cite{mullaney_defining_2011} note, their SED fitting procedure tends to over-estimate the AGN contribution at low luminosities. 
We thus suspect that their lower MIR luminosities are still affected by host emission. 
In addition, the next two works find higher slopes and lower intercepts for the BAT AGN based on higher resolution MIR data.

\cite{matsuta_infrared_2012} and \cite{ichikawa_mid-_2012} also use the \swift/BAT AGN catalogues to investigate the MIR--X-ray correlation at hardest X-rays but using MIR data from the most recent infrared satellites.
\cite{matsuta_infrared_2012} utilizes the \akarii all-sky survey while \cite{ichikawa_mid-_2012} also uses \irass  and \wisee data.
On the other hand, \cite{matsuta_infrared_2012} cross-correlates with the 22-month BAT sample arriving at 158 AGN, while \cite{ichikawa_mid-_2012} starts with the nine-month BAT sample  arriving at 128 objects. 
Compared to \cite{mullaney_defining_2011} and this work, these studies have the advantage of a higher coverage of the flux-limited BAT samples but suffer again by a biases towards non-CT and higher-luminosity sources, as already discussed in Sect.~\ref{sec:BAT}.
Both works find strong correlations of the observed 14-195\,keV luminosities with the 9 and $18\um$ luminosities, and a weaker but significant one with the $90\um$ luminosities.
\cite{matsuta_infrared_2012} uses the bisector method for fitting of the correlation.
If we apply the same method to our reliable sample with $\log \nh < 24$, we get a slope of $0.97\pm 0.04$ and an intercept of $-0.18 \pm 0.05$.
Therefore, the found correlation slopes of \cite{matsuta_infrared_2012} are slightly flatter but fully consistent with our results, while the intercept is $\sim0.3$\,dex higher.
These differences are presumably caused by host contamination affecting their low-resolution MIR data.
Unfortunately, \cite{ichikawa_mid-_2012} does not state the exact fitting algorithm used so an accurate comparison is not possible.
However, their fit is consistent with all our fits, bisector or \texttt{linmix\_err}.
The similarity of the correlations between 9, 18 and $22\um$ as found in \cite{matsuta_infrared_2012} and \cite{ichikawa_mid-_2012}  and 12 and $18\um$ in this work indicate that the difference in wavelengths contributes at most 0.1\,dex in this wavelength region (at least between $\sim8$ to $24\um$).
Furthermore, \cite{matsuta_infrared_2012} also fits the radio-loud sources in their sample only, also finding the trend of a flatter slope similar to our work.

Finally, \cite{sazonov_contribution_2012} used a cleaned AGN sample from the \integrall all-sky survey also observed with \spitzerr  to determine the MIR--X-ray correlation between the $15\um$ and observed 17-60\,keV wavebands.
They find 
\begin{eqnarray}
 \log \left( \frac{\lsf}{10^{43}\textrm{erg}\,\mathrm{s}^{-1}}\right)   &=& ( 0.23 \pm 0.04) \cr &+& ( 0.74 \pm 0.06 )  \log \left( \frac{\lxhi}{10^{43}\textrm{erg}\,\mathrm{s}^{-1}}\right),
\end{eqnarray}
which is significantly flatter than our correlation (Fig.~\ref{fig:fits}) and also the correlations found by \cite{matsuta_infrared_2012} and \cite{ichikawa_mid-_2012} but in good agreement with \cite{mullaney_defining_2011}.
Unfortunately, \cite{sazonov_contribution_2012} does not state which algorithm was used exactly but likely it was ordinary least square fitting with $\lxhi$ as independent variable.
The overlap with our reliable sample is only 33 objects (54 per cent of their clean sample). 
Thus, we can not robustly calculate the $\lnn$--$\lxh$ correlation for their sample.

Similar to the BAT samples, the \integrall sample also under-represents lower luminosities and is dominated by sources with $\log \lxhi \gtrsim 43$.
In fact, the only low-luminosity source ($\log \lxhi < 42$) in their sample that is not star formation dominated is NGC\,4395.
They exclude this object from most analysis because as a high Eddington system with a small black hole mass in a dwarf galaxy it might belong to a physically different class of AGN. 
However our continuous coverage down to the low-luminosity level of NGC\,4395 does not indicate any significant difference between this object and other more massive low-luminosity AGN. 
Including NGC\,4395 into the fitting to their sample results in a formally stronger correlation with a steeper slope that would be consistent with our $\lnn$--$\lxh$ correlation. 
The intercept is still $\sim 0.3\,$dex higher in that case, which is at least partly caused by host contamination of their \spitzerr data as indicated by the median ratio of $\fnn$ over $\fspft$ of  0.65.
This contamination likely also affects the slope of the correlation or decrease of MIR--X-ray ratio with increasing luminosity as Sazonov et al. found (their Fig.~3).
Specifically, all but one (Cygnus\,A) of the eight  sources overlapping with our reliable sample with $\fnn / \fspft < 0.46$ (median - standard deviation) are between $\sim42 < \log \lxhi < 43.5$, which is the region where $\lsf/\lxhi$ is largest in their sample. 

\cite{sazonov_contribution_2012}  pointed out that the correlation slopes might flatten at high luminosities ($\log \lxhi \gtrsim 44$), which is the reason they find flatter slopes compared to \cite{gandhi_resolving_2009}. 
Our coverage at high luminosities is better  compared to both works but still we did not find compelling evidence for a decrease in the MIR--X-ray ratio (Sect.~\ref{sec:ldep}).
Furthermore, a flattening appears to be inconsistent with \cite{fiore_chasing_2009}, unless the 2-10\,keV and 14-195\,keV or $5.8$ and $12\um$ wavebands decouple at high luminosities.

Finally, we note that the four NLS1 sources in \citeauthor{sazonov_contribution_2012} (2012; none in common with our sample) do not show significantly higher MIR--X-ray ratios than the rest of their sample, contrary to what we find for our sample.

In summary, our results are in general consistent with all previous works focusing on low to moderate luminosity AGN while there appears to be significant differences to studies focusing on high luminosity sources, whereas both flatter and steeper slopes are found (mainly for type~I AGN). 
However, the most recent comprehensive study focusing on high luminosities does not find any change in slope, validating our result \citep{mateos_revisiting_2015}.

\subsection{Implications for the torus scenario}\label{sec:tor}
Current clumpy and smooth torus models (e.g., \citealt{nenkova_agn_2008,honig_dusty_2010,stalevski_3d_2012}) predict differences of $\sim 0.3\,$dex in the  MIR--X-ray ratio with changing inclination angles assuming isotropic X-ray/UV emission.
This is however not found in our results neither in Section~\ref{sec:typ} nor in Section~\ref{sec:n_h} which show that differences between unobscured and highly obscured objects are at best small ($<0.1\,$dex).
Only when entering the Compton-thick obscuration regime, the MIR--X-ray ratio seems to further decrease by $\sim 0.15\,$dex owing to strong absorption also in the MIR (intrinsic or foreground).
This small difference could be explained in three ways.
First, the X-ray emission could be anisotropic as indicated by comparison to \oiv emission coming presumably from the narrow line region  \citep{liu_are_2014}.
Then, the anisotropy of the MIR and X-ray emission would have roughly a similar dependence on the orientation angle as recently found by \cite{yang_anisotropy_2015}.
Second, as the most recent interferometric results indicate, the majority of MIR emission is actually produced in the polar outflow region rather than in a canonical torus \citep{honig_parsec-scale_2012,honig_dust_2013,tristram_dusty_2014}.
In that case, the MIR anisotropy between face-on and edge-on systems is possibly much lower.
Third, higher obscured objects could have on average higher covering factors \citep{ramos_almeida_testing_2011, elitzur_unification_2012}, which could mitigate differences. 
However, this effect is expected to be strongest in the high obscuration regime where we do find larger differences in the MIR--X-ray ratio. 
Another interesting claim is that the X-ray reflection component is possibly larger in many objects with $\log \nh > 23$ \citep{ricci_reflection_2011}, which would lead to a lower MIR--X-ray ratio for affected sources, at least in the 14-195\,keV band.

The largest MIR--X-ray ratios and differences to the other subsamples are exhibited by objects with intermediate columns, $22 \lesssim \log \nh \lesssim 23$ (Sect.~\ref{sec:n_h}; Fig.~\ref{fig:rat_bin_nh}).
The surprising increase from low to intermediate $\nh$ can possibly be explained by a deficit of warm dust in the least obscured AGN.
Indeed unobscured objects with a blue MIR spectral energy distribution in \cite{asmus_subarcsecond_2014} display on average lower MIR--X-ray ratios than those with an emission peak at $\sim18\um$.
Many of the latter have intermediate X-ray column densities. 
Such a behaviour is in fact predicted by \cite{honig_dusty_2010} where the contribution of the MIR emission to the bolometric luminosity depends on the distribution of the dust and blue MIR spectral slopes  would come from a compact distribution with most of the emitting dust mass confined close to the sublimation radius (see also \citealt{honig_dusty_2010-1}).
Note that this explanation is independent of the obscuring fraction or opening angle and does not depend on the accretion luminosity.

It is curious that the decrease in the MIR--X-ray ratio from intermediate to high $\nh$ occurs around $10^{23}\,$cm$^{-2}$, which is the value for which the obscuration in the MIR becomes optically thick assuming a standard gas to dust ratio of 100.
Therefore, self-obscuration in the torus or foreground absorption in the host might play a role here. 
In fact, we caution the reader concerning the direct application of these results to torus models because this MIR--X-ray analysis does not separate between AGN-intrinsic and foreground, host obscuration. 
The latter can be a large contributor to the measured column densities (e.g., \citealt{honig_what_2014}). 

The uncertainties in both the host contribution to the obscuration and the isotropy of the MIR and X-ray emission does prevent us from making any statements about the obscuring fraction or opening angle and their dependencies on luminosity or object type.  

At the same time, systematics dominate the uncertainties on the correlation slope and only allow constraining it to between 0.9 and 1 (Sections.~\ref{sec:cor}, \ref{sec:altcor}, and \ref{sec:ldep}). 
Thus, the uncertainties are still too large to draw strong conclusions on bolometric corrections and underlying physics. 
We note however that the MIR--X-ray correlation slope $\le1$  together with the previous findings that $\lxi \propto L(\mathrm{UV})^\beta$ with $\beta \approx 0.85 $ \citep{marchese_optical-uv_2012} imply that $\lnn \propto L(\mathrm{UV})^{\beta '}$ with $\beta ' < 0.85 $, assuming a similar geometry for the MIR, X-ray and UV emitters.

\subsection{The MIR--X-ray correlation as a tool to predict $\nh$ and $\lxi$}\label{sec:tool}
The tightness of the MIR--X-ray correlation for all AGN types  suggests that high angular resolution MIR data can be used to predict or constrain the column densities and luminosities in X-rays irrespective of the object nature.
In particular for highly obscured (Compton-thick) objects or those with low S/N X-ray data, it is often difficult to reliably estimate these properties.

\begin{figure}
%    \centering
%    \sidecaption
   \includegraphics[angle=0,width=8.5cm]{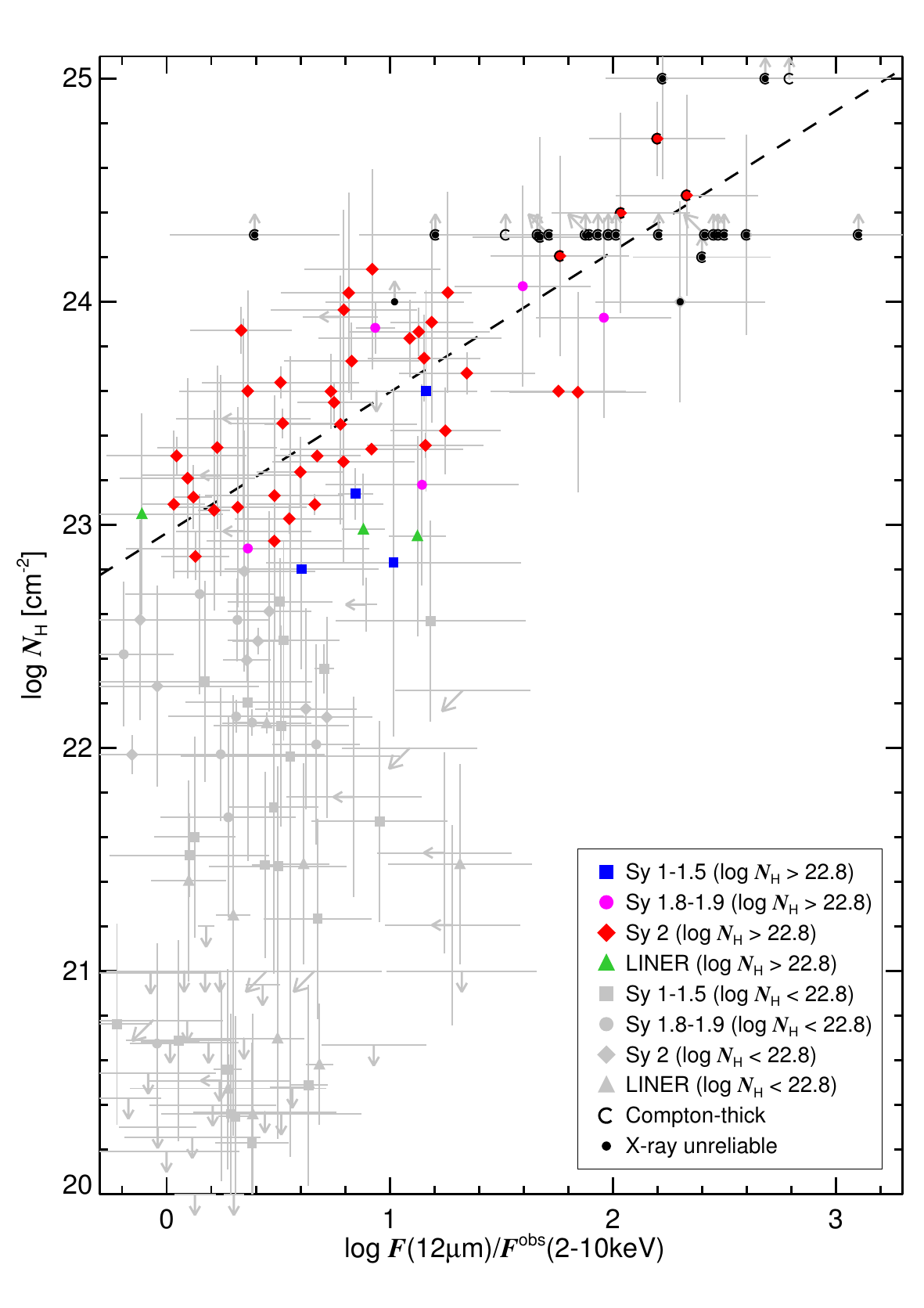}
    \caption{
             X-ray column density diagnostic: $\nh$ versus the observed $12\um$ to 2-10\,keV flux ratio.
             Symbols  are as in Fig.~\ref{fig:f12_f2int}.
             All coloured symbols are reliable AGN with $\log \nh > 22.8$ and have been used for the \texttt{linmix\_err} fit displayed as dashed line. 
             All remaining sources from the reliable sample are plotted in grey.
             In addition, 22 unreliable AGN from the AGN MIR atlas that are Compton-thick or CT candidates are marked with small filled black circles.
             See text for further explanation.
            }
   \label{fig:nh_diag}
\end{figure}

In Fig.~\ref{fig:nh_diag}, $\nh$ is plotted versus the ratio of observed MIR to X-ray fluxes for the reliable sample. 
The MIR--X-ray ratio stays constant with increasing column density up to $\log \nh = 22.8$ and then increases with growing $\nh$. 
The threshold is here determined from the intrinsic to observed 2-10\,keV flux ratio, which increases for higher values, which is visible in the median ratio and through a double-sided KS test.
For the latter, the sub-samples above and below $\log \nh = 22.8$ show maximal difference ($\dks = 0.91; \log \pks = -21.38$).
This fact motivates us to only regard those objects with $\log \nh > 22.8$ for the $\nh$--diagnostic, which conversely is blind to lower column densities than that, i.e. the 2-10\,keV emission is only affected by obscuration above this threshold. 
For the corresponding sub-sample of 53 reliable AGN with higher column densities, $\nh$ and $\log \fnn - \log \fxo$ show a significant correlation ($\rhoS = 0.60; \log \pS = -5.6$).
A corresponding \texttt{linmix\_err} fit yields:
\begin{eqnarray}\label{eq:nh_diag}
 \log \left( \frac{\nh}{10^{22.8}\textrm{cm}^{-2}}\right)   &=& ( 0.14 \pm 0.11) \cr &+& ( 0.67 \pm 0.12 )  \log \left( \frac{\fnn}{\fxo}\right), 
\end{eqnarray}
or for convenient use in directly measured quantities:
\begin{eqnarray}
 \log \left( \frac{\nh}{\textrm{cm}^{-2}}\right)   &=& ( 14.37 \pm 0.11) \cr &+& ( 0.67 \pm 0.11 )  \log \left( \frac{\fnn}{\fxo} \frac{\textrm{erg/s/cm}^2}{\textrm{mJy}} \right), 
\end{eqnarray}
This equation allows in principle predicting $\nh$ to within 0.32\,dex (the observed 1-$\sigma$ scatter from this correlation) for significantly obscured objects.
The caveat is that owing to the constancy of $\log \fnn - \log \fxo$ for $\log \nh < 22.8$, this diagnostic has in practise only predictive power for $\log \fnn - \log \fxo \gtrsim 0.64$ (median + STDDEV), corresponding to $\log \nh \gtrsim 23.36$.
A similar plot but using low-resolution MIR photometry is also discussed in \citeauthor{brightman_xmm-newton_2011} (2011; their Sect.~5, Fig.~12).
The low angular resolution leads to a shift in the observed $12\um$--2-10\,keV distribution by 0.5 to 1\,dex and an increased scatter owing to significant host contamination with respect to our results.
According to the findings of \cite{asmus_subarcsecond_2014}, low angular resolution MIR data can only be reliably used for this $\nh$ diagnostic for objects with $\log \lnn \gtrsim 44$, i.e., where the AGN mostly dominates the total MIR emission.
Another comparable diagnostic is introduced by \cite{severgnini_new_2012}, using the X-ray hardness ratio to better isolate CT objects.

The intrinsic 2-10\,keV luminosity can directly be estimated from Eq.~\ref{eq:inv}.
% inverting the MIR--X-ray correlation of all reliable AGN:
%\begin{eqnarray}\label{eq:lx_diag}
% \log \left( \frac{\lxi}{10^{43}\textrm{erg}\,\mathrm{s}^{-1}}\right)   &=& (- 0.31 \pm 0.03) \cr &+& ( 1.02 \pm 0.03 )  \log \left( \frac{\lnn}{10^{43}\textrm{erg}\,\mathrm{s}^{-1}}\right).
%\end{eqnarray}
The uncertainty of this $\lxi$ prediction is 0.32\,dex (the observed 1-$\sigma$ scatter from the correlation). 
In Appendix~\ref{sec:CT}, we use these derived tools to predict the intrinsic 2-10\,keV luminosities and column densities of objects with uncertain X-ray properties.

Using the MIR--X-ray correlation itself as a diagnostic, we find that 
in most of AGN/starburst composite objects from the atlas the AGN MIR emission is well isolated at subarcsecond scales with the caveat that this diagnostic by itself can not distinguish between Compton-thick obscured AGN and star formation dominated nuclei.
More details are given in Appendix~\ref{sec:CP} including a discussion of the largest outlier, the complex object NGC\,4945 with a very low MIR--X-ray ratio, which can at least be partly explained by particularly high foreground extinction affecting the MIR. 
However, our results indicate that it is unlikely that many objects are significantly affected by this.
Finally, we test the AGN nature of the uncertain objects from the AGN MIR atlas in Appendix~\ref{sec:can}.

\subsection{True type~II AGN and the MIR--X-ray correlation}
The idea of so-called true type~II AGN is based observationally on the absence of broad emission lines in polarized light (but see \citealt{antonucci_panchromatic_2012}) and the absence of significant X-ray absorption in many type~II AGN \citep{ptak_nature_1996,pappa_x-ray_2001,panessa_unabsorbed_2002}.
Theoretically, the disappearance of the broad line region has been predicted to occur at low accretion rates \citep{nicastro_broad_2000, nicastro_lack_2003}.
Many candidates for true type~II AGN were claimed over the years.
However, most of them could later be shown to be either Compton-thick obscured, star formation dominated, or to show in fact broad emission line components \citep{shi_unobscured_2010}.
The best remaining candidates among the local galaxies are NGC\,3147 and possibly NGC\,3660. 
However, the optical classifications for NGC\,3660 are very contradicting ranging from anything between non-AGN to type~I AGN \citep{kollatschny_nuclear_1983,contini_starbursts_1998,
moran_classification_1996,veron-cetty_catalogue_2010}.
Note that \cite{shi_unobscured_2010} also suggests NGC\,4594 but this low-luminosity object is usually classified as LINER or as Sy\,1.9 \citep{veron-cetty_catalogue_2010}. 
While not strictly demanded by definition, many works suggest that not only the broad line region but also the obscuring dusty structure is absent in true type~II AGN. 
Therefore, it is interesting to test whether the candidates found so far exhibit particularly low MIR--X-ray ratios. 
NGC\,3147, NGC\,3660 and NGC\,4594 are all in the AGN MIR atlas.
Their nuclear MIR--X-ray ratios are $\ratmx = 0.35$; 0.38; and 0.1 respectively. 
Therefore, there is no evidence from the MIR point of view for the absence of a dusty obscurer or any other deviation from the other AGN for these objects.

For completeness we also regard two more distant candidates that are not included into the AGN MIR atlas. 
These are GSN\,069 \citep{miniutti_high_2013} and Q2131-427 \citep{panessa_unabsorbed_2009,bianchi_simultaneous_2012}.
For both we use the band~3 fluxes of the Wide-field Infrared Survey Explorer (\wise; \citealt{wright_wide-field_2010}) as a proxy for the nuclear $12\um$ emission. 
The MIR--X-ray ratio of GSN\,069 is $\ratmx \le 1$ (it was not detected in the 2-10\,keV range; \citealt{miniutti_high_2013}).
This is well consistent with the presence of warm dust as in normal type~II AGN with the caveat that the \wisee data might be host contaminated. 
Note that GSN\,069 has an extremely soft X-ray spectrum and thus 2-10\,keV might not be a good intrinsic indicator in this case.

The closest \wisee source to Q2131-427 has a W3 magnitude of 12.645 (distance 2.6\,arcseconds). 
There is no other source within the confusion limit of \wise.
Using the 2-10\,keV  flux stated by \cite{panessa_unabsorbed_2009}, the corresponding MIR--X-ray ratio is $\ratmx = 0.46$.
Since the luminosity is $\approx 10^{44}$\,erg/s, significant star formation contamination is unlikely \citep{asmus_mid-infrared_2011}.

We conclude that all true type~II AGN candidates exhibit MIR--X-ray ratios typical of normal type~II AGN and thus do not show any evidence for the absence of a dusty obscurer.

\subsection{Double AGN and the MIR--X-ray correlation}
During major mergers it is expected that the nuclei of both galaxies become active at some point during the process and temporarily form a binary AGN before finally the super massive black holes merge \citep{begelman_massive_1980}. 
This is one proposed way for black hole growth during the cosmic evolution.
Here, we address the question whether these double and binary AGN also follow the same MIR--X-ray correlation as single AGN. 
One would expect deviations to occur latest as soon as the influence of the other AGN component starts to change the AGN structure. 
But also the infall process during the merger could lead significant changes in the structure already, which should be indicated by deviating MIR--X-ray ratios.

However, only very few double or binary AGN have been found so far (e.g., \citealt{wang_chandra_2010}), with Mrk\,463 \citep{hutchings_double-nucleus_1989} and NGC\,6240 \citep{komossa_discovery_2003} still being the best cases.

The MIR AGN atlas contains three double AGN candidates, Mrk\,266, NGC\,3690 and NGC\,6240.
Their X-ray properties are described in detail in Appendix~\ref{app:obj}.
All are major merger systems, show intense star formation and are heavily obscured along the line of sight.
These complicate extracting the X-ray and MIR properties of the AGN.
For Mrk\,266, only for one nucleus an intrinsic X-ray luminosity is available (Mrk\,266NE), placing it at a relatively high MIR--X-ray ratio ($\ratmx = 1.22$). 
However, the nucleus is presumably CT obscured, which would move it to a much lower ratio.
This is in fact the case for both NGC\,3690 and NGC\,6240 where the more powerful and X-ray brighter nucleus is CT obscured with average to low MIR--X-ray ratios (NGC\,3690W: $\ratmx = 0.49$ and NGC\,6240S: $\ratmx = -0.52$). 
The low ratio of NGC\,6240S might indicate significant absorption also in the MIR, similar to NGC\,4945 (Appendix~\ref{sec:CP}).
For both systems, the fainter nuclei might also be CT obscured.
But while NGC\,6240N still has a nominal MIR--X-ray ratio ($\ratmx = 0.77$), the one of NGC\,3690E is as high as for pure star formation ($\ratmx = 3.57$).
Therefore, the latter object might be star formation dominated also on nuclear scales (see also \citealt{alonso-herrero_uncovering_2013}), unless it is highly CT obscured.

Note that for another object in the AGN MIR atlas, NGC\,3393, a double AGN was claimed based on \chandraa observations \citep{fabbiano_close_2011}.
However, the existence of a double AGN in this regular face-on spiral is disfavoured in a recent multiwavelength analysis also using new \nustarr data \citep{koss_broadband_2015}. 
Only a single nucleus was also detected in the MIR \citep{asmus_subarcsecond_2014}.

In summary, the individual AGN in the double AGN have MIR--X-ray ratios consistent with single AGN but intense star formation and high obscuration makes accurate measurements difficult. 
Additional high angular resolution data, in particular MIR spectra (e.g., \citealt{alonso-herrero_uncovering_2013,alonso-herrero_nuclear_2014}) can help to better isolate the intrinsic AGN power in these systems.

%\subsection{Main driver of the fine structure in the correlation}

\section{Conclusions}\label{sec:concl}
We presented an updated determination of the MIR--X-ray correlation for AGN using the subarcsecond scale 12 and $18\um$ photometry from \cite{asmus_subarcsecond_2014} and averaged X-ray properties from the most recent observations collected from the literature. 
Starting from the atlas sample of 253 AGN, we selected 152 objects with reliable AGN classification, no AGN/starburst composites, and trustworthy X-ray properties. 
This sample is not obviously biased against highly obscured objects, and although we exclude many CT object from the analysis because of unreliable $\lxi$ estimates, their MIR--X-ray ratios agree well with the results summarized below (see App.~\ref{sec:CT}). 
In particular, we obtained the following main results:
\begin{enumerate}
\item An MIR--X-ray correlation is present in flux and luminosity space, extending over the whole probed range from $\sim 10^{40}$ to $\sim 10^{45}\,$erg/s.
The correlation is strong for any combinations of 12 and $18\um$ and 2-10 and 14-195\,keV and likely even extends to shorter (and longer) MIR wavelengths as indicated in the literature already.
Its slope is between 0.9 and 1.0 depending on the exact wavelengths and samples used while the observed scatter is always $<0.4$\,dex.
Therefore, the MIR--X-ray correlation is a useful tool to convert between X-ray and MIR luminosities irrespective of the object nature and underlying physics.

%\item Tests for sample biases by comparison to the uniform flux-limited BAT 9-month sample and volume-limited subsets did not indicate that our results are biased but gave consistent results. 

\item {\it The observed MIR--X-ray ratio can be used to constrain the X-ray column density within $0.32$\,dex.}

\item There is indication for a decrease of the MIR--X-ray ratio at the highest probed X-ray luminosities ($\sim 10^{45}\,$erg/s) but remains statistically insignificant. 
Such a turn over would agree with the decrease of the dust covering factor and the results of \cite{sazonov_contribution_2012} but in contradiction to \cite{fiore_chasing_2009} and \cite{stern_x-ray_2015}.
A well defined sufficiently large high luminosity sample needs to be studied in order to solve this conflict.

\item Differences with optical type are small. 
The correlation for type~I AGN is tightest and type~II objects dominate the scatter in the correlation of all AGN. 
{\it Optical type~I AGN display on average an only 0.15\,dex higher MIR--X-ray ratio compared to optical type~II AGN,} while their correlation slopes are parallel.
LINERs exhibit on average the highest MIR--X-ray ratio and have a flatter correlation slope, which is explained by their radio-loudness (see below).
These differences remain statistically insignificant however.

\item {\it True type~II AGN candidates show the same MIR--X-ray ratios as normal type~II AGN} and, thus, do not indicate any deficit of dust (and obscurer) in those objects.

\item The MIR--X-ray ratios of the individual AGN in double AGN systems are consistent with the MIR--X-ray correlation but intense star formation and heavy obscuration make measurements very uncertain.  

\item {\it Only the NLS1 objects are significantly different with very high MIR--X-ray ratios.}
However, the MIR--X-ray properties of a larger sample of NLS1 have to be studied for more conclusive results.

\item The differences between X-ray obscured and unobscured AGN are smaller ($<0.1\,$dex) than expected from torus models but can be explained by either an anisotropy in the X-ray emission or the MIR emission being dominated by dust in the outflow region. 
{\it Only CT sources have significantly lower MIR--X-ray ratios than unobscured AGN.}
AGN with intermediate obscuration show the highest MIR--X-ray ratio which is possibly related to a deficit of warm dust in many unobscured AGN as indicated by their MIR spectral energy distributions.

\item {\it Radio-loud AGN differ from radio-quiet AGN also in the MIR--X-ray plane in having a significantly flatter correlation slope} leading to higher MIR--X-ray ratios at low luminosities and lower ratios at high luminosities.
This can be explained by an additional MIR contribution or dominance of the jet at low luminosities while the jet contributes much more to the X-rays at higher luminosities.  

\end{enumerate}

\section*{Acknowledgements}
We thank the referee for valuable comments that improved the manuscript.
D.A. acknowledges the JAXA ITYF scheme for visitor support to ISAS in Aug 2012.
P.G. acknowledges support from STFC (grant reference ST/J003697/1).
This research made use of the NASA/IPAC Extragalactic Database
(NED), which is operated by the Jet Propulsion Laboratory, California Institute
of Technology, under contract with the National Aeronautics and Space
Administration. 
This work made use of data supplied by the UK Swift Science Data Centre at the
University of Leicester.
This publication makes use of data products from the Wide-field Infrared Survey Explorer, which is a joint project of the University of California, Los Angeles, and the Jet Propulsion Laboratory/California Institute of Technology, funded by the National Aeronautics and Space Administration.

% 
% for the bibliography, at the end
%\bibliographystyle{aa} % style aa.bst
\bibliographystyle{mn2e} 
\bibliography{my_lib_ref.bib} % your references Yourfile.bib

%\section{Tables}\label{app:tab}
{\onecolumn
%\footnotesize
%\small
\scriptsize
%\tiny
\begin{longtable}{l c c c c c c c c c c l}

\caption{\label{tab:sam}Reliable sample.}\\ % title of Table
\hline\hline
	&		&	Opt.	&		$\log L^\textrm{nuc}$	&		$\log L^\textrm{nuc}$	&		&	$\log L^\textrm{obs}$	&		$\log$	&	$\log L^\textrm{int}$	&		$\log L^\textrm{obs}$	&		&		\\
Object	&	$D$	&	class	&		($12\um$)	&		($18\um$)	&	X-ray	&	(2-10\,keV)	&		$ \nh$	&	(2-10\,keV)	&		(14-195\,keV)	&	radio-	&	2-10\,keV	\\
	&	[Mpc]	&		&		 [erg/s]	&		 [erg/s]	&	epochs	&	 [erg/s]	&		 [cm$^{-2}$]	&	 [erg/s]	&		 [erg/s]	&	loud?	&	Ref.	\\
(1)	&	(2)	&	(3)	&		(4)	&		(5)	&	(6)	&	(7)	&		(8)	&	(9)	&		(10)	&	(11)	&	(12)	\\
\hline\endfirsthead\caption{continued.}\\ \hline\hline																											
	&		&	Opt.	&		$\log L^\textrm{nuc}$	&		$\log L^\textrm{nuc}$	&		&	$\log L^\textrm{obs}$	&		$\log$	&	$\log L^\textrm{int}$	&		$\log L^\textrm{obs}$	&		&		\\
Object	&	$D$	&	class	&		($12\um$)	&		($18\um$)	&	X-ray	&	(2-10\,keV)	&		$ \nh$	&	(2-10\,keV)	&		(14-195\,keV)	&	radio-	&	2-10\,keV	\\
	&	[Mpc]	&		&		 [erg/s]	&		 [erg/s]	&	epochs	&	 [erg/s]	&		 [cm$^{-2}$]	&	 [erg/s]	&		 [erg/s]	&	loud?	&	Ref.	\\
(1)	&	(2)	&	(3)	&		(4)	&		(5)	&	(6)	&	(7)	&		(8)	&	(9)	&		(10)	&	(11)	&	(12)	\\
\hline\endhead\hline\endfoot																									 1H 0419-577	 & 	499.0	 & 	1.5	 & 		44.67	$\pm$	0.08	 & 					 & 	5+	 & 	44.54	$\pm$	0.16	 & 		21.6	$\pm$	0.45	 & 	44.89	$\pm$	0.56	 & 		44.87	$\pm$	0.05	 & 	0	 & 1, 2, 3 \\
 1RXS J112716.6+	 & 	512.0	 & 	1.8	 & 		44.52	$\pm$	0.05	 & 					 & 	2	 & 	44.12	$\pm$	0.30	 & 	$\le$	19.9			 & 	44.07	$\pm$	0.30	 & 		44.69	$\pm$	0.13	 & 	?	 & 4, 3 \\
 2MASX J03565655	 & 	351.0	 & 	1.9	 & 		44.07	$\pm$	0.15	 & 					 & 	2	 & 	43.92	$\pm$	0.30	 & 		22.7	$\pm$	0.45	 & 	43.78	$\pm$	0.30	 & 		44.46	$\pm$	0.10	 & 	?	 & 5, 3 \\
 2MASX J09180027	 & 	781.0	 & 	2	 & 		44.51	$\pm$	0.08	 & 					 & 	1	 & 	44.03	$\pm$	0.30	 & 		23.1	$\pm$	0.45	 & 	44.11	$\pm$	0.30	 & 	$\le$	44.99			 & 	?	 & 6, 3 \\
 3C 29	 & 	202.0	 & 	L	 & 	$\le$	42.57			 & 					 & 	1	 & 	41.15	$\pm$	0.30	 & 	$\le$	22.3			 & 	41.32	$\pm$	0.30	 & 	$\le$	43.82			 & 	1	 & 7 \\
 3C 33	 & 	273.0	 & 	2	 & 		43.89	$\pm$	0.18	 & 					 & 	2	 & 	43.38	$\pm$	0.30	 & 		23.6	$\pm$	0.07	 & 	43.91	$\pm$	0.30	 & 		44.39	$\pm$	0.07	 & 	1	 & 8, 9, 10 \\
 3C 78	 & 	127.0	 & 	1	 & 	$\le$	42.76			 & 					 & 	3	 & 	42.30	$\pm$	0.35	 & 	$\le$	21.0			 & 	42.37	$\pm$	0.28	 & 	$\le$	43.41			 & 	1	 & 11, 12, 13 \\
 3C 93	 & 	1967.0	 & 	1	 & 	$\le$	44.51			 & 					 & 	1	 & 	44.62	$\pm$	0.30	 & 	$\le$	21.3			 & 	44.58	$\pm$	0.30	 & 	$\le$	45.79			 & 	1	 & 14 \\
 3C 98	 & 	137.0	 & 	2	 & 		43.17	$\pm$	0.11	 & 					 & 	3	 & 	42.60	$\pm$	0.22	 & 		23.0	$\pm$	0.07	 & 	42.79	$\pm$	0.25	 & 		43.18	$\pm$	0.20	 & 	1	 & 15, 16, 17 \\
 3C 105	 & 	421.0	 & 	2	 & 		43.72	$\pm$	0.15	 & 					 & 	3	 & 	43.49	$\pm$	0.22	 & 		23.3	$\pm$	0.37	 & 	43.93	$\pm$	0.35	 & 		44.77	$\pm$	0.08	 & 	1	 & 18, 3, 6 \\
 3C 120	 & 	150.0	 & 	1.5	 & 		44.26	$\pm$	0.01	 & 					 & 	5+	 & 	44.08	$\pm$	0.04	 & 	$\le$	21.2			 & 	44.11	$\pm$	0.08	 & 		44.40	$\pm$	0.02	 & 	1	 & 19, 20, 21, 3 \\
 3C 227	 & 	414.0	 & 	1.5	 & 		44.22	$\pm$	0.04	 & 					 & 	3	 & 	43.71	$\pm$	0.30	 & 		22.1	$\pm$	0.45	 & 	43.52	$\pm$	0.38	 & 		44.69	$\pm$	0.09	 & 	1	 & 22, 23, 24 \\
 3C 273	 & 	792.0	 & 	1	 & 		45.73	$\pm$	0.07	 & 					 & 	5+	 & 	45.89	$\pm$	0.06	 & 	$\le$	19.7			 & 	45.87	$\pm$	0.12	 & 		46.52	$\pm$	0.00	 & 	1	 & 25, 19, 26 \\
 3C 285	 & 	378.0	 & 	2	 & 	$\le$	43.42			 & 					 & 	1	 & 	43.07	$\pm$	0.30	 & 		23.5	$\pm$	0.45	 & 	43.36	$\pm$	0.30	 & 	$\le$	44.36			 & 	1	 & 27, 17 \\
 3C 293	 & 	211.0	 & 	L	 & 	$\le$	42.84			 & 					 & 	1	 & 	42.50	$\pm$	0.30	 & 		23.0	$\pm$	0.45	 & 	42.78	$\pm$	0.30	 & 	$\le$	43.85			 & 	1	 & 18, 28 \\
 3C 317	 & 	160.0	 & 	2/L	 & 	$\le$	42.31			 & 					 & 	2	 & 	41.65	$\pm$	0.30	 & 	$\le$	21.0			 & 	41.65	$\pm$	0.30	 & 	$\le$	43.61			 & 	1	 & 29, 13 \\
 3C 353	 & 	138.0	 & 	2/L	 & 		42.68	$\pm$	0.11	 & 					 & 	2	 & 	42.31	$\pm$	0.30	 & 		22.8	$\pm$	0.45	 & 	42.46	$\pm$	0.30	 & 	$\le$	43.48			 & 	1	 & 30, 31 \\
 3C 382	 & 	267.0	 & 	1	 & 		44.33	$\pm$	0.04	 & 					 & 	5+	 & 	44.55	$\pm$	0.15	 & 		20.8	$\pm$	0.45	 & 	44.54	$\pm$	0.17	 & 		44.87	$\pm$	0.02	 & 	1	 & 32, 33, 3 \\
 3C 390.3	 & 	259.0	 & 	1.5	 & 		44.46	$\pm$	0.15	 & 					 & 	4	 & 	44.41	$\pm$	0.19	 & 		20.7	$\pm$	0.45	 & 	44.39	$\pm$	0.21	 & 		44.91	$\pm$	0.02	 & 	1	 & 10, 34, 3 \\
 3C 403	 & 	271.0	 & 	2	 & 		44.32	$\pm$	0.06	 & 					 & 	2	 & 	42.97	$\pm$	0.30	 & 		23.7	$\pm$	0.09	 & 	43.73	$\pm$	0.42	 & 		44.48	$\pm$	0.08	 & 	1	 & 3, 35, 36 \\
 3C 445	 & 	254.0	 & 	1.5	 & 		44.54	$\pm$	0.04	 & 					 & 	3	 & 	43.69	$\pm$	0.07	 & 		23.1	$\pm$	0.12	 & 	44.11	$\pm$	0.16	 & 		44.51	$\pm$	0.05	 & 	1	 & 37, 38, 39 \\
 3C 452	 & 	378.0	 & 	2	 & 		44.25	$\pm$	0.08	 & 					 & 	3	 & 	43.52	$\pm$	0.05	 & 		23.6	$\pm$	0.17	 & 	43.96	$\pm$	0.06	 & 		44.75	$\pm$	0.06	 & 	1	 & 10, 40, 6, 3 \\
 4C +73.08	 & 	271.0	 & 	2	 & 		43.69	$\pm$	0.13	 & 					 & 	1	 & 	42.89	$\pm$	0.30	 & 		24.0	$\pm$	0.45	 & 	43.79	$\pm$	0.30	 & 		44.13	$\pm$	0.11	 & 	1	 & 41 \\
 Ark 120	 & 	149.0	 & 	1	 & 		44.22	$\pm$	0.02	 & 					 & 	3	 & 	43.91	$\pm$	0.04	 & 	$\le$	20.0			 & 	43.90	$\pm$	0.06	 & 		44.27	$\pm$	0.03	 & 	0	 & 42, 43, 3 \\
 Cen A	 & 	3.8	 & 	2	 & 		41.82	$\pm$	0.04	 & 		41.81	$\pm$	0.05	 & 	4	 & 	41.61	$\pm$	0.06	 & 		23.1	$\pm$	0.45	 & 	41.95	$\pm$	0.10	 & 		42.38	$\pm$	0.00	 & 	1	 & 44, 45 \\
 Circinus	 & 	4.2	 & 	2	 & 		42.65	$\pm$	0.05	 & 		42.68	$\pm$	0.04	 & 	3	 & 	40.44	$\pm$	0.30	 & 		24.7	$\pm$	0.17	 & 	42.26	$\pm$	0.28	 & 		41.76	$\pm$	0.01	 & 	0	 & 46, 47, 48 \\
 Cygnus A	 & 	257.0	 & 	2	 & 		44.05	$\pm$	0.09	 & 					 & 	3	 & 	44.01	$\pm$	0.30	 & 		23.3	$\pm$	0.08	 & 	44.34	$\pm$	0.04	 & 		45.04	$\pm$	0.01	 & 	1	 & 10, 49, 17 \\
 ESO 5-4	 & 	22.4	 & 	2	 & 		41.64	$\pm$	0.05	 & 		41.82	$\pm$	0.13	 & 	1	 & 	40.82	$\pm$	0.30	 & 		24.0	$\pm$	0.45	 & 	41.77	$\pm$	0.30	 & 		42.30	$\pm$	0.06	 & 	?	 & 50 \\
 ESO 33-2	 & 	82.3	 & 	2	 & 		43.55	$\pm$	0.13	 & 					 & 	3	 & 	42.83	$\pm$	0.16	 & 		22.1	$\pm$	0.45	 & 	42.89	$\pm$	0.15	 & 		43.24	$\pm$	0.08	 & 	0	 & 51, 52, 14 \\
 ESO 103-35	 & 	59.5	 & 	2	 & 		43.71	$\pm$	0.19	 & 					 & 	3	 & 	43.04	$\pm$	0.06	 & 		23.3	$\pm$	0.03	 & 	43.22	$\pm$	0.35	 & 		43.68	$\pm$	0.02	 & 	0	 & 53, 16, 3 \\
 ESO 121-28	 & 	187.0	 & 	2	 & 		43.19	$\pm$	0.04	 & 					 & 	1	 & 	43.09	$\pm$	0.30	 & 		23.2	$\pm$	0.45	 & 	43.41	$\pm$	0.30	 & 		44.07	$\pm$	0.06	 & 	?	 & 3 \\
 ESO 141-55	 & 	169.0	 & 	1.2	 & 		44.11	$\pm$	0.09	 & 					 & 	3	 & 	43.85	$\pm$	0.10	 & 	$\le$	20.8			 & 	43.85	$\pm$	0.10	 & 		44.27	$\pm$	0.04	 & 	0	 & 54, 55, 14 \\
 ESO 198-24	 & 	208.0	 & 	1	 & 		43.64	$\pm$	0.11	 & 					 & 	3	 & 	43.67	$\pm$	0.13	 & 	$\le$	20.3			 & 	43.67	$\pm$	0.11	 & 		44.18	$\pm$	0.06	 & 	?	 & 56, 57, 58 \\
 ESO 209-12	 & 	189.0	 & 	1.5	 & 		44.23	$\pm$	0.06	 & 					 & 	2	 & 	43.62	$\pm$	0.30	 & 		19.0	$\pm$	0.45	 & 	43.62	$\pm$	0.30	 & 		43.96	$\pm$	0.08	 & 	?	 & 53, 59 \\
 ESO 263-13	 & 	158.0	 & 	2	 & 		43.62	$\pm$	0.03	 & 					 & 	3	 & 	42.87	$\pm$	0.16	 & 		23.5	$\pm$	0.09	 & 	43.42	$\pm$	0.03	 & 		44.02	$\pm$	0.06	 & 	?	 & 53, 60, 61 \\
 ESO 297-18	 & 	111.0	 & 	2	 & 		43.04	$\pm$	0.07	 & 					 & 	1	 & 	42.69	$\pm$	0.30	 & 		23.6	$\pm$	0.45	 & 	43.60	$\pm$	0.30	 & 		44.01	$\pm$	0.03	 & 	?	 & 50 \\
 ESO 323-32	 & 	76.4	 & 	1.9	 & 		42.97	$\pm$	0.05	 & 					 & 	2	 & 	41.40	$\pm$	0.30	 & 		24.1	$\pm$	0.45	 & 	42.50	$\pm$	0.30	 & 		43.03	$\pm$	0.19	 & 	?	 & 53, 60, 25 \\
 ESO 323-77	 & 	71.8	 & 	1.2	 & 		43.74	$\pm$	0.09	 & 					 & 	5+	 & 	42.57	$\pm$	0.21	 & 		23.6	$\pm$	0.45	 & 	42.76	$\pm$	0.12	 & 		43.31	$\pm$	0.10	 & 	?	 & 62 \\
 ESO 362-18	 & 	56.5	 & 	1.5	 & 		43.17	$\pm$	0.05	 & 					 & 	3	 & 	42.62	$\pm$	0.49	 & 		22.0	$\pm$	1.80	 & 	42.73	$\pm$	0.31	 & 		43.27	$\pm$	0.04	 & 	?	 & 19, 43, 63 \\
 ESO 416-2	 & 	272.0	 & 	1.9	 & 		43.73	$\pm$	0.06	 & 					 & 	2	 & 	43.59	$\pm$	0.30	 & 	$\le$	21.6			 & 	43.58	$\pm$	0.30	 & 		44.31	$\pm$	0.09	 & 	1	 & 64, 14 \\
 ESO 506-27	 & 	119.0	 & 	2	 & 		43.88	$\pm$	0.04	 & 					 & 	3	 & 	42.69	$\pm$	0.18	 & 		23.9	$\pm$	0.45	 & 	43.37	$\pm$	0.44	 & 		44.20	$\pm$	0.03	 & 	?	 & 65, 16, 61 \\
 ESO 511-30	 & 	105.0	 & 	1	 & 		43.24	$\pm$	0.04	 & 					 & 	2	 & 	43.31	$\pm$	0.30	 & 	$\le$	21.0			 & 	43.28	$\pm$	0.30	 & 		43.72	$\pm$	0.06	 & 	?	 & 66, 43, 3 \\
 ESO 548-81	 & 	63.5	 & 	1	 & 		43.02	$\pm$	0.14	 & 					 & 	3	 & 	42.85	$\pm$	0.09	 & 	$\le$	21.0			 & 	42.87	$\pm$	0.12	 & 		43.33	$\pm$	0.04	 & 	0	 & 63, 43, 63 \\
 Fairall 9	 & 	215.0	 & 	1.2	 & 		44.59	$\pm$	0.04	 & 					 & 	5	 & 	44.03	$\pm$	0.09	 & 	$\le$	20.5			 & 	43.99	$\pm$	0.11	 & 		44.44	$\pm$	0.03	 & 	?	 & 37, 67, 43, 3 \\
 Fairall 49	 & 	90.1	 & 	2	 & 		43.96	$\pm$	0.20	 & 					 & 	4	 & 	43.33	$\pm$	0.11	 & 		22.2	$\pm$	0.45	 & 	43.34	$\pm$	0.11	 & 		43.14	$\pm$	0.12	 & 	?	 & 68, 57 \\
 Fairall 51	 & 	64.1	 & 	1.5	 & 		43.68	$\pm$	0.04	 & 		43.76	$\pm$	0.01	 & 	3	 & 	42.98	$\pm$	0.01	 & 		22.4	$\pm$	0.12	 & 	43.05	$\pm$	0.03	 & 		43.30	$\pm$	0.05	 & 	0	 & 69, 70, 14 \\
 Fornax A	 & 	19.0	 & 	L	 & 		41.27	$\pm$	0.12	 & 					 & 	2	 & 	39.96	$\pm$	0.30	 & 		21.5	$\pm$	0.45	 & 	39.73	$\pm$	0.41	 & 	$\le$	41.76			 & 	1	 & 19, 11, 71 \\
 H 0557-385	 & 	156.0	 & 	1.2	 & 		44.49	$\pm$	0.04	 & 					 & 	5+	 & 	43.47	$\pm$	0.57	 & 		22.8	$\pm$	0.78	 & 	43.86	$\pm$	0.23	 & 		43.93	$\pm$	0.05	 & 	?	 & 72, 73, 74, 3 \\
 H1143-182	 & 	156.0	 & 	1.5	 & 		43.69	$\pm$	0.04	 & 					 & 	2	 & 	43.77	$\pm$	0.30	 & 	$\le$	20.5			 & 	43.75	$\pm$	0.30	 & 		44.19	$\pm$	0.04	 & 	0	 & 75, 3 \\
 Hydra A	 & 	260.0	 & 	L	 & 	$\le$	42.98			 & 					 & 	3	 & 	42.09	$\pm$	0.03	 & 		22.6	$\pm$	0.12	 & 	42.25	$\pm$	0.08	 & 	$\le$	44.03			 & 	1	 & 76, 11, 77 \\
 IC 1459	 & 	30.3	 & 	L	 & 	$\le$	41.67			 & 					 & 	1	 & 	40.83	$\pm$	0.30	 & 		21.8	$\pm$	0.45	 & 	40.71	$\pm$	0.30	 & 	$\le$	42.17			 & 	1	 & 76, 78, 79 \\
 IC 4329A	 & 	76.5	 & 	1.2	 & 		44.31	$\pm$	0.04	 & 		44.41	$\pm$	0.03	 & 	5+	 & 	43.87	$\pm$	0.05	 & 		21.5	$\pm$	0.42	 & 	43.85	$\pm$	0.09	 & 		44.31	$\pm$	0.01	 & 	0	 & 80, 19, 81, 3 \\
 IC 5063	 & 	49.1	 & 	2	 & 		43.77	$\pm$	0.03	 & 		43.95	$\pm$	0.01	 & 	4	 & 	42.61	$\pm$	0.26	 & 		23.4	$\pm$	0.06	 & 	42.86	$\pm$	0.10	 & 		43.32	$\pm$	0.03	 & 	1	 & 82, 83, 36 \\
 IRAS 09149-6206	 & 	269.0	 & 	1	 & 		45.00	$\pm$	0.06	 & 					 & 	2	 & 	44.05	$\pm$	0.30	 & 		21.7	$\pm$	0.45	 & 	44.05	$\pm$	0.30	 & 		44.44	$\pm$	0.05	 & 	?	 & 84, 85 \\
 IRAS 13349+2438	 & 	522.0	 & 	1n	 & 		45.59	$\pm$	0.05	 & 					 & 	3	 & 	43.96	$\pm$	0.11	 & 		21.6	$\pm$	0.45	 & 	43.95	$\pm$	0.09	 & 	$\le$	44.64			 & 	?	 & 86, 14, 19 \\
 I Zw 1	 & 	269.0	 & 	1n	 & 		44.96	$\pm$	0.07	 & 					 & 	3	 & 	43.68	$\pm$	0.16	 & 	$\le$	21.0		0.45	 & 	43.68	$\pm$	0.17	 & 	$\le$	44.06			 & 	0	 & 14, 19, 87 \\
 LEDA 13946	 & 	166.0	 & 	2	 & 	$\le$	42.98			 & 					 & 	2	 & 	42.74	$\pm$	0.36	 & 		23.2	$\pm$	0.45	 & 	43.20	$\pm$	0.30	 & 		43.84	$\pm$	0.09	 & 	?	 & 88, 3 \\
 LEDA 170194	 & 	173.0	 & 	2	 & 		43.58	$\pm$	0.08	 & 					 & 	2	 & 	43.70	$\pm$	0.30	 & 		22.6	$\pm$	0.45	 & 	43.80	$\pm$	0.30	 & 		44.17	$\pm$	0.06	 & 	?	 & 89, 3 \\
 LEDA 178130	 & 	160.0	 & 	2	 & 		43.60	$\pm$	0.13	 & 					 & 	3	 & 	43.57	$\pm$	0.06	 & 		23.1	$\pm$	0.33	 & 	43.81	$\pm$	0.36	 & 		44.27	$\pm$	0.03	 & 	0	 & 88, 5, 3 \\
 LEDA 549777	 & 	284.1	 & 	2	 & 		43.89	$\pm$	0.06	 & 					 & 	2	 & 	43.23	$\pm$	0.30	 & 		23.1	$\pm$	0.05	 & 	43.29	$\pm$	0.30	 & 		44.25	$\pm$	0.12	 & 	?	 & 90, 91, 3 \\
 M51a	 & 	8.1	 & 	2	 & 		40.70	$\pm$	0.19	 & 					 & 	5+	 & 	39.18	$\pm$	0.31	 & 	$\ge$	24.3			 & 	40.59	$\pm$	0.60	 & 	$\le$	41.02			 & 	0	 & 92, 83, 19 \\
 M81	 & 	3.6	 & 	L	 & 		40.73	$\pm$	0.05	 & 		40.70	$\pm$	0.05	 & 	5+	 & 	40.23	$\pm$	0.06	 & 		20.7	$\pm$	0.45	 & 	40.23	$\pm$	0.05	 & 		40.39	$\pm$	0.10	 & 	1	 & 92, 93, 94, 3 \\
 M87	 & 	16.7	 & 	L	 & 		41.25	$\pm$	0.06	 & 	$\le$	41.39			 & 	3	 & 	40.63	$\pm$	0.10	 & 		21.5	$\pm$	0.45	 & 	40.66	$\pm$	0.16	 & 	$\le$	41.65			 & 	1	 & 76, 78, 29 \\
 MCG-1-5-47	 & 	73.4	 & 	2	 & 		42.94	$\pm$	0.15	 & 					 & 	3	 & 	42.33	$\pm$	0.13	 & 		23.2	$\pm$	0.16	 & 	42.81	$\pm$	0.13	 & 		43.17	$\pm$	0.09	 & 	?	 & 95, 61, 96 \\
 MCG-1-13-25	 & 	71.2	 & 	1.2	 & 		42.57	$\pm$	0.13	 & 					 & 	2	 & 	42.76	$\pm$	0.30	 & 	$\le$	19.6			 & 	42.76	$\pm$	0.30	 & 		43.28	$\pm$	0.08	 & 	?	 & 64, 64 \\
 MCG-1-24-12	 & 	93.8	 & 	2	 & 		43.46	$\pm$	0.04	 & 					 & 	1	 & 	42.98	$\pm$	0.30	 & 		22.9	$\pm$	0.45	 & 	43.36	$\pm$	0.30	 & 		43.63	$\pm$	0.05	 & 	?	 & 97, 3 \\
 MCG-2-8-14	 & 	72.5	 & 	2	 & 		42.84	$\pm$	0.07	 & 					 & 	2	 & 	42.52	$\pm$	0.30	 & 		23.1	$\pm$	0.45	 & 	42.90	$\pm$	0.30	 & 		43.21	$\pm$	0.07	 & 	?	 & 95, 98 \\
 MCG-2-8-39	 & 	133.0	 & 	2	 & 		44.09	$\pm$	0.07	 & 		44.12	$\pm$	0.05	 & 	1	 & 	42.24	$\pm$	0.30	 & 		23.6	$\pm$	0.45	 & 	42.82	$\pm$	0.30	 & 	$\le$	43.45			 & 	?	 & 19, 99 \\
 MCG-3-34-64	 & 	79.3	 & 	1.8/2	 & 		44.00	$\pm$	0.05	 & 					 & 	1	 & 	42.24	$\pm$	0.30	 & 		23.6	$\pm$	0.01	 & 	43.33	$\pm$	0.48	 & 		43.36	$\pm$	0.07	 & 	?	 & 100, 19 \\
 MCG-5-23-16	 & 	42.8	 & 	1.9	 & 		43.59	$\pm$	0.04	 & 		43.73	$\pm$	0.02	 & 	3	 & 	43.28	$\pm$	0.30	 & 		22.1	$\pm$	0.07	 & 	43.28	$\pm$	0.30	 & 		43.64	$\pm$	0.01	 & 	0	 & 101, 102, 103 \\
 MCG-6-30-15	 & 	38.8	 & 	1.5	 & 		43.19	$\pm$	0.07	 & 					 & 	5+	 & 	42.80	$\pm$	0.15	 & 		20.2	$\pm$	0.45	 & 	42.80	$\pm$	0.14	 & 		43.06	$\pm$	0.03	 & 	0	 & 104, 105, 57 \\
 MR 2251-178	 & 	293.0	 & 	1.5	 & 		44.36	$\pm$	0.06	 & 					 & 	5	 & 	44.26	$\pm$	0.35	 & 		21.5	$\pm$	0.19	 & 	44.41	$\pm$	0.19	 & 		45.02	$\pm$	0.02	 & 	0	 & 106, 107, 108, 3 \\
 Mrk 3	 & 	60.6	 & 	2	 & 		43.71	$\pm$	0.10	 & 					 & 	3	 & 	42.43	$\pm$	0.02	 & 		24.0	$\pm$	0.45	 & 	43.55	$\pm$	0.34	 & 		43.79	$\pm$	0.01	 & 	?	 & 109, 110, 111 \\
 Mrk 304	 & 	301.0	 & 	1	 & 		44.15	$\pm$	0.17	 & 					 & 	1	 & 	43.54	$\pm$	0.30	 & 		22.8	$\pm$	0.45	 & 	43.67	$\pm$	0.30	 & 		44.02	$\pm$	0.19	 & 	?	 & 112, 113 \\
 Mrk 509	 & 	153.0	 & 	1.5	 & 		44.24	$\pm$	0.05	 & 		44.34	$\pm$	0.02	 & 	5+	 & 	44.07	$\pm$	0.13	 & 	$\le$	20.7			 & 	44.12	$\pm$	0.09	 & 		44.43	$\pm$	0.02	 & 	0	 & 114, 115, 57 \\
 Mrk 590	 & 	116.0	 & 	1	 & 		43.60	$\pm$	0.04	 & 					 & 	3	 & 	42.98	$\pm$	0.09	 & 	$\le$	19.4			 & 	42.98	$\pm$	0.09	 & 		43.43	$\pm$	0.13	 & 	?	 & 64, 116, 117 \\
 Mrk 841	 & 	170.0	 & 	1.5	 & 		44.15	$\pm$	0.11	 & 					 & 	5+	 & 	43.71	$\pm$	0.30	 & 	$\le$	20.4			 & 	43.71	$\pm$	0.30	 & 		44.09	$\pm$	0.06	 & 	0	 & 118, 119, 87 \\
 Mrk 915	 & 	104.0	 & 	1.9	 & 		43.44	$\pm$	0.04	 & 		43.64	$\pm$	0.03	 & 	5+	 & 	43.21	$\pm$	0.30	 & 		21.7	$\pm$	0.45	 & 	43.21	$\pm$	0.30	 & 		43.60	$\pm$	0.09	 & 	?	 & 70, 14 \\
 Mrk 926	 & 	210.0	 & 	1.5	 & 		44.16	$\pm$	0.10	 & 					 & 	4	 & 	44.32	$\pm$	0.11	 & 	$\le$	20.4			 & 	44.30	$\pm$	0.13	 & 		44.78	$\pm$	0.02	 & 	0	 & 119, 120, 14, 3 \\
 Mrk 937	 & 	129.0	 & 	1	 & 	$\le$	42.86			 & 					 & 	1	 & 	42.62	$\pm$	0.30	 & 		20.5	$\pm$	0.45	 & 	42.62	$\pm$	0.30	 & 	$\le$	43.43			 & 	?	 & 55, 121 \\
 Mrk 1014	 & 	807.0	 & 	1.5	 & 		45.29	$\pm$	0.15	 & 					 & 	2	 & 	43.96	$\pm$	0.30	 & 	$\le$	21.0			 & 	44.09	$\pm$	0.30	 & 	$\le$	45.02			 & 	?	 & 87, 122, 123 \\
 Mrk 1018	 & 	191.0	 & 	1	 & 		43.76	$\pm$	0.04	 & 					 & 	3	 & 	43.63	$\pm$	0.08	 & 	$\le$	20.0			 & 	43.62	$\pm$	0.06	 & 		44.16	$\pm$	0.07	 & 	0	 & 124, 43, 3 \\
 Mrk 1239	 & 	95.4	 & 	1n	 & 		44.21	$\pm$	0.07	 & 		44.21	$\pm$	0.02	 & 	1	 & 	42.13	$\pm$	0.30	 & 		23.5	$\pm$	0.45	 & 	43.32	$\pm$	0.30	 & 	$\le$	43.16			 & 	?	 & 125, 126 \\
 NGC 235A	 & 	96.2	 & 	2	 & 		43.29	$\pm$	0.16	 & 					 & 	1	 & 	42.53	$\pm$	0.30	 & 		23.5	$\pm$	0.45	 & 	43.18	$\pm$	0.30	 & 		43.72	$\pm$	0.04	 & 	?	 & 3 \\
 NGC 454E	 & 	52.3	 & 	2	 & 		43.05	$\pm$	0.09	 & 					 & 	3	 & 	41.80	$\pm$	0.23	 & 		23.4	$\pm$	0.20	 & 	42.29	$\pm$	0.23	 & 		42.76	$\pm$	0.11	 & 	?	 & 127, 5, 3 \\
 NGC 526A	 & 	82.8	 & 	1.9	 & 		43.68	$\pm$	0.05	 & 					 & 	4	 & 	43.30	$\pm$	0.06	 & 		22.1	$\pm$	0.06	 & 	43.28	$\pm$	0.10	 & 		43.72	$\pm$	0.03	 & 	0	 & 37, 19, 57 \\
 NGC 612	 & 	132.0	 & 	2	 & 	$\le$	43.24			 & 					 & 	4	 & 	42.46	$\pm$	0.16	 & 		23.9	$\pm$	0.18	 & 	43.13	$\pm$	0.42	 & 		44.06	$\pm$	0.04	 & 	1	 & 128, 129, 90, 3 \\
 NGC 788	 & 	57.2	 & 	2	 & 		43.12	$\pm$	0.05	 & 					 & 	2	 & 	42.29	$\pm$	0.30	 & 		23.7	$\pm$	0.17	 & 	42.96	$\pm$	0.30	 & 		43.50	$\pm$	0.02	 & 	?	 & 130, 89, 3 \\
 NGC 985	 & 	195.0	 & 	1.5	 & 		44.31	$\pm$	0.06	 & 					 & 	3	 & 	43.63	$\pm$	0.23	 & 		21.2	$\pm$	0.45	 & 	43.63	$\pm$	0.23	 & 		44.16	$\pm$	0.05	 & 	0	 & 131, 132, 43 \\
 NGC 1052	 & 	19.4	 & 	L	 & 		42.19	$\pm$	0.06	 & 		42.41	$\pm$	0.04	 & 	5+	 & 	41.07	$\pm$	0.11	 & 		23.0	$\pm$	0.45	 & 	41.52	$\pm$	0.08	 & 		42.12	$\pm$	0.07	 & 	1	 & 76, 133 \\
 NGC 1068	 & 	14.4	 & 	1.8/2	 & 		43.80	$\pm$	0.13	 & 		43.90	$\pm$	0.01	 & 	5+	 & 	41.01	$\pm$	0.08	 & 	$\ge$	25.0			 & 	43.64	$\pm$	0.30	 & 		41.94	$\pm$	0.06	 & 	0	 & 53, 134, 110 \\
 NGC 1097	 & 	17.0	 & 	L	 & 		41.17	$\pm$	0.06	 & 		41.41	$\pm$	0.09	 & 	1	 & 	40.78	$\pm$	0.30	 & 		20.4	$\pm$	0.45	 & 	40.78	$\pm$	0.30	 & 	$\le$	41.67			 & 	1	 & 135 \\
 NGC 1144	 & 	128.0	 & 	2	 & 		43.09	$\pm$	0.20	 & 					 & 	3	 & 	42.76	$\pm$	0.11	 & 		23.9	$\pm$	0.10	 & 	43.52	$\pm$	0.10	 & 		44.24	$\pm$	0.02	 & 	?	 & 65, 16, 65 \\
 NGC 1194	 & 	58.2	 & 	1.9	 & 		43.45	$\pm$	0.04	 & 					 & 	1	 & 	41.49	$\pm$	0.30	 & 		23.9	$\pm$	0.45	 & 	42.47	$\pm$	0.30	 & 		43.17	$\pm$	0.06	 & 	0	 & 19, 136 \\
 NGC 1275	 & 	76.8	 & 	1.5/L	 & 		44.21	$\pm$	0.04	 & 					 & 	3	 & 	43.03	$\pm$	0.42	 & 		22.6	$\pm$	0.45	 & 	43.02	$\pm$	0.43	 & 		43.71	$\pm$	0.02	 & 	1	 & 92, 13, 137 \\
 NGC 1365	 & 	17.9	 & 	1.8	 & 		42.54	$\pm$	0.04	 & 					 & 	5+	 & 	41.39	$\pm$	0.43	 & 		23.2	$\pm$	0.45	 & 	42.12	$\pm$	0.20	 & 		42.39	$\pm$	0.02	 & 	?	 & 138, 19, 139, 3 \\
 NGC 1553	 & 	16.4	 & 	L	 & 	$\le$	41.34			 & 					 & 	1	 & 	40.09	$\pm$	0.30	 & 		21.5	$\pm$	0.45	 & 	39.90	$\pm$	0.30	 & 	$\le$	41.63			 & 	?	 & 140, 141 \\
 NGC 1566	 & 	14.3	 & 	1.5	 & 		41.56	$\pm$	0.18	 & 		41.76	$\pm$	0.08	 & 	3	 & 	40.96	$\pm$	0.17	 & 		21.7	$\pm$	0.45	 & 	40.97	$\pm$	0.17	 & 		41.72	$\pm$	0.08	 & 	?	 & 142, 143, 14 \\
 NGC 2110	 & 	35.9	 & 	2	 & 		43.09	$\pm$	0.06	 & 		43.14	$\pm$	0.02	 & 	4	 & 	42.68	$\pm$	0.10	 & 		22.5	$\pm$	0.06	 & 	42.67	$\pm$	0.10	 & 		43.69	$\pm$	0.01	 & 	?	 & 80, 144, 144, 3 \\
 NGC 2992	 & 	39.7	 & 	1.5/2	 & 		42.95	$\pm$	0.13	 & 		43.23	$\pm$	0.02	 & 	5	 & 	42.71	$\pm$	0.45	 & 		22.0	$\pm$	0.30	 & 	42.53	$\pm$	0.59	 & 		42.71	$\pm$	0.08	 & 	?	 & 19, 145, 146, 3 \\
 NGC 3081	 & 	40.9	 & 	2	 & 		42.87	$\pm$	0.07	 & 		42.90	$\pm$	0.03	 & 	3	 & 	41.73	$\pm$	0.31	 & 		23.9	$\pm$	0.11	 & 	42.53	$\pm$	0.29	 & 		43.22	$\pm$	0.03	 & 	0	 & 53, 128, 147 \\
 NGC 3147	 & 	30.1	 & 	2	 & 		41.72	$\pm$	0.12	 & 					 & 	3	 & 	41.37	$\pm$	0.24	 & 	$\le$	20.7			 & 	41.37	$\pm$	0.22	 & 	$\le$	42.16			 & 	1	 & 92, 148, 149 \\
 NGC 3169	 & 	18.7	 & 	L	 & 		40.91	$\pm$	0.20	 & 					 & 	1	 & 	41.02	$\pm$	0.30	 & 		23.0	$\pm$	0.45	 & 	41.33	$\pm$	0.30	 & 	$\le$	41.75			 & 	1	 & 150 \\
 NGC 3227	 & 	22.1	 & 	1.5	 & 		42.47	$\pm$	0.10	 & 		42.89	$\pm$	0.03	 & 	4	 & 	42.10	$\pm$	0.26	 & 		22.2	$\pm$	0.71	 & 	42.14	$\pm$	0.21	 & 		42.81	$\pm$	0.02	 & 	0	 & 151, 19, 57, 3 \\
 NGC 3393	 & 	61.6	 & 	2	 & 		42.88	$\pm$	0.08	 & 					 & 	6	 & 	41.20	$\pm$	0.20	 & 		24.3	$\pm$	0.08	 & 	43.27	$\pm$	0.24	 & 		43.07	$\pm$	0.10	 & 	?	 & 53, 152, 153 \\
 NGC 3718	 & 	17.0	 & 	L	 & 		41.29	$\pm$	0.09	 & 					 & 	3	 & 	40.84	$\pm$	0.18	 & 		22.1	$\pm$	0.05	 & 	40.87	$\pm$	0.17	 & 		41.62	$\pm$	0.13	 & 	1	 & 154, 154, 154 \\
 NGC 3783	 & 	48.4	 & 	1.5	 & 		43.69	$\pm$	0.03	 & 		43.83	$\pm$	0.03	 & 	5+	 & 	43.25	$\pm$	0.07	 & 	$\le$	20.9			 & 	43.24	$\pm$	0.07	 & 		43.71	$\pm$	0.01	 & 	0	 & 75, 155, 156, 3 \\
 NGC 3998	 & 	14.1	 & 	L	 & 		41.63	$\pm$	0.04	 & 		41.78	$\pm$	0.04	 & 	3	 & 	41.33	$\pm$	0.06	 & 		21.3	$\pm$	0.99	 & 	41.33	$\pm$	0.06	 & 		41.55	$\pm$	0.11	 & 	1	 & 76, 154, 157 \\
 NGC 4051	 & 	12.2	 & 	1n	 & 		42.33	$\pm$	0.04	 & 					 & 	5+	 & 	41.55	$\pm$	0.17	 & 	$\le$	20.1			 & 	41.55	$\pm$	0.17	 & 		41.85	$\pm$	0.03	 & 	0	 & 158, 159, 160, 3 \\
 NGC 4074	 & 	107.0	 & 	2	 & 		43.35	$\pm$	0.11	 & 					 & 	1	 & 	42.58	$\pm$	0.30	 & 		23.3	$\pm$	0.45	 & 	42.96	$\pm$	0.30	 & 		43.61	$\pm$	0.07	 & 	0	 & 16 \\
 NGC 4111	 & 	15.0	 & 	L	 & 		40.40	$\pm$	0.10	 & 					 & 	2	 & 	39.46	$\pm$	0.26	 & 		23.6	$\pm$		 & 	39.90	$\pm$	0.47	 & 	$\le$	41.56			 & 	?	 & 76, 140, 157 \\
 NGC 4138	 & 	13.8	 & 	1.9	 & 		41.09	$\pm$	0.06	 & 					 & 	2	 & 	40.71	$\pm$	0.54	 & 		22.9	$\pm$	0.45	 & 	41.24	$\pm$	0.30	 & 		41.83	$\pm$	0.06	 & 	?	 & 161, 162 \\
 NGC 4151	 & 	13.3	 & 	1.5	 & 		42.84	$\pm$	0.07	 & 					 & 	5+	 & 	42.32	$\pm$	0.23	 & 		22.7	$\pm$	0.19	 & 	42.52	$\pm$	0.29	 & 		43.06	$\pm$	0.00	 & 	0	 & 155, 163, 164 \\
 NGC 4235	 & 	41.2	 & 	1.2	 & 		42.26	$\pm$	0.07	 & 		42.28	$\pm$	0.05	 & 	1	 & 	41.76	$\pm$	0.30	 & 		21.5	$\pm$	0.45	 & 	41.77	$\pm$	0.30	 & 		42.82	$\pm$	0.06	 & 	?	 & 165, 124, 4 \\
 NGC 4258	 & 	7.6	 & 	2	 & 		41.26	$\pm$	0.05	 & 					 & 	5+	 & 	40.80	$\pm$	0.18	 & 		22.6	$\pm$	0.45	 & 	40.99	$\pm$	0.15	 & 		41.18	$\pm$	0.07	 & 	?	 & 166, 167 \\
 NGC 4261	 & 	31.7	 & 	L	 & 		41.60	$\pm$	0.09	 & 		41.63	$\pm$	0.02	 & 	4	 & 	40.72	$\pm$	0.04	 & 		23.0	$\pm$	0.25	 & 	41.02	$\pm$	0.04	 & 	$\le$	42.21			 & 	1	 & 76, 168, 133 \\
 NGC 4278	 & 	16.1	 & 	L	 & 		40.29	$\pm$	0.10	 & 					 & 	5+	 & 	40.01	$\pm$	0.43	 & 		20.5	$\pm$	0.15	 & 	40.06	$\pm$	0.45	 & 	$\le$	41.62			 & 	1	 & 150, 169, 133 \\
 NGC 4374	 & 	17.1	 & 	2/L	 & 	$\le$	40.84			 & 					 & 	1	 & 	39.56	$\pm$	0.30	 & 		21.2	$\pm$	0.45	 & 	39.55	$\pm$	0.55	 & 	$\le$	41.67			 & 	1	 & 76, 10, 140 \\
 NGC 4388	 & 	19.2	 & 	2	 & 		42.32	$\pm$	0.07	 & 		42.75	$\pm$	0.01	 & 	5	 & 	41.80	$\pm$	0.24	 & 		23.5	$\pm$	0.07	 & 	42.26	$\pm$	0.23	 & 		43.09	$\pm$	0.01	 & 	?	 & 53, 16, 170, 3 \\
 NGC 4395	 & 	4.3	 & 	1.8	 & 		39.71	$\pm$	0.08	 & 					 & 	3	 & 	39.92	$\pm$	0.21	 & 		22.4	$\pm$	0.32	 & 	39.84	$\pm$	0.11	 & 		40.75	$\pm$	0.06	 & 	0	 & 161, 171, 172 \\
 NGC 4507	 & 	57.5	 & 	2	 & 		43.78	$\pm$	0.04	 & 					 & 	5+	 & 	42.64	$\pm$	0.25	 & 		23.7	$\pm$	0.19	 & 	43.21	$\pm$	0.17	 & 		43.87	$\pm$	0.01	 & 	?	 & 173, 16, 174, 3 \\
 NGC 4579	 & 	16.8	 & 	L	 & 		41.80	$\pm$	0.03	 & 		41.74	$\pm$	0.05	 & 	4	 & 	41.12	$\pm$	0.06	 & 		20.6	$\pm$	0.27	 & 	41.16	$\pm$	0.03	 & 		41.48	$\pm$	0.22	 & 	1	 & 157, 161, 175 \\
 NGC 4593	 & 	45.6	 & 	1	 & 		43.15	$\pm$	0.07	 & 		43.18	$\pm$	0.01	 & 	5+	 & 	42.86	$\pm$	0.31	 & 		20.4	$\pm$	0.45	 & 	42.86	$\pm$	0.31	 & 		43.34	$\pm$	0.02	 & 	0	 & 37, 155, 176, 3 \\
 NGC 4594	 & 	9.1	 & 	L	 & 		40.04	$\pm$	0.12	 & 					 & 	4	 & 	39.94	$\pm$	0.12	 & 		21.4	$\pm$	0.45	 & 	39.94	$\pm$	0.12	 & 	$\le$	41.12			 & 	1	 & 76, 93, 177 \\
 NGC 4941	 & 	21.2	 & 	2	 & 		42.00	$\pm$	0.05	 & 		42.30	$\pm$	0.03	 & 	4	 & 	40.92	$\pm$	0.41	 & 		23.8	$\pm$	0.17	 & 	41.53	$\pm$	0.18	 & 		42.04	$\pm$	0.11	 & 	?	 & 147, 143, 157 \\
 NGC 4992	 & 	119.0	 & 	1/2/L/N	 & 		43.51	$\pm$	0.09	 & 					 & 	3	 & 	42.60	$\pm$	0.01	 & 		23.9	$\pm$	0.12	 & 	43.39	$\pm$	0.18	 & 		43.96	$\pm$	0.04	 & 	?	 & 60, 4, 178 \\
 NGC 5033	 & 	18.1	 & 	1.2	 & 		41.21	$\pm$	0.06	 & 					 & 	3	 & 	40.99	$\pm$	0.28	 & 	$\le$	20.4			 & 	41.00	$\pm$	0.28	 & 		41.34	$\pm$	0.10	 & 	?	 & 161, 165, 93 \\
 NGC 5135	 & 	66.0	 & 	2	 & 		43.23	$\pm$	0.08	 & 					 & 	2	 & 	41.20	$\pm$	0.30	 & 		24.4	$\pm$	0.45	 & 	43.21	$\pm$	0.60	 & 	$\le$	42.84			 & 	1	 & 179, 180 \\
 NGC 5252	 & 	109.0	 & 	1.9	 & 		43.39	$\pm$	0.04	 & 					 & 	3	 & 	43.07	$\pm$	0.13	 & 		22.6	$\pm$	0.19	 & 	43.11	$\pm$	0.12	 & 		44.22	$\pm$	0.02	 & 	?	 & 181, 4, 3 \\
 NGC 5273	 & 	15.3	 & 	1.5	 & 		41.15	$\pm$	0.09	 & 					 & 	2	 & 	40.98	$\pm$	0.47	 & 		22.3	$\pm$	0.45	 & 	40.97	$\pm$	0.46	 & 		41.59	$\pm$	0.15	 & 	0	 & 161, 182 \\
 NGC 5506	 & 	31.6	 & 	2	 & 		43.41	$\pm$	0.03	 & 		43.52	$\pm$	0.04	 & 	5+	 & 	43.06	$\pm$	0.10	 & 		22.4	$\pm$	0.10	 & 	43.12	$\pm$	0.11	 & 		43.46	$\pm$	0.01	 & 	0	 & 83, 83, 155 \\
 NGC 5548	 & 	80.7	 & 	1.5	 & 		43.39	$\pm$	0.21	 & 					 & 	5+	 & 	43.38	$\pm$	0.25	 & 	$\le$	20.2			 & 	43.38	$\pm$	0.25	 & 		43.79	$\pm$	0.02	 & 	0	 & 183, 184, 155 \\
 NGC 5728	 & 	45.4	 & 	1.9/2	 & 		42.48	$\pm$	0.06	 & 		42.70	$\pm$	0.11	 & 	2	 & 	41.56	$\pm$	0.30	 & 		24.1	$\pm$	0.45	 & 	42.82	$\pm$	0.60	 & 		43.34	$\pm$	0.03	 & 	?	 & 60, 185 \\
 NGC 5995	 & 	117.0	 & 	1.9	 & 		44.13	$\pm$	0.06	 & 					 & 	3	 & 	43.46	$\pm$	0.19	 & 		22.0	$\pm$	0.45	 & 	43.45	$\pm$	0.15	 & 		43.85	$\pm$	0.06	 & 	?	 & 186, 24, 14 \\
 NGC 6251	 & 	112.0	 & 	1/2/L	 & 		42.75	$\pm$	0.10	 & 					 & 	5+	 & 	42.77	$\pm$	0.06	 & 		20.7	$\pm$	0.45	 & 	42.77	$\pm$	0.06	 & 	$\le$	43.30			 & 	1	 & 187, 188, 189 \\
 NGC 6300	 & 	14.3	 & 	2	 & 		42.53	$\pm$	0.11	 & 					 & 	3	 & 	41.61	$\pm$	0.39	 & 		23.3	$\pm$	0.01	 & 	42.09	$\pm$	0.44	 & 		42.39	$\pm$	0.02	 & 	?	 & 190, 191, 137 \\
 NGC 6814	 & 	20.1	 & 	1.5	 & 		42.06	$\pm$	0.10	 & 		42.08	$\pm$	0.04	 & 	5+	 & 	41.82	$\pm$	0.31	 & 	$\le$	20.5			 & 	41.82	$\pm$	0.31	 & 		42.57	$\pm$	0.03	 & 	0	 & 53, 43, 14, 3 \\
 NGC 6860	 & 	65.8	 & 	1.5	 & 		43.42	$\pm$	0.05	 & 		43.42	$\pm$	0.02	 & 	5+	 & 	42.94	$\pm$	0.20	 & 		22.0	$\pm$	0.46	 & 	42.94	$\pm$	0.19	 & 		43.44	$\pm$	0.04	 & 	0	 & 192, 192, 14, 3 \\
 NGC 7172	 & 	34.8	 & 	2	 & 		42.83	$\pm$	0.04	 & 					 & 	4	 & 	42.70	$\pm$	0.15	 & 		22.9	$\pm$	0.11	 & 	42.79	$\pm$	0.16	 & 		43.39	$\pm$	0.01	 & 	0	 & 101, 16, 19, 3 \\
 NGC 7213	 & 	23.0	 & 	1.5/L	 & 		42.51	$\pm$	0.04	 & 		42.62	$\pm$	0.06	 & 	4	 & 	42.20	$\pm$	0.07	 & 		20.3	$\pm$	0.45	 & 	42.19	$\pm$	0.06	 & 		42.43	$\pm$	0.05	 & 	1	 & 193, 194, 195, 3 \\
 NGC 7314	 & 	18.3	 & 	1.9/2	 & 		41.79	$\pm$	0.08	 & 		42.06	$\pm$	0.03	 & 	4	 & 	41.95	$\pm$	0.35	 & 		22.0	$\pm$	0.09	 & 	41.98	$\pm$	0.27	 & 		42.32	$\pm$	0.04	 & 	?	 & 114, 196, 197, 3 \\
 NGC 7469	 & 	67.9	 & 	1/1.5	 & 		43.83	$\pm$	0.04	 & 		44.03	$\pm$	0.05	 & 	5+	 & 	43.19	$\pm$	0.07	 & 	$\le$	20.7		0.45	 & 	43.19	$\pm$	0.07	 & 		43.57	$\pm$	0.03	 & 	?	 & 14, 75, 155 \\
 NGC 7626	 & 	45.4	 & 	L:	 & 	$\le$	42.06			 & 					 & 	1	 & 	40.97	$\pm$	0.30	 & 	$\le$	22.0			 & 	40.97	$\pm$	0.30	 & 	$\le$	42.52			 & 	1	 & 198 \\
 NGC 7674	 & 	126.0	 & 	2	 & 		44.26	$\pm$	0.05	 & 		44.43	$\pm$	0.01	 & 	3	 & 	42.07	$\pm$	0.09	 & 	$\ge$	24.4			 & 	44.02	$\pm$	0.55	 & 	$\le$	43.41			 & 	0	 & 199, 200, 19 \\
 PG 0026+129	 & 	691.0	 & 	1.2	 & 		44.66	$\pm$	0.04	 & 					 & 	3	 & 	44.57	$\pm$	0.16	 & 		21.2	$\pm$		 & 	44.57	$\pm$	0.16	 & 		44.85	$\pm$	0.15	 & 	0	 & 201, 202, 14 \\
 PG 0052+251	 & 	759.0	 & 	1.2	 & 		44.68	$\pm$	0.12	 & 					 & 	3	 & 	44.69	$\pm$	0.02	 & 	$\le$	20.7			 & 	44.69	$\pm$	0.02	 & 		44.98	$\pm$	0.14	 & 	0	 & 203, 201, 14 \\
 PG 0844+349	 & 	302.0	 & 	1	 & 		44.03	$\pm$	0.15	 & 					 & 	5+	 & 	43.52	$\pm$	0.33	 & 	$\le$	20.4			 & 	43.52	$\pm$	0.33	 & 		43.63	$\pm$	0.33	 & 	?	 & 87, 204, 205 \\
 PG 2130+099	 & 	288.0	 & 	1.5	 & 		44.67	$\pm$	0.05	 & 					 & 	3	 & 	43.74	$\pm$	0.23	 & 	$\le$	20.7			 & 	43.74	$\pm$	0.23	 & 	$\le$	44.12			 & 	0	 & 206, 207, 202 \\
 Pictor A	 & 	161.0	 & 	1.5/L	 & 		43.76	$\pm$	0.04	 & 		43.83	$\pm$	0.06	 & 	4	 & 	43.49	$\pm$	0.05	 & 		20.6	$\pm$	0.45	 & 	43.48	$\pm$	0.04	 & 		44.08	$\pm$	0.04	 & 	1	 & 208, 43, 22, 3 \\
 PKS 1417-19	 & 	586.0	 & 	1.5	 & 		44.31	$\pm$	0.18	 & 					 & 	2	 & 	44.27	$\pm$	0.30	 & 	$\le$	21.0			 & 	44.27	$\pm$	0.30	 & 	$\le$	44.74			 & 	1	 & 52, 14 \\
 PKS 1814-63	 & 	302.0	 & 	2	 & 		43.88	$\pm$	0.08	 & 					 & 	2	 & 	43.92	$\pm$	0.45	 & 		22.3	$\pm$	0.45	 & 	43.90	$\pm$	0.43	 & 		44.00	$\pm$	0.22	 & 	1	 & 101, 209 \\
 Z 41-20	 & 	170.0	 & 	2	 & 		43.40	$\pm$	0.04	 & 					 & 	3	 & 	43.28	$\pm$	0.07	 & 		23.1	$\pm$	0.14	 & 	43.52	$\pm$	0.17	 & 		43.86	$\pm$	0.09	 & 	0	 & 63, 25, 3 \\

\end{longtable}
\normalsize
{\it -- Notes:} 
(1), (2), (3), (4), and (5) short object name, distance, optical class, nuclear 12 and 18$\um$ luminosities from \cite{asmus_subarcsecond_2014};
(6) number of 2-10\,keV epochs used;
(7) average observed 2-10\,keV luminosity;
(8) average X-ray column density;
(9) average absorption-corrected 2-10\,keV luminosity;
(10) average observed 14-195\,keV luminosity from \swift/BAT by combining the data of the 54 and 70\,month source catalogues \citep{cusumano_palermo_2010,baumgartner_70_2013};
(11) radio-loudness flag (see Sect.~\ref{sec:rad} for explanation);
(12) references for the 2-10\,keV luminosities and column densities:
1: \cite{fabian_x-ray_2005}; 2: \cite{turner_suzaku_2009}; 3: \cite{winter_x-ray_2009}; 4: \cite{vasudevan_x-ray_2013}; 5: \cite{vasudevan_three_2013}; 6: \cite{fioretti_x-ray_2013}; 7: \cite{massaro_chandra_2012}; 8: \cite{kraft_radio_2007}; 9: \cite{torresi_3c_2009}; 10: \cite{evans_chandra_2006}; 11: \cite{rinn_x-ray_2005}; 12: this work (CSC); 13: \cite{balmaverde_chandra_2006}; 14: this work (XSPEC); 15: \cite{isobe_xmm-newton_2005}; 16: \cite{noguchi_new_2009}; 17: \cite{hodges-kluck_chandra_2010}; 18: \cite{massaro_chandra_2010}; 19: \cite{brightman_xmm-newton_2011-1}; 20: \cite{cowperthwaite_central_2012}; 21: \cite{ogle_multiwavelength_2005}; 22: \cite{hardcastle_chandra_2007}; 23: \cite{cusumano_palermo_2010}; 24: \cite{warwick_xmm-newton_2012}; 25: \cite{fukazawa_fe-k_2011}; 26: \cite{chernyakova_2003-2005_2007}; 27: \cite{hardcastle_x-ray_2006}; 28: \cite{hardcastle_active_2009}; 29: \cite{donato_obscuration_2004}; 30: \cite{kataoka_chandra_2008}; 31: \cite{goodger_inverse_2008}; 32: \cite{gliozzi_nature_2007}; 33: \cite{sambruna_suzaku_2011}; 34: \cite{sambruna_structure_2009}; 35: \cite{kraft_chandra_2005}; 36: \cite{tazaki_suzaku_2011}; 37: \cite{horst_mid_2008}; 38: \cite{braito_evidence_2011}; 39: \cite{reeves_chandra_2010}; 40: \cite{shelton_dynamics_2011}; 41: \cite{evans_xmm-newton_2008}; 42: \cite{vasudevan_simultaneous_2009}; 43: \cite{winter_swift_2012}; 44: \cite{evans_chandra_2004}; 45: \cite{markowitz_suzaku_2007}; 46: \cite{smith_chandra_2001}; 47: \cite{arevalo_2-79_2014}; 48: \cite{yang_suzaku_2009}; 49: \cite{young_chandra_2002}; 50: \cite{ueda_suzaku_2007}; 51: \cite{vignali_asca_1998}; 52: \cite{saxton_first_2008}; 53: \cite{gandhi_resolving_2009}; 54: \cite{gondoin_xmm-newton_2003}; 55: \cite{ueda_asca_2001}; 56: \cite{porquet_xmm-newton_2004}; 57: \cite{shu_cores_2010}; 58: \cite{laha_origin_2014}; 59: \cite{panessa_broad-band_2008}; 60: \cite{comastri_suzaku_2010}; 61: \cite{landi_agn_2007}; 62: \cite{miniutti_properties_2014}; 63: \cite{winter_x-ray_2008}; 64: \cite{gallo_xmm-newton_2006}; 65: \cite{winter_suzaku_2009}; 66: \cite{tombesi_evidence_2010}; 67: \cite{emmanoulopoulos_xmm-newton_2011}; 68: \cite{lobban_iron_2014}; 69: \cite{ricci_multi-zone_2010}; 70: \cite{beckmann_second_2009}; 71: \cite{kim_chandra_2003}; 72: \cite{ashton_xmm-newton_2006}; 73: \cite{longinotti_obscuring_2009}; 74: \cite{coffey_absorption_2014}; 75: \cite{nandra_xmm-newton_2007}; 76: \cite{gonzalez-martin_x-ray_2009}; 77: \cite{russell_radiative_2013}; 78: \cite{satyapal_joint_2004}; 79: \cite{fabbiano_x-ray-faint_2003}; 80: \cite{dadina_bepposax_2007}; 81: \cite{brenneman_broad-band_2014}; 82: \cite{risaliti_ubiquitous_2002}; 83: \cite{lamassa_uncovering_2011}; 84: \cite{vasudevan_optical--x-ray_2009}; 85: \cite{malizia_swift_2007}; 86: \cite{holczer_absorption_2007}; 87: \cite{piconcelli_xmm-newton_2005}; 88: \cite{eguchi_suzaku_2009}; 89: \cite{de_rosa_x-ray_2008}; 90: \cite{parisi_accurate_2009}; 91: Balokovic et al. (in prep); 92: \cite{panessa_x-ray_2006}; 93: \cite{ho_detection_2001}; 94: \cite{markoff_results_2008}; 95: \cite{baumgartner_70_2013}; 96: \cite{trippe_xmm_2011}; 97: \cite{malizia_bepposax/pds_2002}; 98: \cite{rodriguez_swift_2010}; 99: \cite{noguchi_scattered_2010}; 100: \cite{miniutti_iras_2007}; 101: \cite{awaki_variability_2006}; 102: \cite{reeves_revealing_2007}; 103: \cite{balestra_reprocessing_2004}; 104: \cite{miniutti_suzaku_2007}; 105: \cite{ponti_mapping_2004}; 106: \cite{gofford_broad-band_2011}; 107: \cite{ramirez_chandra_2008}; 108: \cite{kaspi_properties_2004}; 109: \cite{awaki_wide-band_2008}; 110: \cite{levenson_penetrating_2006}; 111: \cite{bianchi_xmm-newton_2005}; 112: \cite{piconcelli_evidence_2004}; 113: \cite{teng_x-quest:_2010}; 114: \cite{shinozaki_spectral_2006}; 115: \cite{ponti_xmm-newton_2009}; 116: \cite{longinotti_x-ray_2007}; 117: \cite{rivers_suzaku_2012}; 118: \cite{petrucci_unveiling_2007}; 119: \cite{bianchi_x-ray_2004}; 120: \cite{rivers_suzaku_2011}; 121: \cite{turner_asca_1997}; 122: \cite{teng_chandra_2005}; 123: \cite{boller_mrk_2002}; 124: \cite{bianchi_caixa:_2009}; 125: \cite{grupe_markarian_2004}; 126: \cite{miyazawa_broad-band_2009}; 127: \cite{marchese_ngc_2012}; 128: \cite{eguchi_suzaku_2011}; 129: \cite{sambruna_x-ray_1999}; 130: \cite{marinucci_link_2012}; 131: \cite{krongold_xmm-newton_2009}; 132: \cite{krongold_ionized_2005}; 133: \cite{hernandez-garcia_x-ray_2013}; 134: \cite{bauer_nustar_2014}; 135: \cite{nemmen_radiatively_2006}; 136: \cite{greenhill_prevalence_2008}; 137: \cite{beckmann_first_2006}; 138: \cite{risaliti_strong_2009}; 139: \cite{brenneman_examination_2013}; 140: \cite{flohic_central_2006}; 141: \cite{blanton_diffuse_2001}; 142: \cite{levenson_isotropic_2009}; 143: \cite{kawamuro_broadband_2013}; 144: \cite{evans_probing_2007}; 145: \cite{yaqoob_precision_2007}; 146: \cite{shu_ngc_2010}; 147: \cite{maiolino_heavy_1998}; 148: \cite{matt_suzaku_2012}; 149: \cite{bianchi_ngc_2008}; 150: \cite{terashima_chandra_2003}; 151: \cite{markowitz_comprehensive_2009}; 152: \cite{fabbiano_close_2011}; 153: \cite{koss_broadband_2015}; 154: \cite{younes_study_2011}; 155: \cite{rivers_spectral_2011}; 156: \cite{brenneman_spin_2011}; 157: \cite{terashima_x-ray_2002}; 158: \cite{king_distinctive_2011}; 159: \cite{mchardy_combined_2004}; 160: \cite{vaughan_rapid_2011}; 161: \cite{cappi_x-ray_2006}; 162: \cite{zhang_census_2009}; 163: \cite{wang_revisiting_2010}; 164: \cite{lubinski_extreme_2010}; 165: \cite{papadakis_x-ray_2008}; 166: \cite{fruscione_x-ray_2005}; 167: \cite{reynolds_probing_2009}; 168: \cite{satyapal_link_2005}; 169: \cite{younes_x-ray_2010}; 170: \cite{shirai_detailed_2008}; 171: \cite{gonzalez-martin_x-ray_2006}; 172: \cite{moran_extreme_2005}; 173: \cite{braito_decoupling_2013}; 174: \cite{marinucci_x-ray_2013}; 175: \cite{eracleous_three_2002}; 176: \cite{markowitz_suzaku_2009}; 177: \cite{li_x-ray_2011}; 178: \cite{sazonov_identification_2005}; 179: \cite{levenson_accretion_2004}; 180: \cite{singh_suzaku_2012}; 181: \cite{dadina_x-ray_2010}; 182: \cite{capetti_host_2005}; 183: \cite{andrade-velazquez_two-phase_2010}; 184: \cite{krongold_suzaku_2010}; 185: \cite{zhang_extragalactic_2006}; 186: \cite{panessa_unabsorbed_2002}; 187: \cite{evans_chandra_2005}; 188: \cite{gliozzi_xmm-newton_2004}; 189: \cite{gliozzi_long-term_2008}; 190: \cite{matsumoto_xmm-newton_2004}; 191: \cite{guainazzi_formerly_2002}; 192: \cite{winter_complex_2010}; 193: \cite{lobban_evidence_2010}; 194: \cite{bianchi_origin_2003}; 195: \cite{bianchi_broad-line_2008}; 196: \cite{ebrero_probing_2011}; 197: \cite{yaqoob_fe_2003}; 198: \cite{ho_radiatively_2009}; 199: \cite{malaguti_bepposax_1998}; 200: Gandhi et al. (in prep.); 201: \cite{ueda_asca_2005}; 202: \cite{lawson_ginga_1997}; 203: \cite{dammando_xmm-newton_2008}; 204: \cite{brinkmann_xmm-newton_2003}; 205: \cite{gallo_quasar_2011}; 206: \cite{gallo_investigating_2006}; 207: \cite{cardaci_characterization_2009}; 208: \cite{shi_far-infrared_2005}; 209: \cite{mingo_x-ray_2014};
}

\newpage
\appendix
\twocolumn
\section{XRT spectral fits}\label{sec:xrt}
For 19 sources, we analysed \swift/XRT data using  the online data products and tools from the UK Swift Science Data Centre \citep{evans_methods_2009}. 
These were used to build average calibrated 0.3-10\,keV spectra of all XRT observations for each object, which were then fitted with an absorbed power-law in the 1 to 10\,keV range using the \texttt{XSPEC} package v12.8.2 \citep{arnaud_xspec:_1996}.
The fitting results are summarised in the following Table~\ref{tab:xrt}. 
Galactic absorption was taken into account, and in case the fitted intrinsic absorption was smaller, we used the former as upper limit for $\nh$.
Two objects with strong absorption/low S/N are individually discussed in Appendix~\ref{app:obj}.

\begin{table*}
\caption{XRT spectral fits} % title of Table
\label{tab:xrt} % is used to refer this table in the text
\centering % used for centering table
\begin{tabular}{l		c			c		c		c		c		c		c	}

\hline\hline																										
Object	&	$\log N_\mathrm{H,Gal}$	&		$\log N_\mathrm{H,abs}$	&	$\Gamma$	&	$\chi^2$	&	DOF	&	$\log \fxo$	&	$\log \lxi$	\\
	&	[cm$^{-2}$]	&		[cm$^{-2}$]	&		&		&		&	[erg/s/cm$^{2}$]	&	[erg/s]	\\
(1)	&	(2)	&		(3)	&	(4)	&	(5)	&	(6)	&	(7)	&	(8)	\\

\hline																									
3C 93	&	21.29	&	$\le$	21.29	&	1.51	&	898.4	&	432	&	-12.05	&	44.58	\\
ESO 33-2	&	21.10	&		22.27	&	1.62	&	561.7	&	503	&	-11.09	&	42.88	\\
ESO 141-55	&	20.81	&	$\le$	20.81	&	2.1	&	302.2	&	405	&	-10.76	&	43.77	\\
ESO 253-3	&	20.61	&		23.48	&	2.56	&	154.8	&	160	&	-12.26	&	42.93	\\
ESO 416-2	&	20.25	&		21.91	&	2.11	&	192.6	&	182	&	-11.38	&	43.64	\\
Fairall 51	&	20.98	&		22.45	&	1.42	&	352.9	&	425	&	-10.73	&	43.02	\\
IRAS 13349+2438	&	20.02	&		20.85	&	1.49	&	1042.2	&	607	&	-11.54	&	43.92	\\
I Zw 1	&	20.77	&	$\le$	20.77	&	2.12	&	551.1	&	517	&	-11.39	&	43.55	\\
Mrk 915	&	20.82	&		21.65	&	1.26	&	727.8	&	554	&	-10.95	&	43.16	\\
Mrk 926	&	20.52	&	$\le$	20.52	&	1.68	&	619.3	&	572	&	-10.32	&	44.41	\\
NGC 1566	&	19.95	&		21.01	&	1.67	&	806.1	&	623	&	-11.48	&	40.91	\\
NGC 5995	&	21.20	&		22.14	&	1.64	&	882.9	&	677	&	-10.81	&	43.43	\\
NGC 6814	&	21.17	&	$\le$	21.17	&	1.56	&	968.2	&	701	&	-11.18	&	41.50	\\
NGC 6860	&	20.56	&		21.77	&	1.31	&	837.1	&	631	&	-10.80	&	42.91	\\
NGC 7469	&	20.73	&	$\le$	20.73	&	1.74	&	1121.7	&	834	&	-10.63	&	43.12	\\
PG 0026+129	&	20.80	&		21.30	&	2.08	&	693.0	&	490	&	-11.32	&	44.42	\\
PG 0052+251	&	20.71	&	$\le$	20.71	&	1.8	&	868.5	&	440	&	-11.14	&	44.70	\\
PKS 1417-19	&	21.00	&	$\le$	21.00	&	1.86	&	602.5	&	541	&	-11.30	&	44.31	\\
PKS 2158-380	&	20.14	&	$\ge$	22.00	&	\dots	&	\dots	&	\dots	&	-12.66	&	\dots	\\

\hline						                              
\end{tabular}
\end{table*}

\section{The remaining sources from the AGN MIR atlas}\label{app}
In the following sections, the remaining objects from the MIR AGN atlas that were not included in the reliable sample are discussed in three groups:
\begin{description}
   \item -- \textbf{X-ray unreliable AGN}: objects that are known securely to host AGN without significant star formation contamination, but do not have reliable $\lxi$ estimates, see Section~\ref{sec:CT};  
   \item -- \textbf{AGN/starburst composites}: objects that have certain AGN but possibly suffer from significant star formation contamination, see Section~\ref{sec:CP}; 
    \item -- \textbf{uncertain AGN}: objects with controversial or insufficient evidence for the presence of an AGN (similar to \citealt{asmus_subarcsecond_2014}), see Section~\ref{sec:can}.      
\end{description}
In addition, particularly interesting objects are further discussed in App.~\ref{app:obj} individually.

\subsection{The X-ray unreliable and Compton-thick AGN}\label{sec:CT}
\begin{figure}
%   \centering
%    \sidecaption
   \includegraphics[angle=0,width=8.5cm]{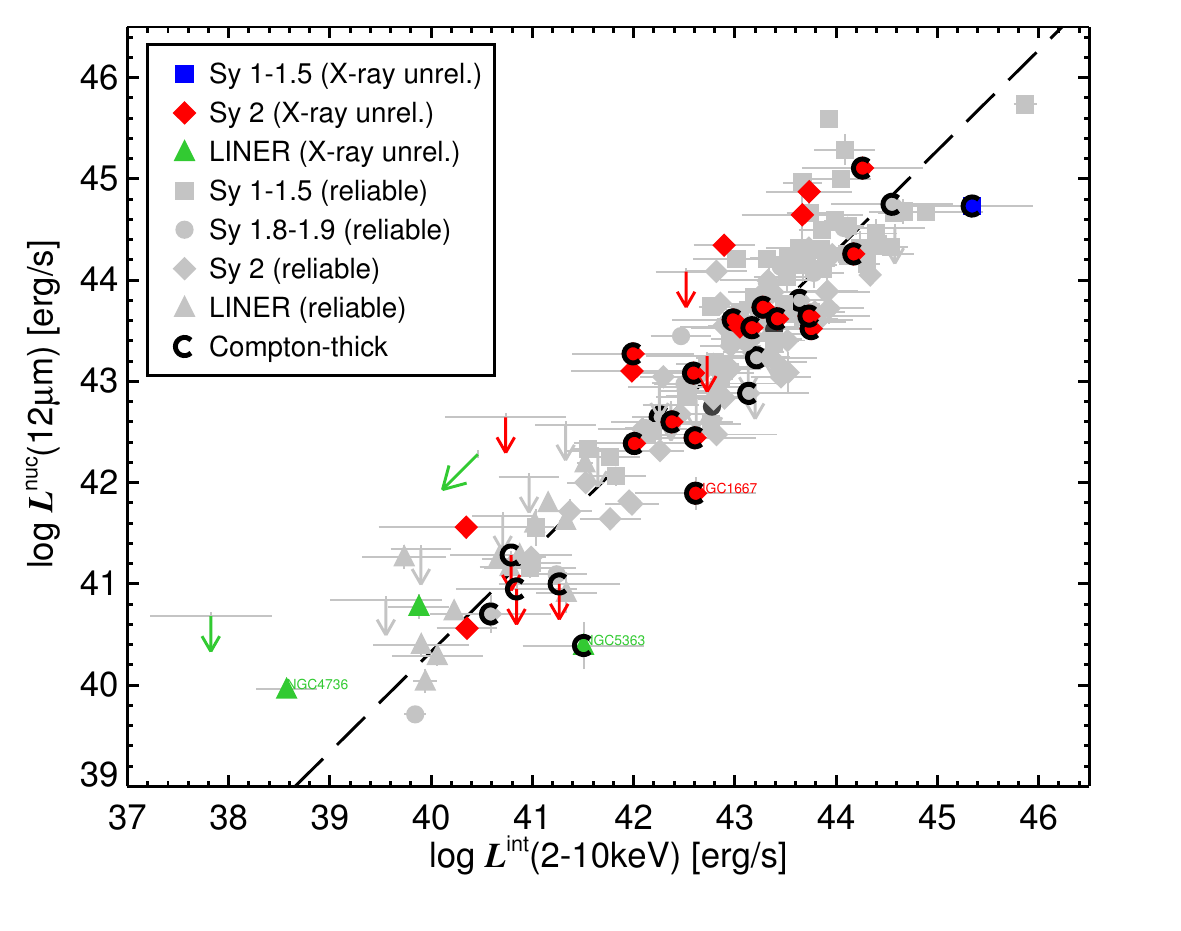}
    \caption{
             Distribution of the X-ray unreliable AGN in the observed $12\um$ to intrinsic 2-10\,keV plane.
  Symbols are as in Fig.~\ref{fig:f12_f2int}.
             All reliable AGN are in light gray and the long dashed line is their \texttt{linmix\_err} fit. 
             X-ray unreliable sources are plotted with colours.
            }
   \label{fig:unrel}
\end{figure}

Using Equations~\ref{eq:nh_diag} and \ref{eq:inv}, we can now predict $\nh$ and $\lxi$ for the 45 AGN from the AGN MIR atlas that we had to discard because of uncertain $\lxi$ and/or $\nh$ values (see Table~\ref{tab:unrel}).
Of these, 22 are previously classified as CT or CT candidates and are shown in Fig.~\ref{fig:nh_diag}. 
Only two of these objects are incompatible with being CT within the uncertainties using our $\nh$ diagnostic (NGC\,1667 and NGC\,5363).
Both objects have more than 1\,dex lower predicted $\lxi$ than stated in the literature assuming they are CT (see Fig.~\ref{fig:unrel}).
NGC\,7743 has not been detected in the MIR but the resulting upper limit on the predicted $\lxi$ is also significantly lower than the value stated in the literature assuming CT obscuration. 

Of the 23 objects previously not classified as Compton-thick, seven are compatible with being CT within the uncertainties (ESO\,253-3, IC\,4518W, IRAS\,01003-2238, IRAS\,05189-2524, NGC\,34, NGC\,3982, and NGC\,4736).
All these objects have indeed predicted $\lxi$ values significantly higher than the values found in the literature for non-CT scenarios. 
Finally, our diagnostics predict $\lxi$ and $\nh$ in good agreement with the literature values for the remaining sources, where the latter is available (apart from one exception: NGC\,5005).
Therefore, it is not surprising that all the other X-ray unreliable objects are lying close to the MIR--X-ray correlation as shown in Fig.~\ref{fig:unrel}.

Note that 13 of the 22 objects listed as bona-fide CT in \cite{gandhi_nustar_2014} are included in the atlas as well (NGC\,7582 is excluded because it turned out to be a changing look AGN; see, e.g., \citealt{asmus_subarcsecond_2014}).
All of them apart from NGC\,4945 (see next section) follow the general MIR--X-ray correlation, which indicates that this will also be the case for the whole local CT population.

\begin{table*}
\caption{Properties of X-ray unreliable AGN.} % title of Table
\label{tab:unrel} % is used to refer this table in the text
\centering % used for centering table
%\footnotesize
%\small
\scriptsize
%\tiny
\begin{tabular}{l	c	c	c	c	c	c	c	c	c	l}

\hline\hline																										
 &	 &	 &	 &	 &	 Lit.	&	Lit.	&	Pred.	& Pred.	&	&	\\
	&		&	Opt.	&		$\log L^\textrm{nuc}$	&	$\log L^\textrm{obs}$	&		$\log$	&	$\log L^\textrm{int}$	&		$\log$			&	$\log L^\textrm{int}$			&		$\log L^\textrm{obs}$	&		\\
Object	&	$D$	&	class	&		($12\um$)	&	(2-10\,keV)	&		$ \nh$	&	(2-10\,keV)	&		$ \nh$			&	(2-10\,keV)			&		(14-195\,keV)	&	2-10\,keV	\\
	&	[Mpc]	&		&		 [erg/s]	&	 [erg/s]	&		 [cm$^{-2}$]	&	 [erg/s]	&		 [cm$^{-2}$]			&	 [erg/s]			&		 [erg/s]	&	Ref.	\\
(1)	&	(2)	&	(3)	&		(4)	&	(5)	&		(6)	&	(7)	&		(8)			&	(9)			&		(10)	&	(11)	\\

\hline																									
 3C 135	 & 	620.0	 & 	2	 & 	$\le$	44.08			 & 		42.47	$\pm$	0.30	 & 		22.3	$\pm$	0.45	 & 		42.52	$\pm$	0.30	 & 					 & 	$\le$	44.08			 & 	$\le$	44.79			 & 1 \\
 3C 305	 & 	192.0	 & 	2	 & 	$\le$	42.64			 & 		40.30	$\pm$	0.30	 & 		21.2	$\pm$	0.34	 & 		40.74	$\pm$	0.60	 & 					 & 	$\le$	42.71			 & 	$\le$	43.77			 & 2, 3 \\
 3C 321	 & 	460.0	 & 	2	 & 		44.65	$\pm$	0.24	 & 		42.34	$\pm$	0.30	 & 		24.0	$\pm$	0.45	 & 		43.67	$\pm$	0.60	 & 		24.5	$\pm$	0.5	 & 		44.24	$\pm$	0.44	 & 	$\le$	44.53			 & 4, 5 \\
 3C 327	 & 	505.0	 & 	1	 & 		44.73	$\pm$	0.11	 & 		42.28	$\pm$	0.30	 & 	$\ge$	24.3			 & 		45.34	$\pm$	0.60	 & 		24.6	$\pm$	0.5	 & 		44.33	$\pm$	0.39	 & 	$\le$	44.61			 & 6, 7 \\
 3C 424	 & 	613.0	 & 	2	 & 	$\le$	43.25			 & 		42.43	$\pm$	0.30	 & 		22.7	$\pm$	0.45	 & 		42.73	$\pm$	0.60	 & 					 & 	$\le$	43.29			 & 	$\le$	44.78			 & 8 \\
 3C 449	 & 	72.4	 & 	L	 & 	$\le$	42.28			 & 	$\le$	40.48			 & 					 & 	$\le$	40.46			 & 					 & 	$\le$	42.37			 & 	$\le$	42.92			 & 4, 9, 10 \\
 ESO 138-1	 & 	41.8	 & 	2	 & 		43.61	$\pm$	0.02	 & 		41.62	$\pm$	0.30	 & 	$\ge$	24.3			 & 		42.98	$\pm$	0.60	 & 		24.3	$\pm$	0.4	 & 		43.26	$\pm$	0.37	 & 		42.61	$\pm$	0.10	 & 11, 12 \\
 ESO 253-3	 & 	196.0	 & 	2	 & 		44.34	$\pm$	0.08	 & 		42.40	$\pm$	0.30	 & 		23.5	$\pm$	0.45	 & 		42.93	$\pm$	0.30	 & 		24.2	$\pm$	0.4	 & 		43.96	$\pm$	0.38	 & 	$\le$	43.79			 & 7 \\
 ESO 428-14	 & 	28.2	 & 	2	 & 		42.44	$\pm$	0.06	 & 		40.56	$\pm$	0.30	 & 	$\ge$	24.3			 & 		42.61	$\pm$	0.60	 & 		24.2	$\pm$	0.4	 & 		42.15	$\pm$	0.37	 & 	$\le$	42.11			 & 13 \\
 IC 3639	 & 	53.6	 & 	2	 & 		43.52	$\pm$	0.04	 & 		40.84	$\pm$	0.56	 & 	$\ge$	25.0			 & 		43.75	$\pm$	0.60	 & 		24.7	$\pm$	0.6	 & 		43.17	$\pm$	0.37	 & 	$\le$	42.66			 & 11, 14, 15 \\
 IC 4518W	 & 	76.1	 & 	2	 & 		43.54	$\pm$	0.06	 & 		42.73	$\pm$	0.79	 & 		23.3	$\pm$	0.11	 & 		43.05	$\pm$	0.59	 & 		23.5	$\pm$	0.9	 & 		43.19	$\pm$	0.38	 & 		43.28	$\pm$	0.07	 & 16, 17, 18 \\
 IRAS 01003-2238	 & 	565.0	 & 	2	 & 		45.31	$\pm$	0.04	 & 		41.58	$\pm$	0.43	 & 					 & 					 & 		25.5	$\pm$	0.5	 & 		44.88	$\pm$	0.37	 & 	$\le$	44.71			 & 19, 20 \\
 IRAS 04103-2838	 & 	567.0	 & 	2	 & 		45.11	$\pm$	0.07	 & 		42.01	$\pm$	0.30	 & 	$\ge$	24.3			 & 		44.26	$\pm$	0.60	 & 		25.0	$\pm$	0.4	 & 		44.68	$\pm$	0.38	 & 	$\le$	44.71			 & 21, 20 \\
 IRAS 05189-2524	 & 	196.0	 & 	2	 & 		44.87	$\pm$	0.15	 & 		42.71	$\pm$	0.76	 & 		22.8	$\pm$	0.02	 & 		43.73	$\pm$	0.43	 & 		24.4	$\pm$	0.8	 & 		44.46	$\pm$	0.40	 & 		43.76	$\pm$	0.15	 & 22, 23, 24 \\
 Mrk 266SW	 & 	128.0	 & 	2	 & 		42.46	$\pm$	0.08	 & 		41.44	$\pm$	0.30	 & 	$\ge$	24.0			 & 					 & 		23.6	$\pm$	0.4	 & 		42.17	$\pm$	0.38	 & 	$\le$	43.42			 & 25 \\
 Mrk 573	 & 	73.1	 & 	2	 & 		43.53	$\pm$	0.07	 & 		41.13	$\pm$	0.30	 & 	$\ge$	24.2			 & 		43.17	$\pm$	0.60	 & 		24.6	$\pm$	0.4	 & 		43.18	$\pm$	0.38	 & 	$\le$	42.93			 & 15, 26, 27 \\
 NGC 34	 & 	83.5	 & 	2	 & 		43.10	$\pm$	0.05	 & 		41.36	$\pm$	0.30	 & 		23.7	$\pm$	0.45	 & 		41.98	$\pm$	0.60	 & 		24.1	$\pm$	0.4	 & 		42.78	$\pm$	0.37	 & 	$\le$	43.05			 & 22, 15, 26 \\
 NGC 424	 & 	49.5	 & 	2	 & 		43.73	$\pm$	0.11	 & 		41.51	$\pm$	0.23	 & 		25.0	$\pm$	0.45	 & 		43.28	$\pm$	0.60	 & 		24.4	$\pm$	0.4	 & 		43.38	$\pm$	0.39	 & 		42.78	$\pm$	0.08	 & 28, 29, 30 \\
 NGC 676	 & 	19.5	 & 	2	 & 	$\le$	41.28			 & 		38.87	$\pm$	0.30	 & 	$\ge$	24.3			 & 		40.79	$\pm$	0.60	 & 					 & 	$\le$	41.42			 & 	$\le$	41.79			 & 31, 32, 33 \\
 NGC 1386	 & 	16.5	 & 	2	 & 		42.39	$\pm$	0.08	 & 		39.92	$\pm$	0.09	 & 	$\ge$	24.3			 & 		42.01	$\pm$	0.60	 & 		24.6	$\pm$	0.3	 & 		42.10	$\pm$	0.38	 & 	$\le$	41.64			 & 13, 22, 34 \\
 NGC 1667	 & 	67.8	 & 	2	 & 		41.90	$\pm$	0.16	 & 		40.69	$\pm$	0.30	 & 	$\ge$	24.3			 & 		42.61	$\pm$	0.60	 & 		23.7	$\pm$	0.5	 & 		41.63	$\pm$	0.40	 & 	$\le$	42.87			 & 22, 31, 35 \\
 NGC 3166	 & 	25.3	 & 	L	 & 		41.05	$\pm$	0.11	 & 					 & 					 & 					 & 					 & 		40.83	$\pm$	0.39	 & 	$\le$	42.01			 &  \\
 NGC 3281	 & 	52.8	 & 	2	 & 		43.62	$\pm$	0.04	 & 		41.93	$\pm$	0.30	 & 		24.3	$\pm$	0.45	 & 		43.42	$\pm$	0.60	 & 		24.1	$\pm$	0.4	 & 		43.27	$\pm$	0.37	 & 		43.46	$\pm$	0.02	 & 11 \\
 NGC 3312	 & 	48.4	 & 	L	 & 		41.74	$\pm$	0.07	 & 	$\le$	41.16			 & 					 & 					 & 	$\le$	23.6			 & 		41.48	$\pm$	0.38	 & 	$\le$	42.57			 & 36 \\
 NGC 3368	 & 	10.6	 & 	L	 & 	$\le$	40.70			 & 		39.31	$\pm$	0.60	 & 					 & 					 & 					 & 	$\le$	40.86			 & 	$\le$	41.26			 & 37, 38, 39 \\
 NGC 3623	 & 	12.8	 & 	L	 & 	$\le$	40.92			 & 		38.75	$\pm$	0.30	 & 					 & 					 & 					 & 	$\le$	41.07			 & 	$\le$	41.42			 & 40, 37 \\
 NGC 3982	 & 	21.4	 & 	2	 & 		41.56	$\pm$	0.06	 & 		39.65	$\pm$	0.31	 & 		23.3	$\pm$	0.45	 & 		40.35	$\pm$	0.87	 & 		24.2	$\pm$	0.5	 & 		41.31	$\pm$	0.38	 & 	$\le$	41.87			 & 31, 41, 22 \\
 NGC 4501	 & 	17.9	 & 	2	 & 		40.56	$\pm$	0.06	 & 		39.59	$\pm$	0.30	 & 		23.3	$\pm$	0.45	 & 		40.35	$\pm$	0.30	 & 		23.6	$\pm$	0.4	 & 		40.36	$\pm$	0.37	 & 	$\le$	41.71			 & 32, 42, 41 \\
 NGC 4636	 & 	15.6	 & 	L	 & 	$\le$	40.68			 & 		37.82	$\pm$	0.60	 & 	$\le$	21.0			 & 		37.82	$\pm$	0.60	 & 					 & 	$\le$	40.85			 & 	$\le$	41.59			 & 40, 43, 44 \\
 NGC 4698	 & 	24.4	 & 	2	 & 	$\le$	40.95			 & 		39.06	$\pm$	0.60	 & 	$\ge$	24.3			 & 		40.84	$\pm$	0.60	 & 					 & 	$\le$	41.10			 & 	$\le$	41.98			 & 40, 32, 45 \\
 NGC 4736	 & 	4.9	 & 	L	 & 		39.96	$\pm$	0.08	 & 		38.57	$\pm$	0.30	 & 		20.6	$\pm$	0.45	 & 		38.57	$\pm$	0.30	 & 		23.9	$\pm$	0.4	 & 		39.79	$\pm$	0.38	 & 	$\le$	40.58			 & 40 \\
 NGC 5005	 & 	16.9	 & 	L	 & 		40.78	$\pm$	0.12	 & 		39.83	$\pm$	0.30	 & 		21.0	$\pm$	0.45	 & 		39.88	$\pm$	0.30	 & 		23.6	$\pm$	0.5	 & 		40.57	$\pm$	0.39	 & 	$\le$	41.66			 & 40, 45, 46 \\
 NGC 5347	 & 	38.1	 & 	2	 & 		43.08	$\pm$	0.04	 & 		40.58	$\pm$	0.30	 & 	$\ge$	24.3			 & 		42.59	$\pm$	0.60	 & 		24.6	$\pm$	0.4	 & 		42.76	$\pm$	0.37	 & 	$\le$	42.37			 & 13 \\
 NGC 5363	 & 	21.0	 & 	L	 & 		40.39	$\pm$	0.23	 & 		40.00	$\pm$	0.30	 & 	$\ge$	24.3			 & 		41.50	$\pm$	0.60	 & 	$\le$	23.7			 & 		40.20	$\pm$	0.44	 & 	$\le$	41.85			 & 40, 45 \\
 NGC 5427	 & 	29.4	 & 	2	 & 	$\le$	41.71			 & 	$\le$	40.06			 & 					 & 					 & 					 & 	$\le$	41.83			 & 	$\le$	42.14			 & 15 \\
 NGC 5643	 & 	20.9	 & 	2	 & 		42.53	$\pm$	0.10	 & 		40.59	$\pm$	0.06	 & 	$\ge$	24.3			 & 					 & 		24.2	$\pm$	0.3	 & 		42.23	$\pm$	0.38	 & 	$\le$	41.97			 & 47, 48, 49 \\
 NGC 6890	 & 	33.8	 & 	1.9/2	 & 		42.60	$\pm$	0.09	 & 		40.59	$\pm$	0.30	 & 	$\ge$	24.3			 & 		42.38	$\pm$	0.60	 & 		24.3	$\pm$	0.4	 & 		42.30	$\pm$	0.38	 & 	$\le$	42.26			 & 22 \\
 NGC 7212	 & 	116.0	 & 	2	 & 		43.64	$\pm$	0.12	 & 		41.98	$\pm$	0.16	 & 	$\ge$	24.3			 & 		43.73	$\pm$	0.60	 & 		24.1	$\pm$	0.4	 & 		43.29	$\pm$	0.39	 & 		43.26	$\pm$	0.20	 & 13, 15, 50 \\
 NGC 7479	 & 	30.0	 & 	2	 & 		43.27	$\pm$	0.06	 & 		40.67	$\pm$	0.36	 & 		24.3	$\pm$	0.45	 & 		41.99	$\pm$	0.60	 & 		24.7	$\pm$	0.5	 & 		42.94	$\pm$	0.37	 & 		42.34	$\pm$	0.12	 & 33, 22, 31 \\
 NGC 7743	 & 	19.2	 & 	2/L	 & 	$\le$	41.00			 & 		39.29	$\pm$	0.22	 & 	$\ge$	24.3			 & 		41.26	$\pm$	0.60	 & 					 & 	$\le$	41.15			 & 	$\le$	41.77			 & 51, 31, 45 \\
 PKS 1932-46	 & 	1191.0	 & 	1.9	 & 	$\le$	43.94			 & 					 & 					 & 					 & 					 & 	$\le$	43.94			 & 	$\le$	45.36			 &  \\
 PKS 2158-380	 & 	149.0	 & 	2	 & 		43.11	$\pm$	0.07	 & 		41.77	$\pm$	0.60	 & 	$\ge$	22.0			 & 					 & 		23.8	$\pm$	0.7	 & 		42.78	$\pm$	0.38	 & 	$\le$	43.55			 & 7 \\
 PKS 2354-35	 & 	222.0	 & 	L	 & 	$\le$	42.47			 & 	$\le$	39.67			 & 					 & 					 & 					 & 	$\le$	42.55			 & 	$\le$	43.90			 & 52 \\
 Superantennae S	 & 	291.0	 & 	2	 & 		44.75	$\pm$	0.11	 & 		42.40	$\pm$	0.30	 & 		24.5	$\pm$	0.45	 & 		44.55	$\pm$	0.60	 & 		24.5	$\pm$	0.5	 & 		44.34	$\pm$	0.39	 & 	$\le$	44.13			 & 53, 54, 55 \\
 UGC 12348	 & 	110.0	 & 	2	 & 		43.55	$\pm$	0.06	 & 					 & 					 & 					 & 					 & 		43.20	$\pm$	0.38	 & 		43.08	$\pm$	0.19	 &  \\

\hline						                              
\end{tabular}
\begin{minipage}{1.0\textwidth}
 \normalsize
{\it -- Notes:} 
(1), (2), (3), and (4), short object name, distance, optical class, and nuclear 12$\um$ luminosities from \cite{asmus_subarcsecond_2014};
(5) average observed 2-10\,keV luminosity;
(6) average X-ray column density from the literature;
(7) average absorption-corrected 2-10\,keV luminosity from the literature;
(8) predicted X-ray column density using Eq.~\ref{eq:nh_diag};
(9) predicted absorption-corrected 2-10\,keV luminosity using Eq.~\ref{eq:inv};
(10) average observed 14-195\,keV luminosity from \swift/BAT by combining the data of the 54 and 70\,month source catalogues \citep{cusumano_palermo_2010,baumgartner_70_2013};
(11) references for the 2-10\,keV luminosities and column densities:
1: \cite{massaro_chandra_2010}; 2: \cite{evans_xmm-newton_2008}; 3: \cite{hardcastle_nature_2012}; 4: \cite{evans_chandra_2006}; 5: \cite{evans_radio_2008}; 6: \cite{hardcastle_chandra_2007}; 7: this work (XSPEC); 8: \cite{massaro_chandra_2012}; 9: \cite{balmaverde_chandra_2006}; 10: \cite{donato_obscuration_2004}; 11: \cite{gandhi_resolving_2009}; 12: \cite{piconcelli_x-ray_2011}; 13: \cite{levenson_penetrating_2006}; 14: \cite{miyazawa_broad-band_2009}; 15: \cite{guainazzi_x-ray_2005}; 16: \cite{de_rosa_x-ray_2008}; 17: \cite{pereira-santaella_x-ray_2011}; 18: \cite{rodriguez_swift_2008}; 19: \cite{nardini_compton-thick_2011}; 20: \cite{teng_chandra_2005}; 21: \cite{teng_xmm-newton_2008}; 22: \cite{brightman_xmm-newton_2011-1}; 23: \cite{teng_x-quest:_2010}; 24: \cite{teng_suzaku_2009}; 25: \cite{mazzarella_investigation_2012}; 26: \cite{shu_investigating_2007}; 27: \cite{paggi_cheers_2012}; 28: \cite{matt_chandra_2003}; 29: \cite{marinucci_x-ray_2011}; 30: \cite{balokovic_nustar_2014}; 31: \cite{panessa_x-ray_2006}; 32: \cite{cappi_x-ray_2006}; 33: \cite{akylas_xmm-newton_2009}; 34: \cite{maiolino_heavy_1998}; 35: \cite{marinucci_link_2012}; 36: \cite{hudaverdi_overdensity_2006}; 37: \cite{satyapal_joint_2004}; 38: \cite{liu_chandra_2011}; 39: \cite{ueda_asca_2001}; 40: \cite{gonzalez-martin_x-ray_2009}; 41: \cite{lamassa_uncovering_2011}; 42: \cite{brightman_nature_2008}; 43: \cite{flohic_central_2006}; 44: \cite{baldi_unusual_2009}; 45: \cite{gonzalez-martin_fitting_2009}; 46: \cite{younes_study_2011}; 47: \cite{bianchi_soft_2006}; 48: \cite{matt_x-ray_2013}; 49: \cite{guainazzi_unveiling_2004}; 50: \cite{awaki_x-ray_2000}; 51: \cite{kharb_examining_2012}; 52: \cite{russell_radiative_2013}; 53: \cite{braito_suzaku_2009}; 54: Teng et al. (in prep.); 55: \cite{braito_xmm-newton_2003};     
\end{minipage}

\end{table*}

\subsection{Contamination of AGN/starburst composites}\label{sec:CP}
The AGN MIR atlas contains 18 AGN/starburst composite  objects with verified AGN \citep{asmus_subarcsecond_2014}. 
They were excluded from the reliable sample because the intense nuclear and/or circumnuclear star formation possibly contaminates or even dominates their high angular resolution MIR photometry.
Indeed, often the nuclear MIR emission is extended on subarcsecond scales, making it difficult to separate AGN and starburst contributions.
Even worse, the star formation can significantly contribute also to the X-ray emission in the case of low-luminosity AGN, and/or obscure the AGN even if it is very powerful.
In fact, at least five of the 18 composite sources are CT obscured (NGC\,2623, NGC\,3690W, NGC\,4945, NGC\,6240S, and NGC\,7130).
Therefore, the composite objects are not well suited for determinations of the MIR--X-ray correlation.
However, we can use the MIR--X-ray correlation to possibly obtain a better understanding of the dominant power source in these objects. 
For this purpose, their X-ray properties are collected from the literature as for the other objects of the atlas (as described in Sect.~\ref{sec:sam}).
The corresponding properties and references are given in Table~\ref{tab:cp}.
For three sources (ESO\,420-13, Mrk\,897 and NGC\,7496), unfortunately, no X-ray observation has been published, while the nucleus of NGC\,5953 remained undetected in hard X-rays during \chandraa snapshot observations.
Therefore, we can predict only their $\lxi$ under the assumption that their subarcsecond MIR luminosities are AGN dominated (listed in Table~\ref{tab:cp}).
Furthermore, three AGN/starburst composites either suffer CT obscuration or have complex X-ray properties that do not allow for any intrinsic X-ray luminosity estimates (III\,Zw\,35, NGC\,7592W and UGC\,5101).
For these, the available observed X-ray luminosities predict CT obscuration using Equation~\ref{eq:nh_diag} and the above assumption of AGN MIR dominance. 

\begin{figure}
%    \centering
%    \sidecaption
   \includegraphics[angle=0,width=8.5cm]{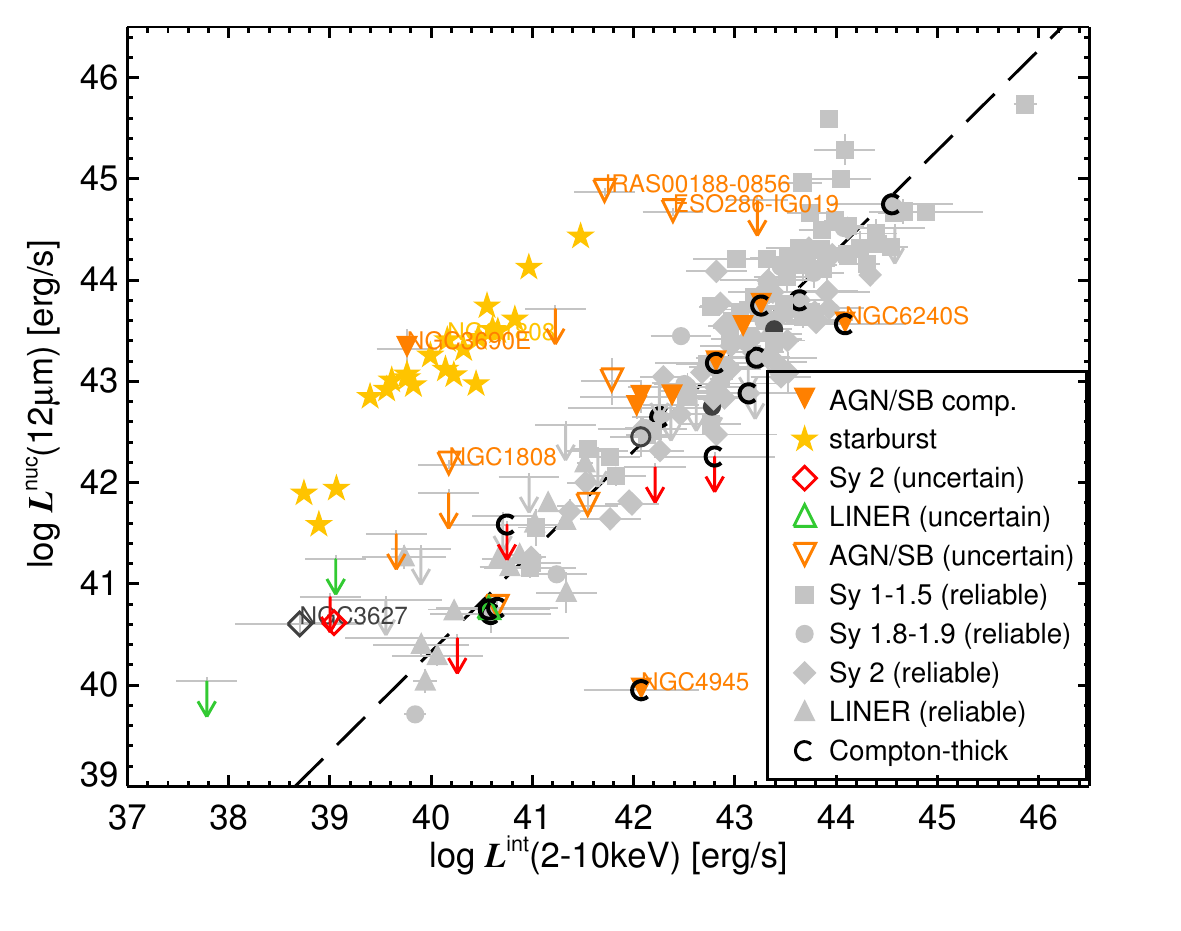}
    \caption{
             Distribution of AGN/SB composite objects and uncertain AGN in the observed $12\um$ to intrinsic 2-10\,keV plane.      
  Symbols are as in Fig.~\ref{fig:f12_f2int}.
             All reliable AGN are in light gray and the long dashed line is their \texttt{linmix\_err} fit. 
             AGN/starburst composites are shown as orange triangles and the starburst galaxies from Ranalli, Comastri \& Setti (2003) as yellow stars.
            Objects with uncertain AGN are shown as open symbols.
            }
   \label{fig:all}
\end{figure}

The remaining 11 AGN/starburst composites are plotted in the $12\um$--2-10\,keV plane next to the reliable AGN in Fig.~\ref{fig:all} using their unresolved nuclear MIR emission.
For comparison, the pure starburst galaxies from \cite{ranalli_2-10_2003} are plotted as well.
For those the total MIR and X-ray luminosities are used in the absence of nuclear compact emission (see also \citealt{asmus_mid-infrared_2011}).
These sources follow their own MIR--X-ray correlation with a similar slope but an MIR--X-ray ratio of $\sim 3$\,dex, which is $\sim 2.7$\,dex higher than that of the reliable AGN.
Therefore, we expect composite sources to also exhibit high MIR--X-ray ratios, lying close to the pure starbursts in the MIR--X-ray plane.
However, most of the verified composite sources, 9 out of 11, are surprisingly close to the correlation of the reliable AGN, indicating that their nuclear MIR emission is not dominated by star formation.
One exception is NGC\,3690E with a very high MIR--X-ray ratio similar to a pure starburst. 
However, this sources is a good candidate for Compton-thick obscuration (see Appendix~\ref{app:obj}), which would mean an order of magnitude higher intrinsic X-ray luminosity and thus lower MIR--X-ray ratio. 
Unfortunately, the available data in the MIR and X-ray are not sufficient to distinguish whether the nuclear MIR emission is star formation contaminated or the nucleus is Compton-thick obscured in X-rays.
At least, deep silicate absorption features in the lower resolution \spitzerr MIR spectrum of NGC\,3690E indicates indeed heavy absorption but also strong PAH emission (typical for star formation).

The other extreme outlier is NGC\,4945, which exhibits instead an extremely low MIR--X-ray ratio ($\ratmx = -2.54$), i.e. being either MIR under-luminous or X-ray over-luminous by almost three orders of magnitude (see also discussion in \citealt{gandhi_compton-thick_2015}).
As discussed in \cite{asmus_subarcsecond_2014}, this very nearby object; $D\sim 4$\,Mpc) shows very extended nuclear MIR emission dominated by star formation with a relatively faint core.
This makes the estimation of the nuclear AGN-related MIR emission very difficult.
However, this complication is not sufficient to alone explain the low MIR--X-ray ratio.

For example, the object distance could also contribute.
Artificially moving NGC\,4945 to the median distance of the AGN MIR atlas (72\,Mpc), all emission within $\sim 7\arcsec$ would be unresolved. 
In that case the $\fnn$ would be between $\sim 400$ to 2000\,mJy (the fluxes in the \textit{Spitzer}/IRS and \textit{ISO}/PHT-S spectra; \citealt{spoon_mid-infrared_2000,perez-beaupuits_deeply_2011}), and thus 20-100 times higher (1.3-2\,dex).
Thus, a possible intrinsic resolution effect can also not explain the MIR--X-ray ratio by itself.
This is, in fact, unlikely because firstly it would mean that the MIR emission of AGN in general is dominated by unresolved circumnuclear star formation contradicting, e.g., \cite{asmus_mid-infrared_2011} and \cite{asmus_subarcsecond_2014}, and secondly because other comparably close and powerful objects like Circinus show even higher than average MIR--X-ray ratios ($\ratmx =0.39$).

On the other hand, NGC\,4945 is Compton-thick obscured in X-rays exhibiting a highly variable and complex spectrum (e.g.,  \citealt{yaqoob_nature_2012,marinucci_x-ray_2012}). 
In fact, the intrinsic 2-10\,keV luminosity estimates span two orders of magnitude \citep{yaqoob_nature_2012}.
However, the most recent NuSTAR observations are consistent with the average luminosity adopted here \citep{puccetti_variable_2014}.
Despite this, the source is possibly beamed, in which case the isotropic emission estimates would be an order of magnitude too high \citep{yaqoob_nature_2012}.
The X-ray spectral analysis also indicates that the CT obscurer has a small filling factor of only 0.1-0.15 and thus is relatively compact (e.g., \citealt{yaqoob_nature_2012, puccetti_variable_2014}).
In that case, the MIR emission of that obscurer would also be significantly lower compared to typical obscurers with filling factors of $\sim0.5$. 

Finally, data at multiple wavelengths provide ample evidence for unusually high foreground extinction towards the AGN in NGC\,4945 even covering the narrow-line region  \citep{moorwood_starburst_1996,spoon_mid-infrared_2000,perez-beaupuits_deeply_2011}.
This is, for example, evident from the extremely deep silicate absorption feature at $9.7\um$, implying an attenuation factor of $\sim25$ (1.4\,dex) at $12\um$ with large uncertainties of the order of 1\,dex \citep{perez-beaupuits_deeply_2011}.
A similar correction factor based on the ratio of the 6.2 to 7.7$\um$ PAH features was already used by \cite{krabbe_n-band_2001} to correct the MIR emission of NGC\,4945.

In conclusion, none of the effects described above can explain by themselves the extraordinary low MIR--X-ray of NGC\,4945. 
Because of this high extinction also towards the narrow line region,  corresponding intrinsic power estimates based on the emission lines \oiii, \oiv, or \nev can not be used to constrain the intrinsic MIR and X-ray luminosities either.
It is beyond the scope of this work to further investigate the obviously complicated situation of this source.
Note however, that interestingly, the ratio of the observed MIR to observed 2-10\,keV ratio is consistent with the MIR--X-ray correlation. 

Finally, this discussion suggests that other objects could also be affected by high foreground extinction which would be indicated by an extremely deep silicate $9.7\um$ absorption feature. 
Indeed, the average MIR--X-ray ratio of sources with deep silicate features is lower than those of the type~II and X-ray obscured AGN ($\aratmx=0.15, \sigmx=0.35$).
However, of the three with the deepest silicate absorption in the reliable sample (ESO\,506-27, NGC\,4992, and NGC\,7172), ESO\,506-27 exhibits a high MIR--X-ray ratio ($\ratmx = 0.52$).
Furthermore, only NGC\,5728 of the four objects with $\ratmx < -0.3$  exhibits deep silicate $9.7\um$ absorption (NGC\,1144, NGC\,3169, ESO\,297-18 and NGC\,5728; \citealt{asmus_subarcsecond_2014}).
Because of this ambiguity and the large difficulties to derive reliable attenuation factors, we refrain from applying an MIR absorption correction to the objects in this work.
The numbers above demonstrate that such a correction would not alter the overall results derived here significantly.

\begin{table*}
\caption{Properties of confirmed AGN/starburst composites.} % title of Table
\label{tab:cp} % is used to refer this table in the text
\centering % used for centering table
\scriptsize
\begin{tabular}{l	c	c	c	c	c	c	c	c	c	l}

\hline\hline																										
 &	 &	 &	 &	 &	 Lit.	&	Lit.	&	Pred.	& Pred.	&	&	\\
	&		&	Opt.	&		$\log L^\textrm{nuc}$	&	$\log L^\textrm{obs}$	&		$\log$	&	$\log L^\textrm{int}$	&		$\log$			&	$\log L^\textrm{int}$			&		$\log L^\textrm{obs}$	&		\\
Object	&	$D$	&	class	&		($12\um$)	&	(2-10\,keV)	&		$ \nh$	&	(2-10\,keV)	&		$ \nh$			&	(2-10\,keV)			&		(14-195\,keV)	&	2-10\,keV	\\
	&	[Mpc]	&		&		 [erg/s]	&	 [erg/s]	&		 [cm$^{-2}$]	&	 [erg/s]	&		 [cm$^{-2}$]			&	 [erg/s]			&		 [erg/s]	&	Ref.	\\
(1)	&	(2)	&	(3)	&		(4)	&	(5)	&		(6)	&	(7)	&		(8)			&	(9)			&		(10)	&	(11)	\\

\hline																									
 3C 459	 & 	1125.0	 & 	Cp	 & 	$\le$	44.79			 & 		43.10	$\pm$	0.30	 & 		22.8	$\pm$	0.45	 & 		43.22	$\pm$	0.30	 & 					 & 	$\le$	44.75			 & 	$\le$	45.31			 & 1, 2 \\
 ESO 420-13	 & 	52.7	 & 	Cp	 & 		43.20	$\pm$	0.08	 & 					 & 					 & 					 & 					 & 		42.87	$\pm$	0.38	 & 	$\le$	42.65			 &  \\
 III Zw 35N	 & 	121.0	 & 	Cp	 & 		43.36	$\pm$	0.17	 & 		39.98	$\pm$	0.30	 & 					 & 					 & 		25.2	$\pm$	0.5	 & 		43.02	$\pm$	0.41	 & 	$\le$	43.37			 & 3, 4 \\
 Mrk 520	 & 	115.0	 & 	Cp	 & 		43.53	$\pm$	0.11	 & 		43.00	$\pm$	0.30	 & 		22.4	$\pm$	0.45	 & 		43.08	$\pm$	0.30	 & 	$\le$	23.7			 & 		43.18	$\pm$	0.39	 & 		43.68	$\pm$	0.07	 & 5 \\
 Mrk 897	 & 	115.0	 & 	Cp	 & 		42.51	$\pm$	0.12	 & 					 & 					 & 					 & 					 & 		42.22	$\pm$	0.39	 & 	$\le$	43.33			 &  \\
 NGC 2623	 & 	87.3	 & 	Cp	 & 		43.61	$\pm$	0.20	 & 		40.96	$\pm$	0.30	 & 	$\ge$	24.3			 & 					 & 		24.7	$\pm$	0.5	 & 		43.26	$\pm$	0.42	 & 	$\le$	43.09			 & 6, 7 \\
 NGC 3690E	 & 	49.1	 & 	Cp	 & 		43.32	$\pm$	0.19	 & 		39.72	$\pm$	0.30	 & 		22.1	$\pm$	0.45	 & 		39.76	$\pm$	0.30	 & 		25.4	$\pm$	0.5	 & 		42.99	$\pm$	0.42	 & 	$\le$	42.59			 & 8, 9 \\
 NGC 3690W	 & 	48.2	 & 	Cp	 & 		43.75	$\pm$	0.21	 & 		41.49	$\pm$	0.30	 & 		24.6	$\pm$	0.45	 & 		43.26	$\pm$	0.60	 & 		24.5	$\pm$	0.5	 & 		43.39	$\pm$	0.43	 & 	$\le$	42.57			 & 10 \\
 NGC 4945	 & 	3.7	 & 	Cp	 & 		39.95	$\pm$	0.12	 & 		39.85	$\pm$		 & 	$\ge$	24.3			 & 		42.07	$\pm$	0.57	 & 	$\le$	23.3			 & 		39.78	$\pm$	0.39	 & 		41.66	$\pm$	0.01	 & 11, 12 \\
 NGC 5953	 & 	31.4	 & 	Cp	 & 	$\le$	41.94			 & 	$\le$	38.85			 & 					 & 					 & 					 & 	$\le$	42.04			 & 	$\le$	42.20			 & 13 \\
 NGC 6240N	 & 	114.0	 & 	Cp	 & 		42.84	$\pm$	0.15	 & 		42.07	$\pm$	0.30	 & 				0.45	 & 		42.07	$\pm$	0.60	 & 		23.4	$\pm$	0.5	 & 		42.53	$\pm$	0.40	 & 	$\le$	44.00			 & 14 \\
 NGC 6240S	 & 	113.0	 & 	Cp	 & 		43.56	$\pm$	0.05	 & 		42.33	$\pm$	0.30	 & 	$\ge$	24.3			 & 		44.09	$\pm$	0.60	 & 		23.8	$\pm$	0.4	 & 		43.22	$\pm$	0.37	 & 		44.04	$\pm$	0.04	 & 15, 4, 16 \\
 NGC 7130	 & 	68.9	 & 	Cp	 & 		43.18	$\pm$	0.08	 & 		40.84	$\pm$	0.30	 & 	$\ge$	24.3			 & 		42.81	$\pm$	0.60	 & 		24.5	$\pm$	0.4	 & 		42.85	$\pm$	0.38	 & 		42.96	$\pm$	0.18	 & 4, 17 \\
 NGC 7496	 & 	21.1	 & 	Cp	 & 		42.35	$\pm$	0.03	 & 					 & 					 & 					 & 					 & 		42.07	$\pm$	0.37	 & 	$\le$	41.85			 &  \\
 NGC 7582	 & 	23.0	 & 	Cp	 & 		42.85	$\pm$	0.07	 & 		41.42	$\pm$	0.27	 & 		23.1	$\pm$	0.25	 & 		42.38	$\pm$	0.25	 & 		23.9	$\pm$	0.4	 & 		42.53	$\pm$	0.38	 & 		42.71	$\pm$	0.02	 & 18, 16, 19 \\
 NGC 7592W	 & 	105.0	 & 	Cp	 & 		43.65	$\pm$	0.14	 & 		40.72	$\pm$	0.30	 & 					 & 					 & 		24.9	$\pm$	0.5	 & 		43.30	$\pm$	0.39	 & 	$\le$	43.25			 & 20 \\
 NGC 7679	 & 	71.7	 & 	Cp	 & 		42.74	$\pm$	0.11	 & 		42.04	$\pm$	0.70	 & 		20.3	$\pm$	0.45	 & 		42.03	$\pm$	0.67	 & 		23.4	$\pm$	0.8	 & 		42.43	$\pm$	0.39	 & 		42.97	$\pm$	0.15	 & 21, 22 \\
 UGC 5101	 & 	182.0	 & 	Cp	 & 		44.35	$\pm$	0.07	 & 		41.68	$\pm$	0.30	 & 					 & 					 & 		24.7	$\pm$	0.4	 & 		43.96	$\pm$	0.38	 & 		43.49	$\pm$	0.21	 & 16, 4 \\

\hline						                              
\end{tabular}
\begin{minipage}{1.0\textwidth}
 \normalsize
{\it -- Notes:} 
(1), (2), (3), and (4), short object name, distance, optical class, and nuclear 12$\um$ luminosities from \cite{asmus_subarcsecond_2014};
(5) average observed 2-10\,keV luminosity;
(6) average X-ray column density from the literature;
(7) average absorption-corrected 2-10\,keV luminosity from the literature;
(8) predicted X-ray column density using Eq.~\ref{eq:nh_diag};
(9) predicted absorption-corrected 2-10\,keV luminosity using Eq.~\ref{eq:inv};
(10) average observed 14-195\,keV luminosity from \swift/BAT by combining the data of the 54 and 70\,month source catalogues \citep{cusumano_palermo_2010,baumgartner_70_2013};
(11) references for the 2-10\,keV luminosities and column densities:
1: \cite{massaro_chandra_2012}; 2: \cite{mingo_x-ray_2014}; 3: \cite{gonzalez-martin_x-ray_2009}; 4: \cite{gonzalez-martin_fitting_2009}; 5: \cite{winter_x-ray_2009}; 6: \cite{evans_off-nuclear_2008}; 7: \cite{maiolino_elusive_2003}; 8: \cite{zezas_chandra_2003}; 9: \cite{ballo_arp_2004}; 10: \cite{ptak_focused_2015}; 11: \cite{yaqoob_nature_2012}; 12: \cite{puccetti_variable_2014}; 13: \cite{guainazzi_x-ray_2005}; 14: \cite{komossa_discovery_2003}; 15: Puccetti et al. (in prep); 16: \cite{brightman_xmm-newton_2011-1}; 17: \cite{levenson_deconstructing_2005}; 18: \cite{bianchi_how_2009}; 19: \cite{dong_chandra_2004}; 20: \cite{wang_chandra_2010}; 21: \cite{della_ceca_unveiling_2001}; 22: \cite{yankulova_luminous_2007}.
\end{minipage}

\end{table*}

\subsection{Classification of uncertain AGN}\label{sec:can}
The MIR Atlas of \cite{asmus_subarcsecond_2014} also includes 38 uncertain AGN, galactic nuclei with insufficient or contradicting evidence for an accreting supermassive black hole being the main source of emission from those galactic nuclei.
Most of these are nearby sources with low-luminosities. 
We have searched the literature for X-ray measurements of the objects with the same procedure as in Sect.~\ref{sec:sam}.
The properties of all uncertain AGN relevant here  are summarized in Table~\ref{tab:unc}, including predicted $\lxi$ and $\nh$ from Equations~\ref{eq:inv} and \ref{eq:nh_diag} (whenever possible).

Eleven nuclei do not have any published X-ray detection with sufficient counts for estimating even the observed flux in the 2-10\,keV range (ESO\,500-34, ESO\,602-25, IRAS08572+3915, NGC\,1433, NGC\,3094, NGC\,3607, NGC\,3628, NGC\,4746, NGC\,5258, NGC\,6221 and NGC\,7590).
Therefore, $\lxi$ can only  be predicted, or constrained in case of MIR non-detections, for those objects. 
Note that two of X-ray non-detections, ESO\,500-34 and NGC\,6221, have actually been detected with \swift/BAT at 14-195\,keV.
Their predicted $\lxi$ values agree well with the expectation from the 2-10\,keV--14-195\,keV correlation (Sect.~\ref{sec:altcor}).

Another seven nuclei have no or contradicting intrinsic luminosity estimates (IC\,883, IRAS\,11095-0238, IRAS\,15250+3609, NGC\,3521, NGC\,4418, NGC\,5866 and NGC\,6810).
Three of these (IC\,883, NGC\,3521 and NGC\,5866) are undetected in the MIR at subarcsecond scales, and again only $\lxi$ can be predicted.
For the others also $\nh$ predictions are possible and indicate Compton-thick obscuration for the nuclei in all four cases. 
However, for the three AGN/starburst composite candidates among them ( IRAS\,11095-0238, IRAS\,15250+3609 and NGC\,6810) again our diagnostics can not distinguish between CT obscured and star formation dominated nuclei.

The remaining 20 uncertain AGN have estimates for $\lxi$ directly from the X-ray data and are plotted as well in Fig.~\ref{fig:all}.
Ten of these have only upper limits on their nuclear MIR luminosities, all of which are consistent with the MIR--X-ray correlation and thus with being real AGN.
However, we note that 3C\,264  and NGC\,4785 exhibit quite low MIR--X-ray ratios ($\ratmx < -0.15$), indicating that these sources are possibly not AGN powered.
On the other hand, for NGC\,4785 firm evidence for the presence of a CT AGN is presented by \cite{gandhi_compton-thick_2015} and used here for $\lxi$. 
Thus, it is possible that NGC\,4795 is also extincted in the MIR similar to NGC\,4945.
Note that this scenario is very unlikely for 3C\,264 as there is no evidence for significant obscuration (Appendix~\ref{app:obj}).
From the other ten objects with both nuclear MIR detections and intrinsic X-ray luminosity estimates, six exhibit very high MIR--X-ray ratios ($\ratmx > 1.2$). 
Among those are the AGN/starburst composites candidates IRAS\,00188-0856, ESO\,286-19, Mrk\,266NE and NGC\,1808.
The former three are possibly CT obscured, while the latter, NGC\,1808 is unlikely to host a heavily obscured AGN.
Thus, we conclude that NGC\,1808 is star formation dominated even on subarcsecond scales. 
The other two sources with high MIR--X-ray ratios are NGC\,3627 and NGC\,4303.
Both do not show signs of heavy obscuration in the X-rays, and therefore we can conclude that their nuclear activity is not AGN-dominated. 
This finally leaves four uncertain AGN which exhibit nuclear MIR--X-ray ratios typical of AGN (NGC\,613, NGC\,3660, NGC\,4438 and NGC\,4457).
NGC\,613 and NGC\,3660 are indeed good candidates for AGN-dominated nuclei, while for NGC\,4438 and NGC\,4457 large uncertainties remain owing to their possibly CT nature (see Appendix~\ref{app:obj}).

In summary, we note that the ambiguity in the MIR--X-ray ratio between a star formation dominated nucleus and CT AGN dominated nucleus severely constrain the diagnostic power of the MIR--X-ray correlation for many objects. 
Despite this, we manage to identify four nuclei (3C\,264, NGC\,1808, NGC\,3627 and NGC\,4303) as presumably not AGN powered and confirm AGN in another two (NGC\,613 and NGC\,3660).   

\onecolumn
\begin{table*}
\caption{Prperties of uncertain AGN.} % title of Table
\label{tab:unc} % is used to refer this table in the text
\centering % used for centering table
\scriptsize
\begin{tabular}{l	c	c	c	c	c	c	c	c	c	l}

\hline\hline																										
 &	 &	 &	 &	 &	 Lit.	&	Lit.	&	Pred.	& Pred.	&	&	\\
	&		&	Opt.	&		$\log L^\textrm{nuc}$	&	$\log L^\textrm{obs}$	&		$\log$	&	$\log L^\textrm{int}$	&		$\log$			&	$\log L^\textrm{int}$			&		$\log L^\textrm{obs}$	&		\\
Object	&	$D$	&	class	&		($12\um$)	&	(2-10\,keV)	&		$ \nh$	&	(2-10\,keV)	&		$ \nh$			&	(2-10\,keV)			&		(14-195\,keV)	&	2-10\,keV	\\
	&	[Mpc]	&		&		 [erg/s]	&	 [erg/s]	&		 [cm$^{-2}$]	&	 [erg/s]	&		 [cm$^{-2}$]			&	 [erg/s]			&		 [erg/s]	&	Ref.	\\
(1)	&	(2)	&	(3)	&		(4)	&	(5)	&		(6)	&	(7)	&		(8)			&	(9)			&		(10)	&	(11)	\\

\hline																									
 3C 264	 & 	103.0	 & 	2:	 & 	$\le$	42.15			 & 		42.21	$\pm$	0.31	 & 	$\le$	22.0			 & 		42.21	$\pm$	0.31	 & 					 & 	$\le$	42.25			 & 	$\le$	43.23			 & 1, 2 \\
 ESO 286-19	 & 	195.0	 & 	Cp:	 & 		44.67	$\pm$	0.04	 & 		41.91	$\pm$	0.30	 & 		23.7	$\pm$	0.45	 & 		42.39	$\pm$	0.30	 & 		24.8	$\pm$	0.4	 & 		44.27	$\pm$	0.37	 & 	$\le$	43.79			 & 3, 4, 5 \\
 ESO 500-34	 & 	60.2	 & 	Cp	 & 	$\le$	42.70			 & 					 & 					 & 			$\pm$		 & 					 & 	$\le$	42.77			 & 		42.72	$\pm$	0.20	 &  \\
 ESO 602-25	 & 	109.0	 & 	Cp:	 & 		43.08	$\pm$	0.15	 & 					 & 					 & 			$\pm$		 & 					 & 		42.75	$\pm$	0.40	 & 	$\le$	43.28			 &  \\
 IC 883	 & 	109.0	 & 	Cp	 & 	$\le$	43.90			 & 		40.97	$\pm$	0.30	 & 					 & 			$\pm$		 & 					 & 	$\le$	43.90			 & 	$\le$	43.28			 & 6 \\
 IRAS 00188-0856	 & 	620.0	 & 	Cp	 & 		44.87	$\pm$	0.04	 & 		41.87	$\pm$	0.30	 & 	$\le$	21.0			 & 		41.71	$\pm$	0.30	 & 		25.0	$\pm$	0.4	 & 		44.46	$\pm$	0.37	 & 	$\le$	44.79			 & 7, 8 \\
 IRAS 08572+3915	 & 	275.0	 & 	Cp	 & 		45.13	$\pm$	0.09	 & 	$\le$	41.36			 & 					 & 			$\pm$		 & 		25.5	$\pm$	0.3	 & 		44.71	$\pm$	0.38	 & 	$\le$	44.08			 & 9, 10 \\
 IRAS 11095-0238	 & 	519.0	 & 	Cp:	 & 		44.83	$\pm$	0.13	 & 		41.41	$\pm$		 & 					 & 			$\pm$		 & 		25.2	$\pm$	0.3	 & 		44.42	$\pm$	0.39	 & 	$\le$	44.64			 & 8 \\
 IRAS 15250+3609	 & 	258.0	 & 	Cp:	 & 		44.64	$\pm$	0.04	 & 		40.95	$\pm$	0.36	 & 					 & 			$\pm$		 & 		25.4	$\pm$	0.5	 & 		44.23	$\pm$	0.37	 & 	$\le$	44.03			 & 7, 8 \\
 Mrk 266NE	 & 	130.0	 & 	Cp	 & 		43.00	$\pm$	0.23	 & 		41.24	$\pm$	0.56	 & 		22.9	$\pm$	0.04	 & 		41.78	$\pm$	0.30	 & 		24.1	$\pm$	0.7	 & 		42.68	$\pm$	0.44	 & 	$\le$	43.43			 & 11, 12, 13 \\
 NGC 1433	 & 	8.3	 & 	2:	 & 	$\le$	40.20			 & 					 & 					 & 			$\pm$		 & 					 & 	$\le$	40.39			 & 	$\le$	41.05			 &  \\
 NGC 1614	 & 	71.0	 & 	Cp:	 & 	$\le$	43.72			 & 		41.20	$\pm$	0.30	 & 		21.3	$\pm$	0.45	 & 		41.23	$\pm$	0.30	 & 					 & 	$\le$	43.73			 & 	$\le$	42.91			 & 3 \\
 NGC 1808	 & 	12.3	 & 	Cp:	 & 		42.18	$\pm$	0.04	 & 		40.13	$\pm$	0.30	 & 		22.1	$\pm$	1.16	 & 		40.17	$\pm$	0.30	 & 		24.3	$\pm$	0.4	 & 		41.90	$\pm$	0.37	 & 	$\le$	41.38			 & 14, 3 \\
 NGC 253	 & 	3.2	 & 	Cp:	 & 	$\le$	41.49			 & 		38.80	$\pm$	0.51	 & 		23.5	$\pm$	0.30	 & 		39.66	$\pm$	0.30	 & 					 & 	$\le$	41.62			 & 	$\le$	40.21			 & 15, 16, 17 \\
 NGC 3094	 & 	40.8	 & 	2:	 & 		43.71	$\pm$	0.04	 & 					 & 					 & 			$\pm$		 & 					 & 		43.35	$\pm$	0.37	 & 	$\le$	42.43			 &  \\
 NGC 3185	 & 	20.3	 & 	2:	 & 	$\le$	41.59			 & 		38.99	$\pm$	0.30	 & 	$\ge$	24.3			 & 		40.75	$\pm$	0.60	 & 					 & 	$\le$	41.71			 & 	$\le$	41.82			 & 18 \\
 NGC 3379	 & 	10.6	 & 	L	 & 	$\le$	40.04			 & 		37.80	$\pm$	0.30	 & 	$\le$	22.0			 & 		37.78	$\pm$	0.30	 & 					 & 	$\le$	40.24			 & 	$\le$	41.26			 & 11, 12, 19 \\
 NGC 3486	 & 	13.7	 & 	2/L	 & 	$\le$	40.46			 & 		39.36	$\pm$	0.30	 & 	$\le$	21.5			 & 		40.26	$\pm$	1.11	 & 		40.3	$\pm$	0.6	 & 	$\le$	40.64			 & 	$\le$	41.48			 & 18, 20, 3 \\
 NGC 3521	 & 	11.5	 & 	L/H	 & 	$\le$	40.60			 & 		38.67	$\pm$	0.30	 & 					 & 			$\pm$		 & 					 & 	$\le$	40.77			 & 	$\le$	41.33			 & 21 \\
 NGC 3607	 & 	21.4	 & 	2/L	 & 	$\le$	41.28			 & 	$\le$	37.76			 & 					 & 			$\pm$		 & 					 & 	$\le$	41.41			 & 	$\le$	41.87			 & 11, 12, 19 \\
 NGC 3627	 & 	10.1	 & 	2/L	 & 		40.60	$\pm$	0.11	 & 		38.72	$\pm$	0.62	 & 		20.7	$\pm$	0.45	 & 		38.70	$\pm$	0.64	 & 		24.2	$\pm$	0.7	 & 		40.40	$\pm$	0.39	 & 	$\le$	41.21			 & 18, 22 \\
 NGC 3628	 & 	12.2	 & 	L/H	 & 	$\le$	40.66			 & 	$\le$	37.10		0.30	 & 					 & 			$\pm$		 & 					 & 	$\le$	40.83			 & 	$\le$	41.37			 & 19 \\
 NGC 3660	 & 	60.8	 & 	1.8/2/L/H	 & 		42.45	$\pm$	0.20	 & 		42.07	$\pm$	0.30	 & 		20.5	$\pm$	0.45	 & 		42.07	$\pm$	0.30	 & 	$\le$	23.7			 & 		42.16	$\pm$	0.42	 & 	$\le$	42.77			 & 20, 23 \\
 NGC 4303	 & 	15.2	 & 	2	 & 		40.62	$\pm$	0.05	 & 		39.04	$\pm$	0.30	 & 	$\le$	22.0			 & 		39.04	$\pm$	0.30	 & 		24.0	$\pm$	0.4	 & 		40.42	$\pm$	0.37	 & 	$\le$	41.57			 & 24, 25 \\
 NGC 4418	 & 	37.8	 & 	2	 & 		43.79	$\pm$	0.05	 & 		39.47	$\pm$	0.30	 & 	$\ge$	24.0			 & 			$\pm$		 & 		25.9	$\pm$	0.4	 & 		43.43	$\pm$	0.37	 & 	$\le$	42.36			 & 26 \\
 NGC 4438	 & 	13.7	 & 	L/H	 & 		40.76	$\pm$	0.11	 & 		39.02	$\pm$		 & 	$\ge$	24.3			 & 		40.65	$\pm$	0.60	 & 		24.1	$\pm$	0.3	 & 		40.55	$\pm$	0.38	 & 	$\le$	41.48			 & 27, 12, 19 \\
 NGC 4457	 & 	17.4	 & 	L	 & 		40.75	$\pm$	0.10	 & 		38.89	$\pm$	0.30	 & 	$\ge$	24.3			 & 		40.57	$\pm$	0.60	 & 		24.2	$\pm$	0.5	 & 		40.54	$\pm$	0.38	 & 	$\le$	41.69			 & 11, 12, 19 \\
 NGC 4472	 & 	17.1	 & 	2/L	 & 	$\le$	40.87			 & 		39.00	$\pm$	0.30	 & 					 & 		39.00	$\pm$	0.30	 & 					 & 	$\le$	41.03			 & 	$\le$	41.67			 & 28, 18 \\
 NGC 4746	 & 	33.5	 & 	L/H	 & 	$\le$	41.56			 & 					 & 					 & 			$\pm$		 & 					 & 	$\le$	41.69			 & 	$\le$	42.26			 &  \\
 NGC 4785	 & 	59.0	 & 	2	 & 	$\le$	42.26			 & 		41.00	$\pm$	0.30	 & 	$\ge$	24.3			 & 		42.80	$\pm$	0.60	 & 					 & 	$\le$	42.35			 & 	$\le$	42.75			 & 29 \\
 NGC 5258	 & 	107.0	 & 	L/H	 & 	$\le$	42.82			 & 					 & 					 & 			$\pm$		 & 					 & 	$\le$	42.88			 & 	$\le$	43.26			 &  \\
 NGC 5813	 & 	29.7	 & 	L:	 & 	$\le$	41.25			 & 		39.06	$\pm$	0.30	 & 		21.1	$\pm$	0.45	 & 		39.06	$\pm$	0.30	 & 					 & 	$\le$	41.38			 & 	$\le$	42.15			 & 30, 12 \\
 NGC 5866	 & 	14.1	 & 	L/H	 & 	$\le$	40.72			 & 		38.32	$\pm$	0.14	 & 					 & 			$\pm$		 & 					 & 	$\le$	40.88			 & 	$\le$	41.50			 & 19, 11, 12 \\
 NGC 613	 & 	26.6	 & 	Cp	 & 		41.77	$\pm$	0.11	 & 		40.96	$\pm$	0.30	 & 		23.6	$\pm$	0.45	 & 		41.55	$\pm$	0.30	 & 		23.5	$\pm$	0.5	 & 		41.51	$\pm$	0.39	 & 	$\le$	42.05			 & 31 \\
 NGC 6221	 & 	10.7	 & 	Cp	 & 		41.55	$\pm$	0.08	 & 					 & 					 & 			$\pm$		 & 					 & 		41.30	$\pm$	0.38	 & 		41.44	$\pm$	0.10	 &  \\
 NGC 6810	 & 	28.6	 & 	Cp	 & 		42.04	$\pm$	0.11	 & 		39.89	$\pm$		 & 					 & 			$\pm$		 & 		24.4	$\pm$	0.3	 & 		41.76	$\pm$	0.39	 & 	$\le$	42.12			 & 3 \\
 NGC 7552	 & 	20.4	 & 	L/H	 & 	$\le$	41.90			 & 		40.13	$\pm$	0.04	 & 	$\le$	21.0			 & 		40.17	$\pm$	0.30	 & 					 & 	$\le$	42.00			 & 	$\le$	41.82			 & 3, 21 \\
 NGC 7590	 & 	26.5	 & 	2:	 & 	$\le$	41.35			 & 	$\le$	38.71		0.30	 & 	$\ge$	24.3			 & 			$\pm$		 & 					 & 	$\le$	41.48			 & 	$\le$	42.05			 & 32, 33, 34 \\

\hline						                              
\end{tabular}
\begin{minipage}{1.0\textwidth}
 \normalsize
{\it -- Notes:} 
(1), (2), (3), and (4), short object name, distance, optical class, and nuclear 12$\um$ luminosities from \cite{asmus_subarcsecond_2014};
(5) average observed 2-10\,keV luminosity;
(6) average X-ray column density from the literature;
(7) average absorption-corrected 2-10\,keV luminosity from the literature;
(8) predicted X-ray column density using Eq.~\ref{eq:nh_diag};
(9) predicted absorption-corrected 2-10\,keV luminosity using Eq.~\ref{eq:inv};
(10) average observed 14-195\,keV luminosity from \swift/BAT by combining the data of the 54 and 70\,month source catalogues \citep{cusumano_palermo_2010,baumgartner_70_2013};
(11) references for the 2-10\,keV luminosities and column densities:
1: \cite{evans_chandra_2006}; 2: \cite{donato_obscuration_2004}; 3: \cite{brightman_xmm-newton_2011-1}; 4: \cite{franceschini_xmm-newton_2003}; 5: \cite{misaki_x-ray_1999}; 6: \cite{modica_multi-wavelength_2012}; 7: \cite{teng_chandra_2005}; 8: \cite{teng_x-quest:_2010}; 9: \cite{teng_suzaku_2009}; 10: \cite{lamassa_uncovering_2011}; 11: \cite{gonzalez-martin_x-ray_2009}; 12: \cite{gonzalez-martin_fitting_2009}; 13: \cite{mazzarella_investigation_2012}; 14: \cite{jimenez-bailon_x-ray_2005}; 15: \cite{lehmer_nustar_2013}; 16: \cite{weaver_chandra_2002}; 17: \cite{muller-sanchez_stellar_2010}; 18: \cite{panessa_x-ray_2006}; 19: \cite{flohic_central_2006}; 20: \cite{brightman_nature_2008}; 21: \cite{grier_discovery_2011}; 22: \cite{hernandez-garcia_x-ray_2013}; 23: \cite{bianchi_simultaneous_2012}; 24: \cite{jimenez-bailon_nuclear_2003}; 25: \cite{tzanavaris_searching_2007}; 26: \cite{maiolino_elusive_2003}; 27: \cite{machacek_chandra_2004}; 28: \cite{maccarone_low-mass_2003}; 29: \cite{gandhi_compton-thick_2015}; 30: \cite{hernandez-garcia_x-ray_2014}; 31: \cite{castangia_new_2013}; 32: \cite{bassani_three-dimensional_1999}; 33: \cite{shu_xmm-newton_2010}; 34: \cite{shu_x-ray_2012}.

\end{minipage}

\end{table*}

\twocolumn
\subsection{Notes on selected objects}\label{app:obj}

\subsubsection{3C\,264 -- NGC\,3862}
3C\,264 is a FR\,I radio source harboured in NGC\,3862 with a tentative Sy\,2/LINER classification. 
It has not been detected at subarcsecond resolution in the MIR.
The \xmmm and \chandraa data \citep{donato_obscuration_2004,evans_chandra_2006} suggest an unabsorbed X-ray luminosity that is still consistent with the expected upper limit value from the MIR--X-ray correlation but does not leave much room for the MIR flux of  an actively accreting nucleus in 3C\,264.

\subsubsection{3C\,305}
\cite{evans_xmm-newton_2008} did not detect the nucleus of 3C\,305 in \xmmm data.
\cite{hardcastle_nature_2012} detect the nucleus embedded in extended soft emission based on \chandraa observations.
While they provide an unabsorbed fit ($\log \lxi = 40.3$), they state that the source might be highly obscured.
Therefore, the source is excluded from the reliable sample.
The MIR--X-ray correlation provides an upper limit on $\log \lxi = 42.7$, not constraining the highly obscured scenario.

\subsubsection{3C\,327}
3C\,327 was observed with \chandraa \citep{evans_probing_2007}. 
We analyzed the data and find the source to be most likely Compton-thick obscured with a strong Fe K$\alpha$ line. 
Unfortunately, the S/N of the detection is insufficient for a proper modelling of the X-ray spectrum.
Therefore, we assume a similar geometry of the reflector to other well known CT AGN with an albedo of 0.022 and a reflection fraction of 0.04 (similar to NGC\,1068).
Owing to all these assumptions and uncertainties, 3C\,327 is excluded from the reliable sample.
The resulting intrinsic luminosity estimate from our X-ray analysis ($\log \lxi = 45.3$) is $\sim 1\,$\,dex higher than the expectation from the MIR--X-ray correlation, while the Compton-thickness is verified.

\subsubsection{3C\,424}
The nucleus of 3C\,424 was only faintly detected with \chandraa \citep{massaro_chandra_2012}.
Reliable modelling of the intrinsic X-ray luminosity is not possible, and thus, 3C\,424 is excluded from the reliable sample.

\subsubsection{3C\,449}
The X-ray emission of 3C\,449 is dominated by soft extended emission and no nuclear hard point source was detected during \chandraa observations \citep{balmaverde_chandra_2006}.
Thus, the lower resolution \xmmm data can not be used to analyse the AGN \citep{donato_obscuration_2004}, and 3C\,449 is excluded from the reliable sample.

\subsubsection{ESO\,253-3}
No X-ray properties are published for ESO\,253-3 but it has been observed with \swift/XRT.
The corresponding X-ray spectrum suffers from low S/N and shows heavy obscuration.
It can be fit with the combination of an absorbed and an unabsorbed power-law ($\Gamma_1 = 2.56; \Gamma_2 = 1.75; \log \lxi = 42.9; \log \nh = 23.5$). 
As noted in \cite{asmus_subarcsecond_2014}, ESO\,253-3 possibly hosts a double AGN which would be unresolved with XRT.
Owing to this, the low S/N, and thus ill constrained obscuration, we conservatively exclude ESO\,253-3 from the reliable sample.
Indeed, the MIR--X-ray correlation predicts a value of $\log \lxi$ more than an order of magnitude higher  and possibly CT obscuration.

\subsubsection{ESO\,286-19}
\cite{ptak_chandra_2003} do not find any evidence for an AGN in the relatively deep \chandraa data of the AGN candidate ESO\,286-19 (see also \citealt{iwasawa_c-goals:_2011}).
On the other hand, \cite{franceschini_xmm-newton_2003} and \cite{brightman_xmm-newton_2011-1} claim the detection of an obscured AGN in the shallower \xmmm data. 
This claim is backed up by the analysis of deep \ascaa data also finding an AGN component \citep{misaki_x-ray_1999}.
The average intrinsic X-ray luminosity found is $\log \lxi = 42.39$.
The nuclear MIR emission predicts a much higher $\log \lxi = 44.3$ and Compton-thick obscuration, $\log \nh = 24.8 \pm 0.44$, assuming the $\lxo$ from \cite{brightman_xmm-newton_2011-1}.
Heavy obscuration is also indicated by the very deep silicate absorption features in the MIR spectrum.  
However, significant star formation might be present since ESO\,286-19 is a merger system.
Clearly more data are needed to determine the dominating power source in this system.

\subsubsection{ESO\,500-34}
We did not find any 2-10\,keV X-ray data on the AGN candidate in ESO\,500-34 but it was detected at 14-195\,keV in the 54 month BAT catalogue with a luminosity of $\lxh = 42.72$ (\citealt{cusumano_palermo_2010}; but not in the 70 month).
The nuclear MIR emission upper limit provides an upper limit on the intrinsic X-ray luminosity of $\log \lxi \le 42.8$ through the MIR--X-ray correlation.
This agrees well with the expectation from the 14-195\,keV luminosity.

%\subsubsection{ESO 602-25}
%According to our knowledge the AGN candidate ESO\,602-25 has not been observed in X-rays and also remains undetected in the latest hardest X-ray all-sky surveys so far.
%The nuclear MIR emission together with the MIR--X-ray correlation predicts an intrinsic X-ray luminosity of $\log \lxi = 42.77 \pm 0.4$ for any AGN present in ESO\,602-25. 

\subsubsection{IC\,883}
IC\,883 is an infrared luminous galaxy with an uncertain AGN/starburst composite nucleus. 
For the nuclear MIR flux only an upper limit is available.
\cite{iwasawa_c-goals:_2011} present \chandraa data on this object showing a complex morphology with indication but no clear evidence of an AGN (see also discussion in \citealt{modica_multi-wavelength_2012}). 
The upper limit on the intrinsic X-ray luminosity of any AGN in IC\,883 from the MIR--X-ray correlation is $\log \lxi \le 43.9$, so leaves sufficient room for a highly obscured AGN.

\subsubsection{IC\,4518W}
\cite{de_rosa_x-ray_2008} and \cite{pereira-santaella_x-ray_2011} analyse the \xmmm data on IC\,4518W that were taken only few days apart but find more than one dex different values for $\lxo$  (and also $\lxi$).
The analysis by \cite{rodriguez_swift_2008} of \swift/XRT data yields intermediate results between the above values.
All works agree on a high obscuration ($\log \nh = 23.29$).
Owing to the inconsistency of the different measurements, IC\,4518W is excluded from the reliable sample.
Interestingly, the predictions for $\lxi$ and $\nh$ are in best agreement with \cite{rodriguez_swift_2008} and thus the average of all the above measurements.

\subsubsection{III\,Zw\,35}
\cite{gonzalez-martin_x-ray_2009} analysed the \chandraa and \xmmm data of the AGN/starburst composite III\,Zw\,35 and faintly detected the northern nucleus, which holds the AGN in this merger system.
However, the S/N is insufficient to perform a spectral analysis and the X-ray luminosity can only be estimated ($\log \lxo = 40$).
Based on the very low X-ray--\oiii ratio, \cite{gonzalez-martin_fitting_2009} then suggests that the AGN might be Compton-thick obscured.
This scenario would agree with the prediction from the MIR--X-ray correlation ($\log \lxi = 43.0 \pm 0.4$; $\log \nh = 25.2 \pm 0.5$). 
On the other hand, the spectral information in the MIR at subarcsecond scales is insufficient to exclude significant star formation contamination of the nuclear MIR luminosity estimate.

\subsubsection{IRAS\,00188-0856}
The ultra luminous infrared galaxy IRAS\,00188-0856 possibly contains an AGN/starburst composite nucleus, which was detected in the MIR at subarcsecond scales. 
It was also detected with \chandraa but with an insufficient S/N for spectral analysis \citep{teng_chandra_2005}.
The deeper \xmmm data have a better S/N and are fitted with an unabsorbed model in \cite{teng_x-quest:_2010} resulting in $\log \lxi = 41.71$, without further discussion.
The resultant very high MIR--X-ray ratio of IRAS\,00188-0856  suggests that this source is either completely star formation dominated or contains a Compton-thick AGN with an intrinsic X-ray luminosity of $\log \lxi = 44.5 \pm 0.4$.
Note that the very deep silicate absorption features in the MIR indicate heavy obscuration towards the nucleus consistent with the CT scenario.

\subsubsection{IRAS\,01003-2238}
IRAS\,01003-2238 was only detected below 5\,keV during \xmmm observations \citep{nardini_compton-thick_2011}. 
The detection with \chandraa was too faint for any extraction of AGN properties \citep{teng_chandra_2005}.
\cite{nardini_compton-thick_2011} interpret the data as IRAS\,01003-2238 hosting a Compton-thick obscured AGN. 
Conversely, it is excluded from the reliable sample.
The scenario of heavy obscuration is supported by the predictions from the MIR--X-ray correlations and the deep silicate absorption features \citep{asmus_subarcsecond_2014}.

\subsubsection{IRAS\,04103-2838}
IRAS\,04103-2838 was only very weakly detected with \chandraa \citep{teng_chandra_2005}.
More information can be extracted from \xmmm data which indicate that the AGN is CT obscured \citep{teng_xmm-newton_2008}.
Therefore, no reliable $\lxi$ estimate is possible and the source is excluded from the reliable sample.
The diagnostics based on the MIR--X-ray correlation support the CT scenario.

\subsubsection{IRAS\,05189-2524}
X-ray observations of IRAS\,05189-2524 provide very different results \citep{teng_suzaku_2009,teng_x-quest:_2010,brightman_xmm-newton_2011-1}.
In particular, \cite{teng_suzaku_2009} note that during the latest \suzakuu observations, the observed flux was a factor of 30 lower because of the AGN either turning off or a dramatic increase in obscuration. 
Owing to this, the average intrinsic X-ray luminosity of IRAS\,05189-2524 is very uncertain, and thus the object is excluded from the reliable sample.
The MIR--X-ray correlation in fact predicts a high intrinsic luminosity ($\log \lxi = 44.5 \pm 0.4$) and CT obscuration.
The most recent X-ray observations with \nustarr however suggest that the AGN is not CT obscured (Teng et al., in prep.).

\subsubsection{IRAS\,08572+3915}
IRAS\,08572+3915 is another ultra luminous infrared galaxy with an uncertain AGN/starburst composite nucleus.
While, it was detected in the mid-infrared at subarcsecond scales, no X-ray detection could be achieved with \xmmm and \suzaku.
Only in the soft X-ray band, the source is detected in \chandraa observations \citep{teng_suzaku_2009}.
Under the assumption that the nuclear MIR emission is AGN-dominated, the intrinsic X-ray luminosity is predicted to be $\log \lxi = 44.7 \pm 0.4$ with $\log \nh = 25.0 \pm 0.4$ according to the MIR--X-ray correlation.
This scenario would also be consistent with the deep silicate absorption features present in the MIR.

\subsubsection{IRAS\,11095-0238}
The ultra luminous infrared galaxy IRAS\,11095-0238 was only very weakly detected with \chandra, not allowing a detailed spectral analysis \citep{teng_x-quest:_2010}.
The observed X-ray luminosity is $\log \lxo = 41.41$ while the MIR--X-ray correlation predicts an intrinsic X-ray luminosity of $\log \lxi = 44.55 \pm 0.39$ and Compton-thick obscuration ($\log \nh = 25.24 \pm 0.35$) if a powerful AGN is indeed present in this object. 
Heavy obscuration is also indicated by the extremely deep silicate absorption features in the MIR spectrum. 
However, strong star formation even dominating the MIR can not be excluded with the current data.

\subsubsection{IRAS\,15250+3609}
IRAS\,15250+3609, an ultra luminous infrared galaxy and AGN candidate, was first detected in X-rays with \chandraa \citep{teng_chandra_2005}.
However, the detection was too faint for any further analysis. 
A better detection was achieved with later \xmmm observations, indicating that the object is Compton-thick obscured in X-rays \citep{teng_x-quest:_2010}.
Therefore, no reliable intrinsic luminosity estimate could be given.
The MIR--X-ray correlation indicates $\log \lxi = 44.36 \pm 0.37$ and also Compton-thickness ($\log \nh = 25.43 \pm 0.48$).
The MIR spectrum shows heavy obscuration through deep silicate absorption features as well.
Similar to IRAS\,11095-0238, star formation can not be excluded as main driver of the nuclear MIR emission however, and thus the results are not sufficient to proof the presence of a Compton-thick AGN in IRAS\,15250+3609.

%\subsubsection{M\,51a -- NGC\,5194}
%The Compton-thick obscuration of the low-luminosity AGN in M\,51a was first indicated in \ascaa observations which showed a low photon index and a strong Fe K$\alpha$ line \citep{terashima_asca_1998}.
%Follow-up \beppoo observations found in addition a large excess in emission at hardest X-ray energies \citep{fukazawa_excess_2001}. 
%A large equivalent width Fe K$\alpha$ line was then verified with \chandraa \citep{terashima_chandra_2001}.
%Finally, the CT scenario is consistent with the low X-ray--\oiii ratio \citep{panessa_x-ray_2006} and more recent \xmmm observations \citep{brightman_xmm-newton_2011-1}.
%The corresponding  intrinsic 2-10\,keV luminosity estimate based on the X-ray observations is $\log \lxi = 40.59$, and we assign an uncertainty of 0.6\,dex for our analysis to account for the difficulty to reliably  estimate $\lxi$ in this obscuration regime. 

\subsubsection{Mrk\,266}
Mrk\,266 is a merger system hosting possibly a double AGN, the confirmed Sy\,2 Mrk\,266SW, and the composite candidate Mrk\,266NE.
A comprehensive study of both nuclei based on \xmmm and \chandraa data is presented in \cite{mazzarella_investigation_2012} and is used here.
Only in \chandra, the emission from both nuclei is resolved and can be disentangled.
Mrk\,266SW turns out to be the apparently fainter nucleus in X-rays.
The low S/N of the detection does not allow for a detailed spectral fitting. 
The observed X-ray luminosity is $\log \lxo = 41.44$.
However, the flat spectral slope and presence of a strong Fe K$\alpha$ line indicate that Mrk\,266SW is heavily obscured. 
Thus, the X-ray properties of Mrk\,266SW are not reliable. 
The nuclear MIR luminosity predicts an intrinsic X-ray luminosity of $\log \lxi = 42.14 \pm 0.38$ and heavy but Compton-thin obscuration ($\log \nh = 23.6 \pm 0.4$).
The brighter observed X-ray spectrum of Mrk\,266NE, on the other hand, can be modelled with an absorbed power-law can a moderate column density, $\log \nh = 22.91$, resulting in an intrinsic X-ray luminosity of $\log \lxi = 41.74$.
\cite{gonzalez-martin_x-ray_2009} find a similar $\lxi$ and $\nh$ from analysing the same date despite stating a significantly higher observed 2-10\,keV flux. 
On the other hand, \cite{gonzalez-martin_fitting_2009} mention that Mrk\,266NE might be Compton-thick obscured based on the X-ray--\oiii ratio with a much higher intrinsic X-ray luminosity ($\log \lxi = 43.56$).
The predicted $\log \lxi = 42.69 \pm 0.44$ from the MIR--X-ray correlation is in between both obscuration scenarios while the predicted column density is in fact mildly CT ($\log \nh = 24.12 \pm 0.69$).
Unfortunately there is no MIR spectral information available at subarcsecond scales, which would give information about nuclear star formation and obscuration. 
However, both are strong on larger scales \citep{asmus_subarcsecond_2014}, so that we can not distinguish whether the nuclear MIR emission of Mrk\,266NE is star formation dominated, or whether it is AGN dominated with heavy obscuration.

\subsubsection{NGC\,34}
Unfortunately, only one X-ray observation has been analyzed an published so far for NGC\,34. 
\cite{guainazzi_x-ray_2005} and \cite{brightman_xmm-newton_2011-1} model the object as highly obscured (but Compton-thin), while \cite{shu_investigating_2007} claim that the obscuration is Compton-thick. 
The latter is support by the low X-ray--\oiii ratio but on the other hand no Fe K$\alpha$ line was detected. 
\cite{esquej_starburst-active_2012} argue that the nucleus of NGC\,34 is instead starburst dominated which is consistent with the MIR data (see also \citealt{asmus_subarcsecond_2014}).
Therefore, NGC\,34 is excluded from the reliable sample. 
Owing to the possible star formation contamination of the subarcsecond MIR luminosity, the prediction for $\lxi$ from  MIR--X-ray correlation is not reliable.

\subsubsection{NGC\,253}
The presence of an AGN in the nearby starburst galaxy NGC\,253 has been controversially discussed throughout the literature.
In X-rays, an obscured point source, X-1, was detected close to the nucleus with \chandraa \citep{weaver_chandra_2002}.
While it does not coincide with the compact non-thermal radio source it is within the range of possible location for the dynamical centre of NGC\,253 \citep{muller-sanchez_stellar_2010}.
Interestingly, during the deep \chandra/\nustarr monitoring in 2012, X-1 was not detected but a similarly bright source $\sim1$\,arcsec away \citep{lehmer_nustar_2013}.
This new source however, has a different X-ray spectrum and its location is incompatible with X-1.
Its properties are typical for an ultraluminous X-ray binary. 
In addition, the high energy emission detected by \nustarr does not match expectations from an AGN as source.
Combining all results, it is not clear whether X-1 is an X-ray binary or obscured AGN. 
Since no compact MIR source was detected at subarcsecond scales at the centre of NGC\,253, its AGN nature is still unclear.
The corresponding MIR upper limit puts through the MIR--X-ray correlation an upper limit on the intrinsic X-ray luminosity of $\log \lxi \le 41.52$, well compatible with the estimates from \cite{weaver_chandra_2002} and \cite{muller-sanchez_stellar_2010}.

\subsubsection{NGC\,613}
The AGN candidate in NGC\,613 has been observed with \xmmm and shows a heavily obscured but Compton-thin X-ray spectrum with an intrinsic luminosity of $\log \lxi = 41.55$. 
The predicted value from the MIR--X-ray correlation is in good agreement, as well as the predicted X-ray column density. 
Together with the compact MIR morphology at subarcsecond scales, this strongly favours the presence of an AGN in NGC\,613 in addition to the circumnuclear star formation.

%\subsubsection{NGC\,1433}
%NGC\,1433 has not been observed in X-rays so far. 
%The upper limit in the MIR provides through the MIR--X-ray correlation an expected upper limit on the X-ray luminosity of $\log \lxi \le 40.2$.

\subsubsection{NGC\,1614}
\cite{risaliti_hard_2000-1} were the first to claim the existence of an AGN in the starburst galaxy NGC\,1614 from \beppoo observations.
\cite{brightman_xmm-newton_2011-1} provide the analysis of newer \xmmm data and find a merely obscured AGN with $\log \lxi = 41.23$. 
However, \cite{herrero-illana_multi-wavelength_2014} argue using recent \chandraa data that there is no evidence for an AGN in NGC\,1614.
An AGN was also not clearly detected in the subarcsecond MIR observations.
The corresponding upper limit from the MIR--X-ray correlation is $\log \lxi \le 43.79$ and thus not constraining the presence of an AGN.

\subsubsection{NGC\,1667}
NGC\,1667 appears to be Compton-thick obscured in the \asca, \xmmm and \suzakuu observations \citep{panessa_x-ray_2006,brightman_xmm-newton_2011-1,marinucci_link_2012}.
However, no \chandraa data have been published for this low-luminosity AGN and it remains uncertain how strongly the lower resolution data of the other telescopes are affected by extended host emission.
No reliable intrinsic X-ray luminosity estimate is available for NGC\,1667 and it is thus excluded from reliable sample.
The predictions from the MIR--X-ray correlation agree better with Compton-thin obscuration but would as well be affected by host contamination of $\lxo$.

\subsubsection{NGC\,1808}
A long-term variable hard X-ray source in the starburst galaxy NGC\,1808 was first detected by \cite{awaki_asca_1996} using \asca.
\cite{jimenez-bailon_x-ray_2005} then used \chandraa and \xmmm to resolve the individual sources in this galaxy and find a central ultra luminous X-ray source or low-luminosity AGN without significant obscuration.
\cite{brightman_xmm-newton_2011-1} analysed only the \xmmm data and find a slightly more powerful and moderately obscured AGN.
In the MIR, a compact but resolved source was detected at subarcsecond scales.
As indicated by the corresponding MIR spectrum, the emission is still star formation dominated at these scales (see also \citealt{gonzalez-martin_dust_2013}).
This explains the offset position of NGC\,1808 in the MIR--X-ray plane close to pure starburst galaxies.
Conversely, the predicted intrinsic X-ray luminosity from the MIR--X-ray correlation of $\log \lxi = 41.85$ and obscuring column density $\log \nh = 24.31$  have to be regarded as upper limits for the AGN in NGC\,1808.

%\subsubsection{NGC\,3094}
%NGC\,3094 has been observed with \xmmm in 2010 but no results have been published yet.
%There are no other X-ray data available.
%From the MIR--X-ray correlation, we expect an intrinsic X-ray luminosity of $\lxi = 43.41 \pm 0.37$ if NGC\,3094 indeed harbours and AGN.
%In that case, the observed X-ray luminosity will be much lower owing to heavy obscuration indicated by the deep silicate $\sim10\um$ absorption feature \citep{roche_silicate_2007}.

\subsubsection{NGC\,2623}
The AGN/starburst composite NGC\,2623 harbours most likely a Compton-thick obscured AGN.
This is indicated by the observed X-ray spectral properties, like the extremely hard spectral index, as found from \chandraa data \citep{maiolino_elusive_2003}. 
Unfortunately, the \xmmm observations were not sufficiently deep for a detailed spectral analysis \citep{evans_off-nuclear_2008}.
The CT scenario is supported by the predictions from the MIR--X-ray correlation ($\log \lxi = 43.31 \pm 0.42$; $\log \nh = 24.72 \pm 0.48$), which are close to the intrinsic luminosity estimate from \cite{maiolino_elusive_2003} assuming one per cent reflection efficiency. 
We note however that the nuclear MIR emission might still be significantly star formation contaminated \citep{asmus_subarcsecond_2014}.
Therefore, our predictions could be overestimations.

\subsubsection{NGC\,3185}
The optically borderline AGN in NGC\,3185 has so far only been observed with \xmm. 
The corresponding data were analysed in \cite{cappi_x-ray_2006} and \cite{panessa_x-ray_2006} who claim a Compton-thick source based on the data.
It was not detected in the MIR at subarcsecond resolution but the upper limit on the intrinsic X-ray luminosity estimated from the MIR--X-ray correlation is consistent with the $\lxi$ estimate in \cite{panessa_x-ray_2006}.
From the current data no further conclusion about the AGN nature of NGC\,3185 can be drawn.

\subsubsection{NGC\,3312}
The only publication of X-ray data for NGC\,3312 is still \cite{hudaverdi_overdensity_2006} who state the global luminosity only.
Therefore, we remove NGC\,3312 from the reliable sample.

\subsubsection{NGC\,3379}
The uncertain LINER AGN in NGC\,3379 shows several comparable X-ray point sources in the nuclear region, and \cite{flohic_central_2006} identifies the most central but not brightest one as low-luminosity AGN with an observed X-ray luminosity of $\log \lxo = 37.23$.
\cite{gonzalez-martin_x-ray_2009} estimate a higher X-ray luminosity ($\log \lxo = 38.1$) from the same data but note that the low S/N makes reliable estimates difficult.
Based on the low X-ray--\oiii ratio, \cite{gonzalez-martin_fitting_2009} argue that the LLAGN in NGC\,3379 is in fact Compton-thick obscured with $\log \lxi = 39.91$.
The upper limit on the intrinsic X-ray luminosity from the MIR--X-ray correlation would be compatible with this scenario ($\log \lxi = 40.04$).

%\subsubsection{NGC\,3393}
%NGC\,3393 hosts a highly obscured AGN, which is possibly a close double AGN in fact \citep{fabbiano_close_2011}. 
%The early X-ray results were already discussed and summarised in \cite{gandhi_resolving_2009}.
%Since then, NGC\,3393 has been observed with \suzakuu \citep{miyazawa_broad-band_2009,fukazawa_fe-k_2011} and with \chandraa \citep{fabbiano_close_2011}.
%The combination of all data shows a varying X-ray column-density around the CT threshold. 
%\cite{fabbiano_close_2011} estimate the an intrinsic 2-10\,keV luminosity of $\log \lxi \approx 43$ for the brighter south-western nucleus corrected to the distance used here. 
%The north-eastern nucleus is less than a factor two fainter. 
%Owing to the complex borderline CT absorption and unclear dual nature, we assign an increased uncertainty of 0.6\,dex to $\lxi$ of NGC\,3393 during our analysis.

\subsubsection{NGC\,3486}
NGC\,3486 hosts a borderline Sy\,2/LINER nucleus with uncertain power source. 
The nucleus remaine undetected in a Chandra snapshot survey \citep{ho_detection_2001} and only appears in the much deeper \xmmm data \citep{cappi_x-ray_2006}.
While \cite{brightman_nature_2008} favours a Compton-thick obscuration scenario for the source, \cite{brightman_xmm-newton_2011-1} provide only an unobscured fit to the \xmmm data without further discussion.
The luminosity estimates range from $ 39.63 \le \log \lxi \le 41.5$ depending on the assumed model.
The subarcsecond scale MIR upper limit together with the MIR--X-ray correlation predicts $\lxi \ge 40.47$, which is inconsistent with the CT scenario if NGC\,3486 is indeed AGN powered.

\subsubsection{NGC\,3521}
\cite{grier_discovery_2011} list the observed 0.3-8\,keV flux for the uncertain LINER AGN in NGC\,3521 using the deeper of the two \chandraa observations, unfortunately without providing any further details on the X-ray analysis. 
We assume a power-law with $\Gamma = 1.7$ to calculate an $\log \lxo =  38.67$.
The upper limit on the intrinsic X-ray luminosity of any AGN in NGC\,3521 is $\log \lxi \le 40.61$ as predicted from the MIR--X-ray correlation.

\subsubsection{NGC\,3607}
As already reported in \cite{asmus_subarcsecond_2014}, the AGN candidate in NGC\,3607 has not been detected in X-rays \citep{terashima_x-ray_2002, flohic_central_2006}, nor in the sub-arcsecond MIR observations. 
\cite{gonzalez-martin_fitting_2009} argue that a Compton-thick AGN could be present indicated by the X-ray to \oiii flux ratio. 
Unfortunately, the upper limit in the MIR and the MIR--X-ray correlation are not constraining enough to favour any scenario for NGC\,3607.

\subsubsection{NGC\,3627}
NGC\,3627 contains a low-luminosity AGN candidate, which remained undetected in the first snapshot \chandraa observations \citep{ho_detection_2001}.
Furthermore, \cite{panessa_x-ray_2006} argue that the \xmmm data on this source is significantly contaminated by off-nuclear emission and thus not usable to extract information about the nucleus.
They derive an upper limit on the observed X-ray luminosity of $\log \lxo \le 38.25$. 
On the other hand, \cite{hernandez-garcia_x-ray_2013} analyse a deeper \chandraa observation from 2008 and find $\log \lxo = 39.15$.
They attribute this $\sim1\,$dex difference to the different model used but do not comment on the fact that the earlier value was also an upper limit.   
The nuclear MIR flux suggests an even higher intrinsic X-ray luminosity and heavy obscuration ($\log \lxi =40.25 \pm 0.39; \log \nh = 24.2 \pm 0.71$).
However, in \cite{asmus_subarcsecond_2014}, we point out that even the subarcsecond measurement of the nuclear MIR flux is presumably contaminated by nuclear star formation (see also \citealt{masegosa_nature_2011}).
While our results here are consistent with the presence of an AGN in NGC\,3627, it presumably is not energetically important even on nuclear scales.

\subsubsection{NGC\,3628}
\cite{flohic_central_2006} analysed the deepest of the \chandraa observations of the AGN candidate NGC\,3628 and does not detect a nuclear point source but only diffuse emission due to star formation.
They derive an upper limit on the observed X-ray luminosity of $\log \lxo = 37.1$.
Owing to the a number of relatively bright circumnuclear sources, the \xmmm data are not usable to determine the properties of the putative AGN.
The MIR--X-ray relation puts an upper limit on the intrinsic X-ray emission of any AGN to $\log \lxi = 40.67$.

\subsubsection{NGC\,3660}
Based on \ascaa observations which show an unabsorbed and variable powerful source in NGC\,3660, \cite{brightman_nature_2008} classified this objects as a true Seyfert\,2 candidate. 
This finding was supported by new \xmmm data presented in \cite{bianchi_simultaneous_2012}. 
However, in \cite{asmus_subarcsecond_2014}, we conservatively classified NGC\,3660 only as candidate AGN because of the controversial optical classifications spanning everything between H\,II region and Sy\,1.8 (see also \cite{shi_unobscured_2010}.
This controversy is likely caused by a nuclear starburst that dilutes the optical spectra. 
Its position directly on the MIR--X-ray correlation supports the AGN nature but would favour the existence of a dusty nuclear structure s well. 
However, as pointed out in \cite{asmus_subarcsecond_2014}, the nuclear MIR emission is likely contaminated or even dominated by a nuclear starburst. 
While this could mean the absence of a torus in NGC\,3660, \cite{shi_unobscured_2010} argue that broad optical emission lines are actually present and thus NGC\,3660 would not be a true Seyfert\,2 in any case.

\subsubsection{NGC\,3690 -- Arp\,299}
NGC\,3690 is a merger system with two AGN, NGC\,3690E (Arp\,299A; sometimes wrongly called IC\, 694; see NED and \citealt{yamaoka_supernova_1998}) and NGC\,3690W (Arp\,299B).
First evidence for the presence of an highly obscured AGN came from \beppoo observations \citep{della_ceca_enshrouded_2002}, while the presence of a double AGN was then later postulated on the basis of \chandraa and \xmmm observations \citep{zezas_chandra_2003,ballo_arp_2004}.
Recently, NGC\,3690 was observed with \nustarr, where ony NGC\,3690W was detected above 10\,keV and thus contributes at least 90 per cent of the observed hard X-rays \citep{ptak_focused_2015}.
Its intrinsic luminosity is estimated to be $\log \lxi = 43.26$ with an obscuring column density of $\log \nh = 24.6$.
NGC\,3690E, on the other hand, has to be much fainter or even more obscured (see also \citealt{alonso-herrero_uncovering_2013}).
The predicted X-ray properties from the MIR--X-ray correlation ($\log \lxi = 43.45 \pm 0.43$; $\log \nh = 24.46 \pm 0.49$) for NGC\,3690W match very well the direct estimates from \nustar, while those for NGC\,3690E ($\log \lxi = 43.02 \pm 0.43$; $\log \nh = 25.37 \pm 0.49$) agree with the highly CT scenario.
However, as pointed out in \cite{asmus_subarcsecond_2014}, the nuclear MIR fluxes of both nuclei might still be significantly contaminated by star formation. 
In particular, NGC\,3690E might be completely star formation dominated.

\subsubsection{NGC\,3982}
The nucleus of NGC\,3982 was detected only with a few counts above 2\,keV \citep{guainazzi_x-ray_2005,lamassa_uncovering_2011}.
It is embedded in extended host emission, which heavily contaminates lower resolution X-ray data from \ascaa \citep{panessa_x-ray_2006} and \xmmm \citep{brightman_xmm-newton_2011-1}.
Still, the nucleus might be heavily obscured \citep{ghosh_chandra_2007}. 
Therefore, $\lxi$ is highly uncertain, and we have to exclude NGC\,3982 from the reliable sample. 
The predictions from the MIR--X-ray correlation are consistent with heavy obscuration.

\subsubsection{NGC\,4303}
The case of NGC\,4303 was already discussed in \cite{asmus_mid-infrared_2011} in the context of the MIR--X-ray correlation. 
The existence of a low-luminosity AGN in this object as indicated in X-rays \citep{jimenez-bailon_nuclear_2003, tzanavaris_searching_2007}  can not be verified owing to the presence of a nuclear star cluster, which presumably dominates the subarcsecond MIR emission in this object.
We conclude that if an AGN is present in NGC\,4303 it is not energetically important even on nuclear scales.

\subsubsection{NGC\,4418}
Little information is available about the X-ray properties of the Sy\,2 candidate NGC\,4418. 
Apparently it remained undetected during \ascaa observations in 1994. 
\cite{maiolino_elusive_2003} published \chandraa data from 2003 in which the nucleus is faintly detected, insufficient for a detailed spectral analysis.
They conclude that NGC\,4418 is probably Compton-thick and infer an intrinsic 2-10\,keV luminosity of $\sim10^{41.2}\,$erg/s based on the assumption of one per cent reflection sensitivity for a non-stated object distance. 
In addition, \suzaku data were taken in 2006 but remain unpublished so far. 
The observed MIR--X-ray ratio supports the CT scenario predicting an X-ray column density of $\log \nh = 25.85 \pm 0.44$.
The intrinsic 2-10\,keV luminosity estimated from the MIR--X-ray correlation would be $\log \lxi = 43.49 \pm 0.37$ so four orders of magnitude higher than what is observed.
However, owing to the large uncertainty in the X-ray properties, the MIR--X-ray correlation can not be used to verify the AGN nature of NGC\,4418.

\subsubsection{NGC\,4438}
\cite{machacek_chandra_2004} claim the detection of an X-ray nucleus embedded into diffuse emission in the AGN candidate NGC\,4438 from \chandraa observations, which was then verified by \cite{flohic_central_2006} using the same data.
The derived observed X-ray luminosity is $\log \lxo = 39.02$.
On the other hand, \cite{satyapal_link_2005} and \cite{gonzalez-martin_x-ray_2009} claim that the nucleus is not detect again from the same data. 
However, their upper limit is consistent with the above value.
Finally, \cite{gonzalez-martin_fitting_2009} argue that the AGN might in fact be Compton-thick obscured based on the X-ray spectral slope, the low X-ray--\oiii ratio, and the high upper limit on the equivalent width of any Fe K$\alpha$ line. 
While this scenario would be consistent with the MIR--X-ray correlation putting NGC\,4438 very close to it, one has to note that relatively strong circum nuclear star formation is present in this object, which might easily also dominate the subarcsecond MIR luminosity (see also \citealt{mason_nuclear_2012}). 
Better high angular resolution MIR data are required to distinguish between AGN and starburst dominance.

\subsubsection{NGC\,4457}
The LINER AGN candidate in NGC\,4457 shows significant circumnuclear star formation in the MIR and X-rays but also a nuclear X-ray point source \citep{satyapal_link_2005}.
Its observed X-ray luminosity is $\log \lxo = 38.89$  \citep{flohic_central_2006,gonzalez-martin_x-ray_2009}, while \cite{gonzalez-martin_fitting_2009} argue that the source has $\log \lxi = 40.57$ assuming that it is Compton-thick obscured owing to the low X-ray spectral index and X-ray--\oiii ratio.
The expectations from the MIR--X-ray correlation on $\lxi$ and $\nh$ are consistent with this scenario.
However, the nuclear MIR luminosity used might still be affected or even dominated by star formation, so this result is not robust (see also \citealt{mason_nuclear_2012}).

%\subsubsection{NGC\,4472}
%See \cite{asmus_mid-infrared_2011}. 
%No new result since then on the AGN candidate.

\subsubsection{NGC\,4501}
Off-nuclear emission is contaminating lower resolution X-ray data of NGC\,4501 such as from \xmmm \citep{brightman_nature_2008}.
Furthermore, the nucleus is only barely detected at hard energies \citep{satyapal_link_2005,lamassa_uncovering_2011}.
\cite{brightman_nature_2008} note that the AGN could be Compton-thick obscured.
Owing to this uncertain nature of the X-ray properties, NGC\,4501 is 
excluded from the reliable sample.
The $\lxi$ and $\nh$ predictions from the MIR--X-ray diagnostics indicate indeed heavy (but Compton-thin) obscuration for the AGN.

\subsubsection{NGC\,4698}
According to \cite{gonzalez-martin_x-ray_2009}, the \xmmm data of NGC\,4698 are dominated by off nuclear emission (but see \citealt{cappi_x-ray_2006}).
The low X-ray--\oiii ratio based on \chandraa observations indicates that the AGN might be Compton-thick obscured according to \cite{gonzalez-martin_fitting_2009}.
Therefore, $\lxi$ is highly uncertain, and NGC\,4698 is excluded from the reliable sample.
The MIR--X-ray correlation diagnostic does not constrain the nature of the AGN in this object.

\subsubsection{NGC\,4736}
NGC\,4736 possesses many off-nuclear X-ray point sources and the identification of which belongs to the AGN is uncertain \citep{gonzalez-martin_x-ray_2009}. 
For the same reason, the \xmmm data can not be used to measure the AGN properties. 
Therefore, we exclude NGC\,4736 from the reliable sample.

%\subsubsection{NGC\,4746}
%No X-ray observations are available for NGC\,4746. 
%The MIR--X-ray correlation provides an upper limit for the intrinsic X-ray luminosity of any AGN in NGC\,4746 of $\log \lxi = 41.59$.

\subsubsection{NGC\,4785}
NGC\,4785 harbours a Sy\,2 candidate which most likely is CT in X-rays. 
A detailed analysis of the X-ray data and its relation to the high angular resolution MIR data are presented in \cite{gandhi_compton-thick_2015} and thus not repeated here.
Owing to the large uncertainty of the X-ray properties and the non-detection in the MIR, no conclusion about the AGN nature of this object can be drawn from the MIR--X-ray correlation.

\subsubsection{NGC\,5005}
Based on \chandraa and \xmmm data, \cite{gonzalez-martin_x-ray_2009} and \cite{younes_study_2011} claim the detection of an unobscured nucleus in NGC\,5005.
However, \cite{gonzalez-martin_fitting_2009} argue that the source could be Compton-thick obscured based on the low X-ray--\oiii ratio, which is also supported by a low X-ray--\oiv ratio. 
At the same time a strong nuclear starburst is present in NGC\,5005, complicating the analysis of the AGN.
Therefore, $\lxi$ remains uncertain, and the object is excluded from the reliable sample. 
Owing to the likely star formation contamination also to the subarcsecond MIR luminosity \citep{mason_nuclear_2012,asmus_subarcsecond_2014}, the MIR--X-ray correlation does not help to constrain the AGN properties.

\subsubsection{NGC\,5363}
Unfortunately, no \chandraa observations are available for the low-luminosity AGN in NGC\,5363.
The \xmmm data can be fit well with an unobscured power law \citep{gonzalez-martin_x-ray_2009}, while the low X-ray spectral index and X-ray--\oiii ratio indicate Compton-thick obscuration of the nucleus \citep{gonzalez-martin_fitting_2009}.
Thus, $\lxi$ is very uncertain, and we exclude NGC\,5363 from the reliable sample.
The diagnostics based on the MIR--X-ray correlation contradict the CT scenario (see also \citealt{mason_nuclear_2012}).

%\subsubsection{NGC\,5728}
%The X-ray properties of NGC\,5728 were already summarized in \cite{gandhi_resolving_2009}.
%Since then, \cite{winter_x-ray_2009} published another 2-10\,keV luminosity estimate  based on a Compton-thin fit to archival \chandraa data but mention that a CT reflection model better fits the data. 
%Therefore, we use the mildly Compton-thick fit to the \suzakuu data given by \cite{comastri_suzaku_2010} but assign a larger uncertainty of 0.6\,dex to $\lxi$ of NGC\,5728 during our analysis.

\subsubsection{NGC\,5813}
\cite{hernandez-garcia_x-ray_2014} analysed all archival \chandraa and \xmmm data of several epochs for the uncertain LINER AGN in NGC\,5813. 
Bright diffuse X-ray emission renders the \xmmm data unusable to determine the AGN properties. 
The \chandraa data do not show any long-term variation, and a simultaneous fit provides $\log \lxo = 39.07$.
\cite{gonzalez-martin_fitting_2009} identify NGC\,5813 as Compton-thick candidate  based on the low X-ray--\oiii ratio and high upper limit on the equivalent width of an Fe K$\alpha$ line. 
As already mentioned in \cite{asmus_mid-infrared_2011}, the upper limit on the intrinsic X-ray luminosity through the MIR--X-ray correlation of $\log \lxi \le 41.27$ does not help to distinguish the different scenarios for NGC\,5813.

\subsubsection{NGC\,5866}
While \cite{satyapal_link_2005} do not detect any  nuclear source with \chandraa in NGC\,5866, possibly harbouring an AGN, \cite{flohic_central_2006} claim the faint detection of such a source from the same data.
Although too faint for spectral analysis, the estimate the observed X-ray luminosity to be $\log \lxo = 38.42$.
\cite{gonzalez-martin_x-ray_2009} also detect a nuclear source, provide a similar observed luminosity, but classify NGC\,5866 as non-AGN from the X-ray point of view.
On the other hand, \cite{gonzalez-martin_fitting_2009} classify it as a Compton-thick AGN candidate based on the low X-ray--\oiii ratio.
The MIR--X-ray puts an upper limit on the intrinsic X-ray emission of any AGN in NGC\,5866 to $\log \lxi = 40.74$, which is compatible with all the above scenarios.

\subsubsection{NGC\,6221}
So far NGC\,6221 has not been observed at 2-10\,keV but it is detected in the 70 month BAT catalogue with a 14-195\,keV luminosity of $\log \lxh = 41.44$ \citep{baumgartner_70_2013}.
From the MIR--X-ray correlation, we expect an intrinsic 2-10\,keV luminosity of $\log \lxi = 41.21 \pm 0.38$, agreeing well with the expectation from $\lxh$. 
Without more data, we can not prove or verify the AGN nature of the nucleus of NGC\,6221 however.

\subsubsection{NGC\,6240}
NGC\,6240 is an ultra luminous infrared merger system wtih the best case of a double AGN found so far.
The latter was discovered in X-rays by \chandraa observations \citep{komossa_discovery_2003}.
Both nuclei are heavily obscured and surrounded by star formation, which makes precise luminosity estimates difficult.
However, Puccetti et al. (in prep.) utilized the new \nustarr to analyse the hardest part of the X-ray emission. 
While \nustarr can not resolve both nuclei, most of the emission should be associated with the more powerful southern nucleus, NGC\,6240S. 
They find $\log \lxi = 44.09$ and Compton-thick obscuration, which is consistent from more uncertain previous estimates based on \chandraa \citep{gonzalez-martin_x-ray_2009,gonzalez-martin_fitting_2009}.
The MIR--X-ray correlation predicts $\log \lxi = 43.27 \pm 0.37$ and only mild Compton-thickness for NGC\,6240S ($\log \nh = 23.76 \pm 0.44$).

The northern nucleus, NGC\,6240N, appears to be fainter in X-rays \citep{komossa_discovery_2003} but is presumably highly obscured as well, as indicated by the low X-ray spectral index and infrared spectra \citep{risaliti_double_2006}. 
However, no intrinsic luminosity estimate is available and we simply use the observed luminosity from the unobscured fit of \cite{komossa_discovery_2003}, $\log \lxo = 42.07$. 
Here, the MIR--X-ray correlation predicts an intrinsic luminosity of $\log \lxi = 42.53 \pm 0.4$ and at best only mild Compton-thickness for NGC\,6240N ($\log \nh = 23.44 \pm 0.47$).

Therfore the MIR--X-ray correlation suggests that a significant part of the \nustarr detected emission might be coming from the northern nucleus.
However, we can not exclude that the obscuration in both nuclei is in fact so high that the MIR is also significantly absorbed as in NGC\,4945, in which case the MIR--X-ray correlation would lose its predictive power.

\subsubsection{NGC\,6810}
\cite{strickland_new_2007} find no evidence for the presence of an AGN in NGC\,6810 based on X-ray data from \xmm and other data at other wavelengths.
\cite{brightman_xmm-newton_2011-1}, on the other hand, present an unabsorbed AGN spectral fit to the same data with an X-ray luminosity of $\log \lxi = 39.99$, unfortunately without further discussion.
The nuclear MIR luminosity predicts a mucher higher $\log \lxi = 41.71 \pm 0.39$ while assuming the observed 2-10\,keV flux from \cite{brightman_xmm-newton_2011-1} would imply a CT obscuration ($\log \nh = 24.38 \pm 0.34$).
However, the subarcsecond MIR emission used might still be affected or dominated by star formation. 
With the current data it is therefore to distinguish whether indeed a CT AGN is present or the nucleus is pure star formation.

\subsubsection{NGC\,7479}
NGC\,7479 was observed two times with \xmmm and both data sets were modelled with the emission of an absorbed (but Compthon-thin) AGN \citep{panessa_x-ray_2006,akylas_xmm-newton_2009}.
On the other hand, \cite{brightman_xmm-newton_2011-1} show that the AGN might in fact be Compton-thick obscured by using one of the data sets.
Also \cite{diamond-stanic_isotropic_2009} argue for this scenario although \oiv emission remained undetected in NGC\,7479.
On the other hand, the MIR spectrum is highly obscured \citep{asmus_subarcsecond_2014}.
Conclusively, $\lxi$ remains highly uncertain for this source, and it has to be removed from the reliable sample. 
The MIR--X-ray correlation diagnostics favour the Compton-thick scenario (see also \citealt{gonzalez-martin_dust_2013}).

%\subsubsection{NGC\,7552}
%A compact nuclear X-ray source has been detected in the center of the AGN candidate NGC\,7552 with \chandraa \citep{grier_discovery_2011} and \xmmm \citep{brightman_xmm-newton_2011-1}.
%It appears to be unobscured with a luminosity $\log \lxi = 40.17$.

 \subsubsection{NGC\,7590}
NGC\,7590 has a complex nuclear structure with bright star formation and probably a heavily obscured AGN. 
The X-ray emission of NGC\,7590 is in fact dominated by off-nuclear emission \cite{shu_xmm-newton_2010}.
Therefore, only with the most recent \chandraa data, the nucleus can possibly be isolated but remains undetected however.
\cite{shu_x-ray_2012} interprete this as further support for a Compton-thick AGN.
Unfortunately, the nucleus also remained undetected in the MIR at subarcsecond scales (see also \citealt{asmus_mid-infrared_2011}). 
The estimated upper limit on the intrinsic X-ray luminosity from the MIR--X-ray correlation is $\log \lxi \le 41.37$, which is $\sim 2.7$\,dex higher than the upper limit on the observed luminosity. 
Thus, we can not draw any conclusion on the nature of NGC\,7590.

\subsubsection{NGC\,7592W}
The AGN/starburst composite nucleus of NGC\,7592W has only been observed once so far in X-rays, namely with \chandraa \citep{wang_chandra_2010}.
The S/N of the detection is too low for a spectral fitting and only the observed flux can roughly be estimated ($\log \lxo = 40.72$).
The MIR--X-ray correlation predicts a much higher intrinsic luminosity ($\log \lxi = 43.35 \pm 0.39$) and Compton-thick obscuration ($\log \nh = 24.91 \pm 0.46$) assuming that the subarcsecond MIR emission is AGN dominated. 
This, however, remains unsure with the current minimal information on NGC\,7592W.

\subsubsection{PKS\,2158-380}
No X-ray observations of PKS\,2158-380 have been published so far.
The object was only very weakly detected with \swift/XRT and does not allow for spectral fitting analysis. 
However, a strong Fe K$\alpha$ line appears to be present and indicates heavy obscuration. 
This is consistent wit the predicted $\lxi$ and $\nh$ from the MIR--X-ray correlation diagnostics.

\subsubsection{Superatennae\,S -- IRAS\,19254-7245S}
\cite{braito_suzaku_2009} managed to detect the AGN in Superatennae for the first time at energies $> 10$\,keV using \suzaku.
These observations indicated that Superatennae\,S is mildly CT obscured. 
However, the two epochs of \nustarr observations can be well fit with an unabsorbed power-law, leading Teng et al. (in prep.) to disfavour the CT scenario for this source.
Owing to the more than two orders of magnitude differing intrinsic X-ray luminosity estimates  of the two scenarios, we exclude Superatennae\,S from the reliable sample. 
The diagnostics based on the MIR--X-ray correlation are consistent with the CT scenario but it is possible that the nuclear MIR data is star formation contaminated.

\subsubsection{UGC\,5101}
The first detection of the AGN/starburst composite nucleus of UGC\,5101 in X-rays was presented by \cite{ptak_chandra_2003} based on \chandraa and \ascaa observations.
While they state that the nucleus is too faint for spectral analysis and consistent with a pure starburst, \cite{imanishi_compact_2003} find evidence for an AGN in X-rays in the same \chandraa data combined with \xmmm data.
The latter work states that the AGN dominates the emission of UGC\,5101  and is heavily but not Compton-thick obscured.
\cite{gonzalez-martin_x-ray_2009} and \cite{brightman_xmm-newton_2011-1} come to the same conclusion based on the same data ($\log \lxi = 42.5$; $\log \nh = 23.7$).
However, \cite{gonzalez-martin_fitting_2009} assume that the nucleus is Compton-thick obscured based on the flat X-ray spectral slope and low X-ray--\oiii ratio with an intrinsic luminosity of $\log \lxi = 43.97$. 
The Compton-thick scenario would agree well with the prediction of the MIR--X-ray correlation ($\log \lxi = 44.07 \pm 0.38$; $\log \nh = 24.74 \pm 0.33$), assuming that the nuclear subarcsecond-scale MIR emission is AGN-dominated.
Unfortunately, this assumption is not verifiable with the current available MIR data. 
Therefore, it remains unclear whether the AGN in UGC\,5101 is indeed CT obscured or not.

\end{document}